\documentclass[12pt]{article}
\pdfoutput=1
\usepackage[a4paper]{geometry}
\usepackage{jheppub, amsmath,amssymb,amsfonts,amsxtra, mathrsfs, makeidx,graphics,graphicx,amsthm,epsfig, ytableau,bm,longtable,float, color,tikz,mathtools,xfrac,footnote,rotating, lscape, makecell, environ,mathtools, empheq, colortbl, xcolor,enumerate}

\usetikzlibrary{positioning}
\usetikzlibrary{chains}
\usetikzlibrary{arrows, arrows.meta ,fit,decorations.pathreplacing}
\tikzstyle{every picture}+=[remember picture]
\tikzstyle{na} = [baseline]

\usetikzlibrary{arrows, decorations.markings, calc, fadings, decorations.pathreplacing, patterns, decorations.pathmorphing, positioning}
\tikzset{>={Latex[width=1.5mm,length=1.5mm]}}

\usepackage{mdframed}

\def\node#1#2{\overset{#1}{\underset{#2}{{\color{gray} \bullet}}}}

\def\node#1#2{\overset{#1}{\underset{#2}{\circ}}}

\tikzstyle{every picture}+=[remember picture]
\tikzstyle{na} = [baseline=-.5ex]

\usepackage{bbm}
\usepackage{subfigure}

\newcommand{\eg}{\textit{e.g.}}

\newcommand{\ie}{\textit{i.e.}}

\numberwithin{equation}{section}
\newcommand{\bes}[1]{\begin{equation} \begin{split} #1\end{split} \end{equation}}

\newcommand{\be}{\begin{equation}} \newcommand{\ee}{\end{equation}}
\newcommand{\bea}{\begin{equation} \begin{aligned}} \newcommand{\eea}{\end{aligned} \end{equation}}

\def\tilde{\widetilde}

\def\hat{\widehat}

\def\bar{\overline}

\def\rt2{\sqrt{2}}

\def\tr{\mathop{\rm tr}}

\def\CA{{\cal A}}

\def\CH{{\cal H}}
\def\CI{{\cal I}}

\def\CN{{\cal N}}

\def\CU{{\cal U}}

\def\CZ{{\cal Z}}

\def\1{{\ds 1}}

\def\repa{\raise4pt\hbox{$\square$}\mkern-14mu\raise-4pt\hbox{$\square$}}
\def\repab{\overline{\raise4pt\hbox{$\square$}\mkern-14mu\raise-4pt\hbox{$\square$}\mkern-1mu}}

\def\smileface{\ensuremath{\hbox{\large$\bigcirc$}\mkern-15mu\raise-1pt\hbox{\scriptsize$\smallsmile$}%
\mkern-10mu\raise4pt\hbox{..}\mkern4mu}}
\def\frownface{\ensuremath{\hbox{\large$\bigcirc$}\mkern-15mu\raise-1pt\hbox{\scriptsize$\smallfrown$}%
\mkern-10mu\raise4pt\hbox{..}\mkern4mu}}

\newcommand{\ba}{\begin{array}}
\newcommand{\ea}{\end{array}}
\newcommand{\bi}{\begin{itemize}}
\newcommand{\ei}{\end{itemize}}
\def\vec#1{\bm{#1}}
\def\bea#1\eea{\allowdisplaybreaks \begin{align}#1\end{align}}
 \newcommand{\ben}{\begin{enumerate}}
\newcommand{\een}{\end{enumerate}}
\newcommand{\bean}{\begin{eqnarray*}}
\newcommand{\eean}{\end{eqnarray*}}
\newcommand{\eref}[1]{(\ref{#1})}

\newcommand{\PE}{\mathop{\rm PE}}

\newcommand{\tQ}{\widetilde{Q}}

\newcommand{\BC}{\mathbb{C}}

\newcommand{\BZ}{\mathbb{Z}}

\newcommand{\BU}{\mathbf{1}}

\newcommand{\comment}[1]{}
\newcommand{\diag}{\mathrm{diag}}

\newcommand{\Sym}{\mathrm{Sym}}

\definecolor{light-gray}{gray}{0.7}

\newcommand{\purple}{\color{purple}}
\newcommand{\brown}{\color{brown}}
\newcommand{\blue}{\color{blue}}

\newcommand{\red}{\color{red}}

\def\aup#1 {\overset{#1}{\uparrow} \, \overset{\tilde{#1}}{\downarrow}}

\tikzset{snake it/.style={decorate, decoration={snake, amplitude=.4mm, segment length=2mm,
                       post length=0mm,pre length=0mm}}}
                       
\tikzset{->-/.style={decoration={
  markings,
  mark=at position #1 with {\arrow{>}}},postaction={decorate}}}    
\tikzset{-<-/.style={decoration={
  markings,
  mark=at position #1 with {\arrow{<}}},postaction={decorate}}}          
  
\title{Marginal operators and supersymmetry enhancement in 3d $S$-fold SCFTs}
\author[a,b]{Emanuele Beratto,} 
\author[b,c]{Noppadol Mekareeya,}
\author[a,b]{Matteo Sacchi}
\affiliation[a]{Dipartimento di Fisica, Universit\`a di Milano-Bicocca, \\ Piazza della Scienza 3, I-20126 Milano, Italy}
\affiliation[b]{INFN, sezione di Milano-Bicocca, \\Piazza della Scienza 3, I-20126 Milano, Italy}
\affiliation[c]{Department of Physics, Faculty of Science, \\
Chulalongkorn University, Phayathai Road, \\
Pathumwan, Bangkok 10330, Thailand}
\emailAdd{emanuele.beratto@gmail.com}
\emailAdd{n.mekareeya@gmail.com}
\emailAdd{m.sacchi13@campus.unimib.it}
\abstract{The study of exactly marginal deformations of superconformal field theories is a topic that has received considerable attention due to their rich properties. We investigate the $\CN=2$ preserving exactly marginal operators of 3d $S$-fold SCFTs.  Two families of such theories are considered: one is constructed by gauging the diagonal flavour symmetry of the $T(U(2))$ and $T(U(3))$ theories, and the other by gauging the diagonal flavour symmetry of the $T^{[2,1^2]}_{[2,1^2]}(SU(4))$ theory. In both families, it is possible to turn on a Chern--Simons level for each gauge group and to couple to each theory various numbers of hypermultiplets.  The detailed analysis of the exactly marginal operators, along with the superconformal indices, allows us to determine whether supersymmetry gets enhanced in the infrared and to deduce the amount of supersymmetry of the corresponding SCFT.}
\begin{document}
\maketitle

\section{Introduction and conclusion}
The space generated by exactly marginal deformations, also known as the conformal manifold, has been a long-standing subject of study in quantum field theories.  In superconformal field theories (SCFTs), conformal manifolds have several rich structures. For example, as demonstrated in \cite{Kol:2002zt, Green:2010da, Kol:2010ub}, conformal manifolds of 4d $\CN=1$ and 3d $\CN=2$ SCFTs can be described by a symplectic quotient of the space of marginal couplings by the complexified continuous global symmetry group\footnote{See also \cite{Bianchi:2010cx} for the conformal manifold of 3d $\CN=2$ Chern--Simons--matter theories.}.  Moreover, for 4d $\CN=2$ SCFTs, as shown by several recent findings \eg~ \cite{Razamat:2019vfd, Razamat:2020gcc, Razamat:2020pra}, the study of conformal manifolds has led to a number of intriguing dualities; these include 4d $\CN=1$ weakly coupled Lagrangian descriptions of several strongly coupled 4d $\CN=2$ SCFTs.  These provide motivation for studying exactly marginal operators in the SCFTs in this paper.

The main goal is to investigate the operators associated with the $\CN=2$ preserving exactly marginal deformations\footnote{It should be noted that, for any 3d $\CN=3$ SCFT, there is no $\CN=3$ preserving marginal deformation \cite{Cordova:2016xhm}. This statement also holds for $\CN\geq 3$.} in a large class of 3d SCFTs with at least $\CN=3$ supersymmetry, known as the 3d $S$-fold theories \cite{Terashima:2011qi, Ganor:2014pha, Gang:2015wya, Gang:2018wek, Gang:2018huc, Assel:2018vtq, Garozzo:2018kra, Garozzo:2019hbf, Garozzo:2019ejm, Ganor:2019nnv}. Let us first discuss the pure $S$-fold theories. These theories can be realised on D3 branes wrapping a circle with the presence of $SL(2,\BZ)$ duality walls \cite{Gaiotto:2008ak, Gulotta:2011si, Assel:2014awa}, each of which gives rise to a local $SL(2,\BZ)$ action to the worldvolume theory of D3 branes.  For a duality wall associated with the element $J_k = -ST^k$ of $SL(2,\BZ)$, where $S$ and $T$ are the generators of $SL(2,\BZ)$ such that $S^2=-1$ and $(ST)^3=1$, the corresponding theory can be described by the gauging the diagonal $U(N)$ global symmetry of the $T(U(N))$ theory \cite{Gaiotto:2008ak} with Chern-Simons (CS) level $k$ \cite{Terashima:2011qi, Ganor:2014pha, Gang:2015wya, Gang:2018wek, Gang:2018huc, Assel:2018vtq}\footnote{We emphasise that the $S$-fold theories considered in \cite{Terashima:2011qi, Gang:2015wya, Gang:2018wek, Gang:2018huc} were constructed by gauging the diagonal $SU(N)$ global symmetry of the $T(SU(N))$ theory.  These theories were studied in the context of 3d-3d correspondence.  However, the gauge groups of the theories studied in \cite{Assel:2018vtq} were taken to be of the unitary type. Without any further hypermultiplets added to the theory, it was shown in \cite{Garozzo:2019ejm} that the index of these two families of theories are equal. In this paper, we take the gauge groups to be of the unitary type.}.  As a result of this gauging along with the presence of the CS level, the description possesses $\CN=3$ supersymmetry.  However, at the infrared (IR) fixed point, it was shown that for $k\geq 3$ supersymmetry gets enhanced to $\CN=4$ in the case of $N=2$ \cite{Gang:2018huc, Garozzo:2019ejm} and in the large $N$ limit \cite{Assel:2018vtq}.  This result can be generalised to the $S$-fold theories associated with multiple duality walls whose description can be written in terms of a `quiver diagram' with multiple $U(N)$ gauge nodes, possibly with CS levels, connected by $T(U(N))$ links \cite{Assel:2018vtq}.  In addition to the pure $S$-fold theories, we may couple hypermultiplets to $U(N)$ gauge groups in the former.  In terms of the brane configuration, this could be viewed as adding D5 and/or NS5 branes to the aforementioned brane system in the same way as described in \cite{Hanany:1996ie}.  The resulting theories were investigated in \cite{Assel:2018vtq} for vanishing CS levels, where they were dubbed the $S$-flip theories, and in \cite{Garozzo:2019ejm} for general CS levels.  Some of the latter were shown to exhibit supersymmetry enhancement (even up to $\CN=5$) and have interesting dualities that can be regarded a generalisation of 3d mirror symmetry, discovered in \cite{Intriligator:1996ex}.  We shall henceforth refer to the pure $S$-fold theories, constructed as described above, and those coupled to hypermultiplets collectively as $S$-fold theories with $T(U(N))$ building block.  

We also extend our study to cover the case of the $T^{[2,1^2]}_{[2,1^2]} (SU(4))$ building block.  The $T^{[2,1^2]}_{[2,1^2]} (SU(4))$ theory is a 3d $\CN=4$ SCFT with a $G \times G$ global symmetry, where $G= (U(2) \times U(1))/U(1) \cong U(2)$, that admits a Lagrangian description in terms of a linear quiver \cite{Gaiotto:2008ak}.  Similar to the $T(U(N))$ theory, the $T^{[2,1^2]}_{[2,1^2]} (SU(4))$ is also self-mirror.  We can form an $S$-fold theory by gauging the diagonal symmetry $G$ of $T^{[2,1^2]}_{[2,1^2]} (SU(4))$, possibly with a CS level.  As before, we may also couple hypermultiplets to the diagonal symmetry $G$.  In principle, this construction can be applied to a more general $T^{\vec \rho}_{\vec \rho}(SU(N))$ theory.  However, due to various technicalities in the computation, we restrict ourselves to $N=4$ and $\vec \rho =  [2,1^2]$.  We shall henceforth refer to these theories collectively as $S$-fold theories with the $T^{[2,1^2]}_{[2,1^2]} (SU(4))$ building block.

One of the important by-products of the detailed study of the exactly marginal operators in $S$-fold theories is that we can extract the information of the conserved currents, which include $\CN=3$ flavour currents and $\CN=3$ extra SUSY-currents.  From the latter, we can determine whether supersymmetry gets enhanced at the fixed point, and if so we can also deduce the amount of supersymmetry of the SCFT.  This heavily relies on the superconformal index \cite{Bhattacharya:2008zy,Bhattacharya:2008bja, Kim:2009wb,Imamura:2011su, Kapustin:2011jm, Dimofte:2011py, Aharony:2013dha, Aharony:2013kma} of the $S$-fold theory in question.  Let us explain this point in more detail.  It is useful to list the $\CN=2$ multiplets that can non-trivially contribute to the quantity $(1-x^2)(\text{index}-1)$ at order $x^p$ for $p \leq 2$ \cite{Razamat:2016gzx} (see also \cite{Beem:2012yn} for the 4d counterpart). In the following, we follow the notation adopted by \cite{Cordova:2016emh}.
\be \label{N2multiplets}
\begin{tabular}{|c|c|c|}
\hline
Multiplet & {\footnotesize Contribution to $(1-x^2)(\text{index}-1)$} & Type \\
\hline
$A_2 \bar{B}_1[0]^{(1/2)}_{1/2}$ & $+x^{1/2}$ & free fields  \\
$B_1 \bar{A}_2[0]^{(-1/2)}_{1/2}$ & $-x^{3/2}$ & free fields  \\
$L\bar{B}_1[0]^{(1)}_{1}$ & $+x$ & relevant operators \\
$L\bar{B}_1[0]^{(2)}_{2}$ &$+x^2$ & marginal operators\\
$A_2 \bar{A}_2[0]^{(0)}_1$ & $-x^2$ & conserved currents \\ 
\hline
\end{tabular}
\ee
It can be seen that the order $x^2$ of the index corresponds to the marginal operators minus the conserved currents.  However, since the $S$-fold theory has at least $\CN=3$ supersymmetry, we consider the contribution from $\CN=3$ multiplets to the $\CN=2$ index.  In particular, the relevant $\CN=3$ current multiplets and their decomposition to $\CN=2$ multiplets are \cite{Evtikhiev:2017heo}
 \be \label{N3multiplets}
 \scalebox{0.9}{
\begin{tabular}{|l|c|l|}
\hline
Type & $\CN=3$ multiplet & Decomposition into $\CN=2$ multiplets \\
\hline
Flavour current & $B_1[0]^{(2)}_1$ & $L\bar{B}_1[0]^{(1)}_{1} + B_1\bar{L}[0]^{(1)}_{-1}+ A_2 \bar{A}_2[0]^{(0)}_1$ \\
Extra SUSY-current &  $A_2[0]^{(0)}_1$ & $A_2 \bar{A}_2[0]^{(0)}_1 + A_1 \bar{A}_1[1]^{(0)}_{3/2}$ \\
Stress tensor & $A_1[1]^{(0)}_{3/2}$ & $A_1 \bar{A}_1[1]^{(0)}_{3/2}+ A_1 \bar{A}_1[2]^{(0)}_{2}$ \\
\hline
\end{tabular}}
\ee
where it should be noted that the multiplets $A_1 \bar{A}_1[1]^{(0)}_{3/2}$ and $A_1 \bar{A}_1[2]^{(0)}_{2}$ contribute to $(1-x^2)(\text{index}-1)$ as $+x^3$ and $-x^4$ respectively \cite[Table 2]{Razamat:2016gzx}.  From these two tables, we see that orders $x$ and $x^2$ of the index contain the following information:
\be
\scalebox{0.9}{$
\begin{split}
\text{Order $x$:} &\qquad \text{$\CN=3$ flavour currents}~; \\
\text{Order $x^2$:} &\qquad \text{($\CN=2$ preserving exactly marginal operators)} \\
& \qquad - \text{($\CN=3$ flavour currents)} - \text{($\CN=3$ extra SUSY-currents)}~.
\end{split}$}
\ee
This instructs us to study the operators of the $S$-fold theories with $R$-charge up to $2$.  Those with $R$-charge $1$ are in correspondence with the $\CN=3$ flavour currents.  Since the index of the $S$-fold theory can be computed independently using the formula that follows from localisation (see Appendix \ref{app:index}), the information of the $\CN=2$ marginal operators leads to the precise information of the $\CN=3$ extra SUSY-current and, hence, the amount of (enhanced) supersymmetry of the corresponding SCFT.   We emphasise that the detailed analysis of marginal operators in this paper has led us to obtain results on supersymmetry enhancement beyond the scope of our previous work \cite{Garozzo:2019ejm}.

It should be remarked that the problem of enumerating all marginal operators becomes more complicated as the number of the operators with $R$-charges up to $2$ increases.  This is partly due to the fact that not all gauge invariant quantities that one can possibly write down are independent from each other.  They may be subject to various relations.  Some of these relations can actually be derived from the effective superpotential of the theory.  However, as we shall see in the subsequent sections, several $S$-fold theories contain gauge invariant monopole operators and dressed monopole operators in the spectrum, whose existence is indicated by the index.  There can also be relations between these operators that cannot be obtained from the effective superpotential.  In this case, we conjecture the form of such relations based on the index and in analogue of those known in the 3d $\CN=4$ gauge theories presented in Appendix \ref{app:3dN4}.  In this regard, the $S$-fold theories with the $T^{[2,1^2]}_{[2,1^2]} (SU(4))$ building block are much more complicated than those with the $T(U(N))$ building block.  We only present preliminary results for the former theories in this paper.  It would be nice to verify the conjectures using other approaches and complete the understanding of the $S$-fold theories with the $T^{[2,1^2]}_{[2,1^2]} (SU(4))$ building block in the future.

The paper is organised as follows.  In Section \ref{sec:SfoldTUN}, we discuss $S$-fold theories with the $T(U(N))$ building block.  We briefly discuss some properties of the operators in the $T(U(N))$ theory in Section \ref{sec:TUNrev}.  The pure $S$-fold theories are studied in Section \ref{sec:zeroflvTUN} and those coupled to hypermultiplets are studied in Sections \ref{sec:nflvSfoldTUN} and \ref{sec:nflvTUNzeroCS}.  In Section \ref{sec:SfoldT211}, $S$-fold theories with the $T^{[2,1^2]}_{[2,1^2]} (SU(4))$ building block are discussed.  We briefly review  the $T^{[2,1^2]}_{[2,1^2]} (SU(4))$ theory in Section \ref{sec:T211rev}.  The pure $S$-fold theories and those coupled to hypermultiplets are considered in Sections \ref{sec:SfoldT211zeroflv} and \ref{sec:higherflv}, respectively.

\section{$S$-fold theories with the $T(U(N))$ building block} \label{sec:SfoldTUN}
In this section, we consider $S$-folds theories whose building block is the $T(U(N))$ theory.  The $T(U(N))$ theory is briefly reviewed in Section \ref{sec:TUNrev} and Appendix \ref{app:TUN}.  In the subsequent subsections, we investigate $S$-fold theories constructed by commonly gauging the Higgs and Coulomb branch symmetries of $T(U(N))$.  We may also couple such a theory to hypermultiplet matter.  In fact, several aspects of a number of such theories with $N=2$ were studied in \cite{Garozzo:2019ejm}. In this paper, we focus on the cases of $N=2$ and $N=3$ (except in subsection \ref{sec:k2n1TU2} where we discuss only the case of $N=2$) and analyse the operators with $R$-charge up to two in detail.

An important outcome of such an analysis is the precise knowledge of the $\CN=2$ preserving marginal operators, which contribute as the positive terms at order $x^2$ of the index.  Since the index can be computed using the formulae given in Appendix \ref{app:indSfoldTUN}, we know precisely the negative terms, which are the contribution of the $\CN=3$ conserved currents.  The latter consist of $\CN=3$ flavour currents and $\CN=3$ extra SUSY-currents.  The contribution of the former appear at order $x$ of the index.  Hence, in this way, we manage to extract the contribution of extra-SUSY currents and determine whether the IR fixed point of a given $S$-fold theory has enhanced supersymmetry (beyond the scope of \cite{Garozzo:2019ejm}).  If this is the case, this method also allows us to determine the amount of supersymmetry of the corresponding SCFT.

Of course, when there are many operators up to $R$-charge 2, the above analysis can get very complicated due to the relations between them.  Some relations follow from the $F$-terms and algebraic identities but there are also quantum relations, especially between monopole operators.  We establish the latter with the aid of the index and by comparison to the 3d $\CN=4$ gauge theories studied in Appendix \ref{app:3dN4}.

\subsection{The $T(U(N))$ theory} \label{sec:TUNrev}
We briefly discuss some important aspects of the $T(U(N))$ theory in Appendix \ref{app:TUN}.  The indices for $N=2,\, 3$ can be obtained from \eref{indexTUN} and the result is as follows:
\be \label{indexTUN23}
\scalebox{0.95}{$
\begin{split}
N=2: \quad  & 1+ x \left( { d^2 \chi^{SU(2)}_{[2]}(\omega) + d^{-2} \chi^{SU(2)}_{[2]}(f)}   \right) +  x^2 \Big[ d^4 \chi^{SU(2)}_{[4]}(\omega) + d^{-4} \chi^{SU(2)}_{[4]}(f)   \\
& \quad  - { \left( \chi^{SU(2)}_{[2]}(\omega) + \chi^{SU(2)}_{[2]}(f)   \right) } {\brown -1} \Big]  +\ldots \\
N\geq 3: \quad & 1+ x \left( {d^2 \chi^{SU(N)}_{[1,0,\ldots,0,1]}(\vec \omega) + d^{-2} \chi^{SU(N)}_{[1,0,\ldots,0,1]}(\vec f)}   \right) +  x^2 \Big[ d^4 \chi^{SU(N)}_{[2,0,\ldots,0,2]}(\vec \omega) \\
&+ d^4 \chi^{SU(N)}_{[0,1,0,\ldots,0,1,0]}(\vec \omega) + d^4 \chi^{SU(N)}_{[1,0,\ldots,0,1]}(\vec \omega)  +(d \rightarrow d^{-1}, {\vec \omega} \rightarrow {\vec f}) \\
&+\chi^{SU(N)}_{[1,0,\ldots,0,1]}(\vec \omega)  \chi^{SU(N)}_{[1,0,\ldots,0,1]}(\vec f)   \\
& - {\left(  \chi^{SU(N)}_{[1,0,\ldots,0,1]}(\vec \omega) +  \chi^{SU(N)}_{[1,0,\ldots,0,1]}(\vec f)  \right) } {\brown -1} \Big] +\ldots 
\end{split}$}
\ee
where the term $-1$ highlighted in brown is the contribution of the axial $U(1)_d$ symmetry, which can be identified as $U(1)_C-U(1)_H$, where $U(1)_C$ and $U(1)_H$ are the Cartan subalgebras of $SU(2)_C$ and $SU(2)_H$ of the $\CN=4$ $R$-symmetry $SU(2)_C \times SU(2)_H$. 

Since we shall make extensive use of $\CN=3$ supersymmetry in subsequent discussion, it is instructive to view the result from the perspective of the $\CN=3$ index, where $d$ is set to unity.  The terms at order $x$ are the contribution of the $\CN=3$ $SU(N) \times SU(N)$ flavour currents and these terms appear again as negative terms at order $x^2$ and the term $-1$ highlighted in brown is the contribution of the $\CN=3$ extra SUSY-current.  This is as expected since the theory has $\CN=4$ supersymmetry. We also point out the absence of the term $\chi^{SU(2)}_{[2]}(\omega)  \chi^{SU(2)}_{[2]}(f) $ at order $x^2$ for $N=2$.

The operators with $R$-charge $1$ are the Higgs and Coulomb branch moment maps of $T(U(N))$:
\be
(\mu_H)^i_j~, \qquad (\mu_C)^{i'}_{j'}~.
\ee
They are subject to the nilpotent conditions (see \cite[below (3.6)]{Gaiotto:2008ak}):
\be
\mu_H^N = \mu_C^N =0~.
\ee
These imply that all eigenvalues of $\mu_H$ and $\mu_C$ are zero and so
\be \label{trmomentmapzero}
\tr(\mu_H^p) = \tr(\mu_C^p) =0~, \qquad 1\leq p \leq N~. 
\ee
%Due to the tracelessness of $\mu_H$ and $\mu_C$, the flavour symmetry of $T(U(N))$ is $SU(N) \times SU(N)$, and $\mu_H$ and $\mu_C$ transform under the adjoint representation of each $SU(N)$.

There are two types of marginal operators, namely the pure Higgs or Coulomb branch operators and the mixed branch operators.  The pure Higgs or Coulomb branch marginal operators transform in a subrepresentation of
\bes{ \label{sym2adj}
\Sym^2 [1,0,\ldots,0,1] &= [2,0,\ldots,0,2]+ [1,0,\ldots,0,1]\\
&\quad +[0,1,0,\ldots,0,1,0]+[0,\ldots,0]
}
of each $SU(N)$.  Such operators are
\bes{ \label{muHmuHmuCmuC}
(\mu_H)^i_j (\mu_H)^k_l~,\qquad (\mu_C)^{i'}_{j'} (\mu_C)^{k'}_{l'}
}
Since $\tr(\mu_H^2)= \tr(\mu_C^2)=0$, the singlet $[0,\ldots,0]$ in \eref{sym2adj} vanishes.  Thus, each of these operators transform under the representation $[2,0,\ldots,0,2]+ [1,0,\ldots,0,1]+[0,1,0,\ldots,0,1,0]$ of each $SU(N)$ for $N\geq 3$\footnote{For $N=3$, such a representation reduces to $[2,2]+[1,1]$.}.  For $N=2$, we have stronger conditions, namely $\mu_H^2=\mu_C^2=0$, and so each operator in \eref{muHmuHmuCmuC} transforms under $[4]$ of each $SU(2)$.  

Next, we consider the marginal mixed branch operators.  In the case of $N=2$, we have
\be \label{prodHCN2}
(\mu_H)^i_j (\mu_C)^{i'}_{j'} =0~, \qquad \text{for $N=2$}
\ee  
for the following reason.  The $F$-terms with respect to the chiral multiplets $Q$ and $\tQ$ give $Q^i \varphi = 0$ and $\tQ_i \varphi = 0$, and so $(\mu_H)^i_j \varphi = Q^i \tQ_i \varphi=0$.  Since $(V_+, \varphi, V_-)$ transform in a triplet of an unbroken $SU(2)$ global symmetry, we have $(\mu_H)^i_j V_\pm =0$ and so $(\mu_H)^i_j (\mu_C )^{i'}_{j'}=0$.  This explains the absence of the term $\chi^{SU(2)}_{[2]}(\omega)  \chi^{SU(2)}_{[2]}(f) $ at order $x^2$ in the index \eref{indexTUN23} for $N=2$. Note, however, that for $N \geq 3$ the operators
\be
(\mu_H)^i_j (\mu_C)^{i'}_{j'} 
\ee
do not vanish.

\subsection{$U(N)_k$ gauge group and zero flavour} \label{sec:zeroflvTUN}
Let us commonly gauge the Higgs and Coulomb branch symmetries of the $T(U(N))$ theory and obtain the following theory
\be \label{TUNkzeroflv}
\begin{tikzpicture}[baseline]
\tikzstyle{every node}=[font=\footnotesize]
\node[draw, circle] (node1) at (0,0) {$N_k$};
\draw[red,thick] (node1) edge [out=45,in=135,loop,looseness=5, snake it]  (node1);
\node[draw=none] at (1.6,0.7) {{\red $T(U(N))$}};
\end{tikzpicture}
\ee

In the following discussion in this paragraph, we assume that $N\geq 2$ and $k\neq0$.  The case of $N=1$ is discussed in Appendix \ref{app:TU1}. The superpotential is (see also \cite{Gaiotto:2007qi} and \cite[(31)]{Gang:2018huc})
\be \label{WU2ka}
W= -\frac{k}{4 \pi} \tr( \varphi^2) + \tr \left( (\mu_C+\mu_H) \varphi \right)~,
\ee
where $\mu_H$ and $\mu_C$ are the Higgs and Coulomb branch moment maps of $T(U(N))$. For $k \neq0$, we can integrate out $\varphi$ using the $F$-terms with respect to $\varphi$:
\bes{
\varphi^a_b =\frac{2\pi}{k} (\mu_H +\mu_C)^a_b~.
}
Using \eref{trmomentmapzero}, we obtain the effective superpotential
\be \label{WeffTUNzeroflv}
W_{\text{eff}} = \frac{2\pi}{k} \tr (\mu_C \mu_H)~.
\ee 
Since $\mu_C$ and $\mu_H$ carry the axial $U(1)_d$ charges $+2$ and $-2$ respectively, the effective superpotential preserves the axial symmetry $U(1)_d$ in this case. This observation was actually pointed out in \cite{Gang:2018huc}.   In fact, from the perspective of $\CN=3$ supersymmetry, the $U(1)_d$ symmetry plays a role as the extra SUSY-current.  Indeed, $U(1)_d$ commutes with the $\CN=3$ $R$-symmetry $Spin(3)$; the former combines with the latter to become $Spin(4)$ $R$-symmetry of the enhanced $\CN=4$ supersymmetry.  We shall also see this from the perspective of the index, which is given by \eref{indexSfoldTUN}.

Let us consider the case of $|k| \geq 3$.  The indices for $N=2$ are as follows:
\bes{\label{N2kgeq30flv}
N=2,\, |k|=3: &\qquad  1+0 x- 2x^2+2(d^2+d^{-2}) x^3 +\ldots~. \\
N=2,\, |k|\geq 4: &\qquad  1+0 x- x^2+(d^2+d^{-2}) x^3 +\ldots~. \\
}
The case of $|k|=3$ was studied in \cite{Gang:2018huc}, where it was pointed out that the theory in the IR is a product to two copies of the $\CN=4$ SCFTs described by 3d $\CN=2$ $U(1)$ gauge theory with CS level $-3/2$ and one chiral multiplet with charge $+1$, whose supersymmetry gets enhanced to $\CN=4$ in the IR.  The indices for the cases of $|k| \geq 4$ were studied in \cite{Garozzo:2019ejm}, where it was pointed out that supersymmetry gets enhanced to $\CN=4$ in the IR.  For $N=3$, the indices for $|k| \geq 3$ read
\bes{ \label{N3kgeq30flv}
N=3, \, |k|\geq 3: &\qquad  1+0 x+0x^2-2 x^3 +\ldots~. \\
}

The operators up to $R$-charge $2$ are as follows.  Since $\tr \mu_{H}=\tr \mu_{C}=0$, there is no operator with $R$-charge $1$.  The $\CN=3$ flavour symmetry of this theory therefore is empty.   Let us now discuss about the marginal operators.  From \eref{trmomentmapzero}, we have
\be
\tr (\mu_{H}^2)=\tr (\mu_{C}^2) = (\tr \mu_{H})^2=(\tr \mu_{C})^2=0~.
\ee
Furthermore, for $N=2$, we also have  $\tr(\mu_H \mu_C) =0$ due to the relation \eref{prodHCN2}; thus the theory with $N=2$ and $|k| \geq 3$ has no marginal operator.  In this case, we are able to see clearly the contribution of the extra SUSY current at order $x^2$ of the indices \eref{N2kgeq30flv}, since there is no cancellation between the contribution of the conserved currents and that of the marginal operators.  For $N=2$ and $|k| \geq 4$, from the perspective of the $\CN=2$ index $-x^2$ is the contribution of the $U(1)_d$ symmetry, whereas from the perspective of the $\CN=3$ index this is the contribution of the extra SUSY-current.  Indeed, we conclude that $\CN=3$ supersymmetry gets enhanced to $\CN=4$ for $N=2$ and $|k| \geq 4$ \cite{Garozzo:2019ejm}. For $N=2$ and $|k|=3$, there are two extra SUSY conserved currents and this is due to the fact that the theory flows to a product of two $\CN=4$ SCFTs\footnote{For $N=2$ and $|k| \geq 3$, there is no relevant, no marginal and no operator with $R$-charge $3$, since $\tr (\mu_{H,C}^3)= \tr(\mu_H^2 \mu_C) =  \tr(\mu_H \mu_C^2)=0$, etc.  It is thus simple to consider the contribution of the conserved currents at order $x^3$ of index \eref{N2kgeq30flv} with $d=1$.  From Table \ref{N3multiplets} and the remark below, we see that each of the $\CN=3$ extra SUSY-current multiplet $A_2[0]^{(0)}_1$ and the $\CN=3$ stress tensor multiplet $A_1[1]^{(0)}_{3/2}$ contributes $+x^3$ to $(1-x^2)(\text{index}-1)$.  For $N=2$ and $|k|=3$, we have $(1-x^2)(\text{index}-1) = -2x^2+4x^3+\ldots$; the term $+4 x^3$ is indeed in agreement with the claim that there are two $\CN=3$ extra SUSY-currents and two $\CN=3$ stress tensors, since the theory is the product of two $\CN=4$ SCFTs. For $N=2$ and $|k| \geq 4$, we have $(1-x^2)(\text{index}-1) =-x^2+2x^3+\ldots$; the term $+2 x^3$ is indeed in agreement with the claim that there are one $\CN=3$ extra SUSY-current and one $\CN=3$ stress tensor.  Unfortunately, when there are relevant and marginal operators in the theory, the analysis of the index at order $x^p$, with $p\geq 3$, becomes very complicated.  In the rest of the paper, we focus only on the operators with $R$-charges up to $2$.} \cite{Gang:2018huc}.  

For $N = 3$, on the other hand, there is precisely one marginal operator, namely $\tr(\mu_H \mu_C)$, which cancels the contribution of the $U(1)_d$ symmetry in the index; this explains the term $(1-1)x^2= 0x^2$ in \eref{N3kgeq30flv}.  Again, we identify the $U(1)_d$ conserved current with the $\CN=3$ extra SUSY conserved current.  We thus conclude that supersymmetry also gets enhanced to $\CN=4$ for all $|k| \geq 3$.   Although we demonstrated this explicitly for $N=2$ and $N=3$, we conjecture that this statement holds for all $N\geq 2$.

For $|k|=2$, we find that the index of the theory diverges and the theory is `bad' in the sense of \cite{Gaiotto:2008ak}.  In fact, as we shall discuss in more detail in the next subsection, when $n$ flavours of fundamental hypermultiplets are coupled to the theory with $k=2$, there are gauge invariant monopole operators with $R$-charge $n/2$. In the special case of $n=0$, these monopole operators with $R$-charge $0$ render the theory `bad'.

For $|k|=1$ and $k=0$, we find that the index is equal to unity, and it is expected that the theory flows to a topological theory or an empty theory.

\subsection{$U(N)_k$ gauge group with $k\neq0$ and $n \geq 1$ flavours} \label{sec:nflvSfoldTUN}
We can add $n$ flavours of the hypermultiplets in the fundamental representation of $U(N)$ to theory \eref{TUNkzeroflv} and obtain
\be \label{TUNnflv}
\begin{tikzpicture}[baseline]
\tikzstyle{every node}=[font=\footnotesize]
\node[draw, circle] (node1) at (0,0) {$N_k$};
\node[draw, rectangle] (node2) at (2.5,0) {$n$};
\draw[red,thick] (node1) edge [out=45,in=135,loop,looseness=5, snake it]  (node1);
\draw[thick,solid] (node1) to (node2);
\node[draw=none] at (1.6,0.7) {{\red $T(U(N))$}};
\end{tikzpicture}
\ee
%Let us assume that $|k|$ is sufficiently large, say $|k| \gg 1$, in the following analysis.  
We propose that the superpotential for this theory is  
\bes{ \label{suponeflva}
W&= -\frac{k}{4 \pi} \tr( \varphi^2) + \tr \left( (\mu_C+\mu_H) \varphi \right) +\tQ^i_b\varphi^b_a  Q^a_i  \\
&=  -\frac{k}{4 \pi} \tr( \varphi^2) + \tr \left( (\mu_C+\mu_H +\mu_Q) \varphi \right)~,
}
where we define
\be \label{defmuQ}
M^i_j := \tilde{Q}^i_a  Q^a_i~, \qquad (\mu_Q)^a_b = \tilde{Q}^i_b  Q^a_i
\ee
for convenience. The following relations that follow respectively from the $F$-terms with respect to $\tilde{Q}^b_i$, $Q^i_a$ and $\varphi$:
 \bes{ \label{Ftermsmatter}
 \varphi^a_b  Q^i_a =0~, \qquad \varphi^a_b \tilde{Q}^b_i=0 ~, \qquad \varphi^a_b =\frac{2\pi}{k} (\mu_H +\mu_C +\mu_Q)^a_b~,
 }
 We discuss the consequences of these $F$-term on gauge invariant quantities in Appendix \ref{app:Fterms}.
 Using the last equality, we can integrate out $\varphi$ and obtain the effective superpotential
\be \label{Weff1a}
W_{\text{eff}} = \frac{\pi}{k} \tr \left( \mu_C+\mu_H  +\mu_Q \right)^2~.
\ee
From this effective superpotential, the $F$-terms with respect to $\tilde{Q}^b_i$, $Q^i_a$ are
\bes{
(\tr \mu_Q) Q^i_b = 0~, \qquad (\tr \mu_Q) \tQ^a_i = 0~.
}
These imply that
\bes{ \label{trmuQvanishes}
(\tr \mu_Q)\mu_Q  =0 ~, \qquad (\tr \mu_Q)^2=0~.
}
Let us define 
\be
\hat{M}^i_j = M^i_j - \frac{1}{n} (M^k_k) \delta^i_j = M^i_j - \frac{1}{n} (\tr \mu_Q) \delta^i_j ~.
\ee
From \eref{hatM2} and \eref{trmuQvanishes}, we obtain 
\bes{ \label{hatM2TUN}
(\hat{M}^2)^i_j &=  -(\mu_H+\mu_C)^b_a \tQ^i_b Q^a_j -\frac{2}{n}  \hat{M}^i_j (\tr \mu_Q) \\
(\hat{M}^2)^i_i &=  -\tr \left[  ( \mu_H+\mu_C) \mu_Q \right]~. 
}

Apart from the gauge invariant quantities discussed above, there could possibly be gauge invariant monopole operators for some special values of $k$. Subsequently, we perform case by case analyses, with the aid of the index.

\subsubsection{The case of $|k| \geq 3$, with $n \geq 1$}
For $|k| \geq 3$, with $n \geq 1$, the indices can be computed from \eref{indexSfoldTUNwnflv} and the results are as follows.
\bes{ \label{indUNknflv}
n \geq 3: &\quad 1+ x {\blue \left(  1+ \chi^{SU(n)}_{[1,0, \ldots, 0,1]} (\vec f)  \right)} + x^2 \Big[ \chi^{SU(n)}_{[2,0,\ldots,0,2]} (\vec f) + \chi^{SU(n)}_{[0,1,0,\ldots,0,1,0]} (\vec f) +\\
&\qquad + 3\chi^{SU(n)}_{[1,0,\ldots,0,1]} (\vec f) +s -  {\blue \left(  1+ \chi^{SU(n)}_{[1,0, \ldots, 0,1]} (\vec f)  \right)} \Big]+\ldots~, \\
n=2: &\quad 1+ x {\blue \left(  1+ \chi^{SU(2)}_{[2]} (f)  \right)} + x^2 \Big[ \chi^{SU(2)}_{[4]} (f) + 2\chi^{SU(2)}_{[2]} (f) +s  \\
&\quad -  {\blue \left(  1+ \chi^{SU(2)}_{[2]} (f)  \right)} \Big]+\ldots~, \\
n=1: &\quad  1+ {\blue1} x + (s' {-\blue 1})  x^2 +\ldots
}
where
\be
s = \begin{cases} 2 & \qquad N=2 \\ 3& \qquad N=3 \end{cases}
\qquad \quad
s' = \begin{cases} 1 & \qquad N=2 \\ 2& \qquad N=3 \end{cases}
\ee
Let us now analyse the operators with $R$-charge up to $2$ for $n\geq 2$. The operators with $R$-charge 1 are
\be \label{MandMhat}
M^k_k = \tr \mu_Q~, \qquad \hat{M}^i_j
\ee
and so the flavour symmetry of the theory is $U(1) \times SU(n)$.   

The marginal operators are as follows.  For $n\geq 3$, the marginal operators contributing $3\chi^{SU(n)}_{[1,0,\ldots,0,1]} (\vec f)$ to the index \eref{indUNknflv} are
\be
\hat{M}^i_j (\tr \mu_Q) = \hat{M}^i_j  (M^k_k)~, \qquad (\CA_H)^i_j ~, \qquad    (\CA_C)^i_j ~,
\ee  
where we define $(\CA_H)^i_j$ and $ (\CA_C)^i_j$ as in \eref{defCAHC}:
\bes{
 (\CA_H)^i_j &:= (\mu_H)^a_b   \tQ^i_a Q^b_j -\frac{1}{n}\tr (\mu_H \mu_Q) \delta^i_j~, \\  
 (\CA_C)^i_j &:= (\mu_C)^a_b  \tQ^i_a Q^b_j -\frac{1}{n}\tr (\mu_C \mu_Q) \delta^i_j~.
}
However, for $n=2$, we have an extra relation, namely \eref{relAHACMn2}: 
\bes{
 (\CA_H)^i_j+  (\CA_C)^i_j = - \hat{M}^i_j (\tr \mu_Q) = - \hat{M}^i_j  (M^k_k)~, \quad \text{for $n=2$}~.
}
and so there are only two independent quantities of this type.  The marginal operators that contribute to the term $\chi^{SU(n)}_{[0,1,0,\ldots,0,1,0]} (\vec f)$ are
\be \label{marg01010}
\epsilon^{i_1 i_2 \ldots i_n} \epsilon_{j_1 j_2 \ldots j_n} \hat{M}^{j_1} _{i_1} \hat{M}^{j_2}_{i_2} ~.
\ee
Those that contribute to the term $ \chi^{SU(n)}_{[2,0,\ldots,0,2]} (\vec f)$ are
\be \label{marg2002}
R^{ik}_{jl} 
\ee
which is a linear combination $ \hat{M}^i_{j} \hat{M}^{k}_{l}$ and other quantities such that any contraction between an upper index and a lower index yields zero; for example, for $n=2$, where $\hat{M}^2$ satisfies \eref{specialn2}, the marginal operators in the representation $[4]_{\vec f}$ are
\bes{ \label{marginalin4a}
R^{ik}_{jl} := \hat{M}^i_{j} \hat{M}^{k}_{l}  +\frac{1}{6} (\hat{M}^2)^p_p \delta^i_{j} \delta^k_{l} - \frac{1}{3} (\hat{M}^2)^p_p \delta^i_{l} \delta^k_{j} ~, \qquad \text{for $n=2$}~.
}
The marginal operators in the singlet of $SU(n)$ are
\be \label{margTU2singl}
%\begin{array}{ll}
\tr({\mu}_Q {\mu}_H)=  (\mu_H)^a_b   \tQ^i_a Q^b_i ~, \quad \tr({\mu}_Q {\mu}_C)=  (\mu_C)^a_b   \tQ^i_a Q^b_i ~, \quad
\tr(\mu_H \mu_C)~.
%\end{array}
\ee
Thus, there are 3 independent quantities of this type for $N \geq 3$, but for $N=2$ we have $\tr(\mu_H \mu_C)=0$ due to \eref{prodHCN2} and so we have only 2 independent quantities of this type.  Explicitly, the order $x^2$ of the indices in \eref{indUNknflv} for $n\geq 2$ can be written as
\bes{
N=2: &\quad \ldots +x^2 \Big[\ldots + 2  - {\blue \left(1+ \chi^{SU(n)}_{[1,0,\ldots,0,1]} (\vec f) \right)} \Big] + \ldots \\
N=3: &\quad \ldots +x^2 \Big[\ldots + 3  - {\blue \left(1+ \chi^{SU(n)}_{[1,0,\ldots,0,1]} (\vec f) \right)} \Big] + \ldots \\
}
We do not see the presence of an extra SUSY-current.  We thus conclude that, for $n \geq 2$, the theory has $\CN=3$ supersymmetry.   Although we have shown this explicitly for the cases of $N=2$ and $N=3$, we conjecture that this statement holds for any $N \geq 2$.  We point out that, in the above analysis, there is also a symmetry that exchange the quantities with subscripts $H$ and $C$.  We shall shortly see that this symmetry is not present, for example, in the case of $k=2$ and $n = 2$.

The above analysis also applies for $n=1$ with the following extra conditions:
\bes{
\hat{M}=\CA_H=\CA_C=0~. 
}
Moreover, due to \eref{trmuQvanishes}, $M$ is a nilpotent operator satisfying
\bes{
M^2=0~.
}
It then follows from \eref{hatM2TUN} that
\bes{ \label{reloneflv}
\tr(\mu_H \mu_Q) = - \tr(\mu_C \mu_Q) 
}
The operator with $R$-charge $1$ is
\bes{ \label{Rcharge1oneflvTUN}
M= \tr \mu_Q~.
}
The $\CN=3$ flavour symmetry of the theory is therefore $U(1)$.
For $N=2$, there is one marginal operator, given by \eref{reloneflv}, contributing  $+1x^2$ to the index.  For $N=3$, in addition to \eref{reloneflv}, there is another marginal operator $\tr(\mu_H \mu_C)$; these two marginal operators contribute $+2x^2$ to the index. We do not see the presence of an extra SUSY-current for both $N=2$ and $N=3$. Thus, we conclude that the theory has $\CN=3$ supersymmetry.

\subsubsection{The case of $k=2$ and $n\geq 2$}
For $k=2$, there are gauge invariant monopole operators with fluxes $(\pm1, 0,\ldots,0)$, denoted by $X_\pm := X_{(\pm 1,0,\ldots,0)}$, carrying $R$-charge $n/2$ and topological fugacity $\omega^{\pm 1}$.  These operators contribute contribute with the terms $(\omega+\omega^{-1})x^{\frac{n}{2}}$ to the index.   The presence of these operators is analogous to the $T(U(1))$ case presented in Appendix \ref{app:TU1}, where the mixed CS term of $T(U(1))$ after self-gluing cancels with the bare CS level $k=2$.

For $n \geq 5$, the index up to order $x^2$ is the same as the case of $|k|\ge 3$ and  $n\geq 3$ in \eref{indUNknflv}, and so we expect that the operators up to $R$-charge $2$ are as described in \eref{MandMhat}--\eref{margTU2singl}.  For $n=4$, there are additional terms $(\omega+\omega^{-1})x^2$ to the first two lines of \eref{indUNknflv}, and so  the monopole operators $X_\pm$ contribute as the additional marginal operators to those described above.  For $n=3$,  there are additional terms $(\omega+\omega^{-1})x^{\frac{3}{2}}$ to the first two lines of \eref{indUNknflv}, and so $X_\pm$ contribute as the addition operators with $R$-charge $3/2$ to those describe above.  

\subsubsection*{The case of $k=2$ and $n=2$}
Let us now analyse in detail the case of $k=2$ and $n=2$.  From \eref{indexSfoldTUNwnflv}, the indices for $N=2$ and $N=3$ read
\bes{ \label{indexk2n2}
&1+ x \left( {\blue \chi^{SU(2)}_{[2]}(\omega) + \chi^{SU(2)}_{[2]}(f)}   \right) +  x^2 \Big[\Big(2\chi^{SU(2)}_{[4]}(\omega) + \chi^{SU(2)}_{[4]}(f)   \\
& \quad +\chi^{SU(2)}_{[2]}(\omega) \chi^{SU(2)}_{[2]}(f) +\chi^{SU(2)}_{[2]}(f)+s' \Big) - {\blue \left( \chi^{SU(2)}_{[2]}(\omega) + \chi^{SU(2)}_{[2]}(f)   \right) } \Big] \\
& \quad +\ldots~,
}
where $w=\omega^2$ and we highlight the contribution of the $\CN=3$ flavour symmetry in blue and
\be
s' = \begin{cases} 1 & \qquad N=2~, \\ 2& \qquad N=3~. \end{cases}
\ee
Note that the index for $N=2$ was computed in (4.25) of \cite{Garozzo:2019ejm}.  Let us discuss about the operators with $R$-charge up to $2$.  The operators with $R$-charge 1 are
\bes{
[2]_\omega: &\qquad X_+~, \qquad M^k_k = \tr \mu_Q~, \qquad X_-  \\ 
[2]_f: &\qquad \hat{M}^i_j
} 
and so the $\CN=3$ flavour symmetry is $SU(2) \times SU(2)$.  

Let us now discuss the marginal operators, corresponding to order $x^2$ in the index.  The character $2\chi^{SU(2)}_{[4]}(\omega)$ contains the terms $2 \omega^{\pm 4}$. These imply that there are two pairs of marginal operators such that each pair carries topological charges $\pm 2$. One of such pairs is $X_\pm^2$ and we {\it propose} that the other pair consists of the monopole operators with fluxes $\pm(1,1,0,\ldots,0)$, denoted by $X_{++} := X_{(1,1,0,\ldots,0)}$ and $X_{--} := X_{(-1,-1,0,\ldots,0)}$, each carrying $R$-charge $2$.  This proposal is analogous to \eref{Rcharge1U2w1adj1flv} of the 3d $\CN=4$ $U(2)$ gauge theory with one adjoint and one fundamental hypermultiplet.  Moreover, the character $2\chi^{SU(2)}_{[4]}(\omega)$ at order $x^2$ in the index contains the terms $2 \omega^{\pm 2}$.  These imply the existence of two pairs of marginal operators such that each pair carries topological charges $\pm 1$.  One pair can be immediately identified with $ X_\pm (M^k_k)$ and we {\it propose} that the other pair corresponds to the `dressed monopole operators' $X_{\pm ;(0,1)}$, defined in a similar way to \eref{dressedmonopoles} (see \cite{Cremonesi:2013lqa}):
\be
X_{(\pm 1,0); (r,s)} = (\pm 1, 0) m _1^r m_2^s+ (0, \pm 1) m_2^r m_1^s~,
\ee
where $\mu_Q$ is diagonalised as $\diag(m_1, m_2)$\footnote{We dress the bare monopole operators with the components of $\mu_Q$ instead of those of $\varphi$, because for $k\neq 0$ we have integrated out $\varphi$ but $\mu_Q$ remains massless.}.  This proposal is analogous to \eref{Rcharge2U24flv} of the 3d $\CN=4$ $U(2)$ gauge theory with 4 flavours.  In summary, the marginal operators that correspond to the terms $2\chi^{SU(2)}_{[4]}(\omega)+\chi^{SU(2)}_{[2]}(\omega) \chi^{SU(2)}_{[2]}(f)$ are
\be \label{margk2n2}
\scalebox{0.95}{$
\begin{array}{llllll}
~[4]_\omega: &\quad X_+^2~, &\qquad X_+ (M^k_k) ~, &\qquad X_+ X_- ~, &\qquad X_- (M^k_k) ~, &\qquad  X_-^2 \\
~[4]_\omega: &\quad X_{++}~, & \qquad  X_{+;(0,1)}~, & \qquad  (\hat{M}^2)^i_i  & \qquad  X_{-;(0,1)}~, &\qquad X_{--} \\
~[2]_\omega [2]_f: & \quad X_+ \hat{M}^i_j& \qquad \hat{M}^i_j(M^k_k) & \qquad X_- \hat{M}^i_j & & 
\end{array}$}
\ee

The marginal operators that correspond to $\chi^{SU(2)}_{[4]}(f)$ are given in \eref{marginalin4a}.  Noting the relation \eref{relAHACMn2}, we see that the marginal operators corresponding to the term $\chi^{SU(2)}_{[2]}(f)$ can be taken to be either $(\CA_H)^i_j$ or $(\CA_C)^i_j$.  Picking any of these choices necessarily breaks the symmetry that exchanges $H$ and $C$.

Now let us consider the marginal operators that transform as singlets under $SU(n)$. Taking into account of \eref{hatM2TUN}, we can take two of out of three of $(\hat{M}^2)^i_i$, $\tr({\mu}_Q {\mu}_H)$ and $\tr({\mu}_Q {\mu}_C)$ to be independent operators, but since $(\hat{M}^2)^i_i$ has already been listed above, we are left with either $\tr({\mu}_Q {\mu}_H)$ or $\tr({\mu}_Q {\mu}_C)$.  Hence, for $N\geq 3$, we see that the marginal operators in the singlet of $SU(n)$ are similar to \eref{margTU2singl}, namely
\be
\text{either} \,\, \tr({\mu}_Q {\mu}_H) \,\, \text{or} \,\,\tr({\mu}_Q {\mu}_C)~, \qquad \tr(\mu_H \mu_C) ~,
\ee
and so there are {\bf two} operators of this type in this case. For $N=2$, $\tr(\mu_H \mu_C)=0$ due to \eref{prodHCN2} and so we have {\bf one} operators of this type, namely $\tr({\mu}_Q {\mu}_H)$ or $ \tr({\mu}_Q {\mu}_C)$.  We can rewrite the indices for $N=2$ and $N=3$ as
\bes{
N=2: &\quad \ldots +x^2 \Big[\ldots + 1  - {\blue \left( \chi^{SU(2)}_{[2]}(\omega) + \chi^{SU(2)}_{[2]}(f)  \right)} \Big] + \ldots \\
N=3: &\quad \ldots +x^2 \Big[\ldots + 2  - {\blue \left( \chi^{SU(2)}_{[2]}(\omega) + \chi^{SU(2)}_{[2]}(f)   \right)} \Big] + \ldots \\
}
Again we do not see the presence of the extra SUSY-current.  We thus conclude that the theory has $\CN=3$ supersymmetry.   Although we have shown this explicitly for the cases of $N=2$ and $N=3$, we conjecture that this statement holds for any $N \geq 2$.

\subsubsection{The case of $k=2$ and $n=1$} \label{sec:k2n1TU2}
Here we focus only on the case of $N=2$ and postpone the discussion of $N=3$ to future work.  This is due to the complication of the computation of the index in the latter case. For $k=2$ and $n=1$, the index for $N=2$ can be computed from \eref{indexSfoldTUNwnflv} and the result is (see also \cite[(4.20)]{Garozzo:2019ejm}):
\be \label{indexN2k2n1TUN}
\scalebox{0.9}{$
\begin{split}
&1+  x^{\frac{1}{2}} \left(\omega +\frac{1}{\omega }\right)  +x \left(2 \omega ^2+\frac{2}{\omega ^2}+2\right) \\
& \quad +x^{\frac{3}{2}} \left(2 \omega ^3+\frac{2}{\omega ^3}+2 \omega +\frac{2}{\omega }\right)+x^2 \left(3 \omega ^4+\frac{3}{\omega ^4}+2 \omega ^2+\frac{2}{\omega ^2}+1\right)  + \ldots \\
&= 1+x^{\frac{1}{2}}  \chi^{SU(2)}_{[1]}(\omega)+2 x \chi^{SU(2)}_{[2]}(\omega) +x^{\frac{3}{2}} \left[ 2 \chi^{SU(2)}_{[3]}(\omega) + \chi^{SU(2)}_{[1]}(\omega) - \chi^{SU(2)}_{[1]}(\omega) \right] \\
& \qquad +x^2 \left[3 \chi^{SU(2)}_{[4]}(\omega) +  \chi^{SU(2)}_{[2]}(\omega)-  2\chi^{SU(2)}_{[2]}(\omega)  {\brown -1}  \right] +\ldots \\
&=  \CI_{\text{free}}(x; \omega) \times \Big[1+x {\blue \chi^{SU(2)}_{[2]}(\omega)} +x^2 \left( \chi^{SU(2)}_{[4]}(\omega)  - {\blue \chi^{SU(2)}_{[2]}(\omega)} {\brown -1}  \right) +\ldots \Big] 
\end{split}$}
\ee
%{\red whereas, for $N=3$, it reads
%\be\label{indexN3k2n1TUN}
%\scalebox{0.9}{$
%\begin{split}
%&1+  x^{\frac{1}{2}} \left(\omega +\frac{1}{\omega }\right)  +x \left(2 \omega ^2+\frac{2}{\omega ^2}+2\right) \\
%& \quad +x^{\frac{3}{2}} \left(3 \omega ^3+\frac{3}{\omega ^3}+3 \omega +\frac{3}{\omega }\right)+x^2 \left( 3 \omega ^4+\frac{3}{\omega ^4}+3 \omega ^2+\frac{3}{\omega ^2}+4 \right) + \ldots \\
%&= 1+x^{\frac{1}{2}}  \chi^{SU(2)}_{[1]}(\omega)+2 x \chi^{SU(2)}_{[2]}(\omega) +3x^{\frac{3}{2}} \chi^{SU(2)}_{[3]}(\omega) \\
%& \qquad +x^2 \left[3 \chi^{SU(2)}_{[4]}(\omega) + 2  \chi^{SU(2)}_{[2]}(\omega)  +2 - 2\chi^{SU(2)}_{[2]}(\omega) {\brown -1}  \right] +\ldots \\
%&=  \CI_{\text{free}}(x; \omega) \times \Big[1+x {\blue \chi^{SU(2)}_{[2]}(\omega)} +x^{\frac{3}{2}}  \chi^{SU(2)}_{[3]}(\omega)  + x^2  \left( 2 - {\blue \chi^{SU(2)}_{[2]}(\omega)} {\brown -1}  \right) +\ldots \Big]~,
%\end{split}$}
%\ee}
%\NM{For the case of $N=3$, I have some doubts about order $x^2$.  I don't think it's reliable.  We should have $4 \chi^{SU(2)}_{[4]}(\omega)$ at this order.  Please compare this to \eref{U3w1adj1fund}.}

\noindent where the monopole operators $X_\pm$ with fluxes $(\pm 1,0,\ldots,0)$ have $R$-charge $1/2$ and decouple as a free hypermultiplet, which contribute to the index as
\bes{ \label{freeindex}
\CI_{\text{free}}(x; \omega) &=  \frac{(x^{2-\frac{1}{2}} \omega ;x^2)_\infty}{(x^{\frac{1}{2}} \omega^{-1} ;x^2)_\infty}  \frac{(x^{2-\frac{1}{2}} \omega^{-1} ;x^2)_\infty}{(x^{\frac{1}{2}} \omega ;x^2)_\infty} \\
&= 1+\chi^{SU(2)}_{[1]}(\omega) x^{\frac{1}{2}} + \chi^{SU(2)}_{[2]}(\omega) x + \Big[ \chi^{SU(2)}_{[3]}(\omega) \\
& \qquad - \chi^{SU(2)}_{[1]}(\omega)\Big] x^{\frac{3}{2}} + \left[\chi^{SU(2)}_{[4]}(\omega) - \chi^{SU(2)}_{[2]}(\omega)  -1 \right] x^{2}+ \ldots~.
}
Note also that $\hat M =0$ in this case.

We now analyse the operators up to $R$-charge $2$.  The operators with $R$-charge $1/2$ are   
\be
X_{+} ~, \qquad X_{-}~,
\ee
where $X_{\pm}$ denote monopole with fluxes $\pm(1,0,\ldots,0)$.
The operators with $R$-charge $1$ are
\be
\begin{array}{lll}
~[2]_\omega: \quad X_{++}~, &\qquad M = \tr \mu_Q ~, &\qquad X_{--}~,\\
~[2]_\omega: \quad X_+^2~, &\qquad X_+ X_-~, &\qquad X_-^2~, 
\end{array}
\ee
where $X_{++}$ and $X_{--}$ denote monopole with fluxes $\pm(1,1,0,\ldots,0)$.  
%The flavour symmetry of the theory is $SU(2) \times SU(2)$, where one of the $SU(2)$ factors is a symmetry of the free hypermultiplet and the other is the flavour symmetry of the interacting SCFT.    
Upon decoupling the free hypermultiplet containing $X_\pm$, we are left with only the first line, and indeed we see that the $\CN=3$ flavour symmetry of the interacting SCFT is $SU(2)$.

For $N=2$, the operators with $R$-charge $3/2$ are
\be \label{N2Rcharge3half}
\begin{array}{llll}
~[3]_\omega: \quad X_{+}^3~, & \quad X_{+}^2 X_{-}~, & \quad X_{+} X_{-}^2~, & \quad X_{-}^3 \\
~[3]_\omega: \quad X_{++} X_{+}~, & \quad X_{++} X_{-} ~,& \quad X_{--} X_{+}~,  & \quad X_{--} X_{-} \\
~[1]_\omega: \quad X_{+} M ~, & \quad X_{-} M
\end{array}
\ee
where, in the index \eref{indexN2k2n1TUN}, the contribution of the operators in the representation $[1]_\omega$ gets cancelled by the same terms with an opposite sign due to the contribution of the free hypermultiplets; see the first term in the last line of \eref{freeindex}.  It is worth pointing out the similarity between \eref{N2Rcharge3half} and \eref{R3halfU21adj1flv}.  Note that, upon decoupling the free hypermultiplet, we no longer have an operator at order $x^{\frac{3}{2}}$.

%{\red For $N=3$, the operators with $R$-charge $3/2$ are
%\be
%\begin{array}{llll}
%~[3]_\omega: \quad X_{+}^3~, & \quad X_{+}^2 X_{-}~, & \quad X_{+} X_{-}^2~, & \quad X_{-}^3 \\
%~[3]_\omega: \quad X_{++} X_{+}~, & \quad X_{++} X_{-} ~,& \quad X_{--} X_{+} ~,  & \quad X_{--} X_{-} \\
%~[3]_\omega: \quad X_{+++} ~, & \quad X_{+;(0,1)}~,& \quad X_{-;(0;1)} ~, & \quad X_{---} \\
%~[1]_\omega: \quad X_{+} M ~, & \quad X_{-} M
%\end{array}
%\ee
%where we remark that the first two lines and the last line are as the case of $N=2$.  Upon decoupling the free hypermultiplet, we are left with only the third line of the operators for the case of $N=3$.}

For $N=2$, the marginal operators are similar to those presented in \eref{margU21adj1flv}.  It should be noted again that, due to \eref{trmuQvanishes} and \eref{hatM2TUN}, we have
\be
M^2=0~, \qquad \tr(\mu_H \mu_Q) = -\tr(\mu_C \mu_Q) ~.
\ee
Here is the list of the marginal operators:
\be  \label{margk2n1TU2}
\scalebox{0.9}{$
\begin{array}{lllll}
~[4]_\omega: \quad X_{++}^2~, & \quad X_{++}M~, & \quad X_{++} X_{--}= \tr(\mu_H \mu_Q) = -\tr(\mu_C \mu_Q)~,& \\
& \quad X_{--} M~, & \quad X_{--}^2  & & \\
~[4]_\omega: \quad X_{+}^4~, & \quad X_{+}^3 X_{-}~, & \quad X_{+}^2 X_{-}^2~, &   \\
&\quad X_{+} X_{-}^3~, &\quad  X_{-}^4 & & \\
~[4]_\omega: \quad  X_{++} X_{+}^2~, & \quad X_{++} (X_{+} X_{-})~, & \quad X_{++} X_{-}^2 = X_{+}^2 X_{--}\\
& \quad X_{--} (X_{+} X_{-})~, & \quad X_{--} X_{-}^2 & & \\
~[2]_\omega: \quad X_{+}^2 M~, & \quad  X_{+} X_{-} M~, & \quad X_{-}^2M 
\end{array}$}
\ee
where the relation
\be
X_{++} X_{--}  = \tr(\mu_H \mu_Q) = -\tr(\mu_C \mu_Q)
\ee
is analogous to \eref{relmonU2w1flv1adj}, where the quantities on the left and right hand sides both have magnetic flux $(0,0)$. 
Note that, upon decoupling the free hypermultiplet, we are left with only the operators in the first two lines of \eref{margk2n1TU2}.
Due to the quantities as listed in \eref{margk2n1TU2}, we write the index as in \eref{indexN2k2n1TUN}, with the contribution of the extra SUSY conserved current indicated in brown. This leads us to conclude that supersymmetry gets enhanced from $\CN=3$ to $\CN=4$.  This conclusion has in fact been already discussed in \cite{Garozzo:2019ejm}.  

%{\red For $N=3$, the marginal operators are
%\be  \label{margk2n1TU3}
%\scalebox{0.9}{$
%\begin{array}{lllll}
%~[4]_\omega: \quad X_{++}^2  ~, & \quad X_{++}M~, & \quad X_{++} X_{--} ~,& \\
%& \quad X_{--} M~, & \quad X_{--}^2  & & \\
%~[4]_\omega: \quad X_{+}^4~, & \quad X_{+}^3 X_{-}~, & \quad X_{+}^2 X_{-}^2~, &   \\
%&\quad X_{+} X_{-}^3~, &\quad  X_{-}^4 & & \\
%~[4]_\omega: \quad  X_{++} X_{+}^2~, & \quad X_{++} (X_{+} X_{-})~, & \quad X_{++} X_{-}^2 = X_{+}^2 X_{--}\\
%& \quad X_{--} (X_{+} X_{-})~, & \quad X_{--} X_{-}^2 & & \\
%~[2]_\omega: \quad X_{+}^2 M~, & \quad  X_{+} X_{-} M~, & \quad X_{-}^2M \\
%~[2]_\omega: \quad X_{+++} X_- ~, & \quad M^2  ~, & \quad X_{---} X_+  \\
%2[0]_\omega: \quad \tr(\mu_H \mu_C)~, & \quad  \tr(\mu_H \mu_Q) ~\text{or}~   \tr(\mu_C \mu_Q) \\
%\end{array}$}
%\ee
%\NM{There should also be operators like $X_{+++} X_+$ in another $[4]_\omega$.}

%where $M^2$, $\tr(\mu_H \mu_Q)$ and $\tr(\mu_C \mu_Q)$ are are subject to relation \eref{relmargmargk2n1TU2}, and so we can take any two of such operators to be independent ones. Moreover, we remark that $\tr(\mu_H \mu_C)$ does not vanish (unlike the case of $N=2$). The index \eref{indexN3k2n1TUN} indicates that there are one extra SUSY conserved current.  Again, this leads us to conclude that supersymmetry gets enhanced from $\CN=3$ to $\CN=4$.
%
%From these results, we conjecture that the theory has enhanced $\CN=4$ supersymmetry for any $N \geq 2$.}

\subsubsection{Comments on the case of $k=-2$}
From \eref{indexSfoldTUNwnflv} with $\vec{n_f}=0$, we see that the index up to order $x^2$ for $k=-2$ and $n\geq 3$ is equal to that described in \eref{indUNknflv}, and the index for $k=-2$ and $n=2$ reads
\bes{ \label{indSfoldTUNkm2n2}
&1+ x \left( {\blue 1+ \chi^{SU(2)}_{[2]}(f)}   \right) +  x^2 \Big[\Big( \chi^{SU(2)}_{[4]}(f)  +2 \chi^{SU(2)}_{[2]}(f) +s' +2\Big) \\
& \qquad - {\blue \left( 1+ \chi^{SU(2)}_{[2]}(f)   \right) } \Big] +\ldots~,
}
where
\be
s' = \begin{cases} 1 & \qquad N=2~, \\ 2& \qquad N=3~. \end{cases}
\ee

Let us interpret this result.  The monopole operators $(X_+,X_-)$ and $(X_{++}, X_{--})$, discussed in the case of $k=2$, are no longer gauge invariant.  However, the above index suggests that quantities like $X_+ X_-$ and $X_{++}X_{--}$ are gauge invariant.  This can be seen from the observation that the index of the case of $k=-2$ can be obtained from that of $k=2$ by removing the terms involving $\omega^p$ with $p\neq 0$; one can compare \eref{indSfoldTUNkm2n2} for $k=-2, \, n=2$ with \eref{indexk2n2} for $k=2,\, n=2$.

We focus on the case of $k=-2$ and $n=2$.  The operators with $R$-charge $1$ are \eref{MandMhat}.  The $\CN=3$ flavour symmetry is $SU(2)\times U(1)$. The marginal operators are as follows.  Those in $[4]_f$ are \eref{marginalin4}.  Those in $2[2]_f$ are either $(\CA_H)^i_j$ or $(\CA_C)^i_j$, and $\hat{M}^i_j(M^k_k)$.  Those contribute $s'$ are either $\tr({\mu}_Q {\mu}_H)$ or $\tr({\mu}_Q {\mu}_C)$, and $\tr(\mu_H \mu_C)$, which is present for $N=3$ and absent for $N=2$.  Finally, those contribute $+2$ are  $X_+ X_-$ and $(\hat{M}^2)^i_i$.  We do not see the presence of the extra SUSY-current. 

Let us now turn to the case of $k=-2$ and $n=1$.  From \eref{indexSfoldTUNwnflv}, the index is
\bes{
N=2: \qquad & 1+{\blue 2} x +x^2 \left(4 {\blue -  2}  {\brown -1}  \right) +\ldots \\
N=3: \qquad & 1+{\blue 2} x +x^2 \left(5 {\blue -  2}  {\brown -1}  \right) +\ldots \\
}
This, again, can be obtain from \eref{indexN2k2n1TUN} with $\omega^p$ ($p \neq 0$) removed.  We propose that the operators with $R$-charge $1$ are $M=\tr \mu_Q$ and $X_+ X_-$. The $\CN=3$ flavour symmetry is therefore $U(1)^2$. The four marginal operators of the case of $N=2$ are as follows: $X_{++} X_{--} =\tr(\mu_H \mu_Q)=-\tr(\mu_C \mu_Q)$, $X_{+}^2 X_-^2$, $X_{++} X_-^2 = X_+^2 X_{--}$ and $X_+ X_- M$.  For $N=3$, there is an additional marginal operator $\tr(\mu_H \mu_C)$. There is one extra SUSY-current, indicated in brown.  Hence supersymmetry gets enhanced to $\CN=4$.

\subsubsection{The case of $k=1$ and $n=1$}
Here we focus only on the case of $N=2$ and postpone the discussion of $N=3$ to future work, due to the technicality of the index in the latter case.  From \eref{indexSfoldTUNwnflv}, the index is
\bes{ \label{indTU2k1n1}
N=2: \quad &1+ {\blue 1} x +(1 {\blue-1} {\brown -\omega q^{-1} - \omega^{-1} q}) x^2 - (\omega q^{-1} + \omega^{-1} q) x^3+ \ldots \\
%N=3: \quad & 1+x +(4 {\blue-1} +\omega^2 q^{-2} + \omega^{-2} q^2 {\brown - 2}) x^2 -  2(1+\omega q^{-1} + \omega^{-1} q) x^3 + \ldots~.
}
This case was in fact studied in \cite[Section 4.3]{Garozzo:2019ejm}.  In the following we discuss the operators with $R$-charge up to $2$.
In this case, the operator with $R$-charge $1$ corresponds to
\bes{
M = \tr \mu_Q~.
}
The $\CN=3$ flavour symmetry is therefore $U(1)$.  We indicate the contribution of the flavour current to the index \eref{indTU2k1n1} in blue.  Due to \eref{trmuQvanishes}, $M$ is a nilpotent operator satisfying $M^2=0$. From the relation \eref{hatM2TUN}, namely $M^2 = -\tr(\mu_H \mu_Q)-\tr(\mu_C \mu_Q)$, we have
\be \label{onemargTU2k1n1}
\tr(\mu_H \mu_Q)=-\tr(\mu_C \mu_Q)~.
\ee 
This is precisely the marginal operator that contributes to the positive term $+1$ at order $x^2$ in \eref{indTU2k1n1}.

As can be seen from the brown terms in \eref{indTU2k1n1}, there are two extra SUSY conserved currents. This leads to the conclusion that supersymmetry gets enhanced from $\CN=3$ to $\CN=5$ in the IR \cite{Garozzo:2019ejm}.  Note that \eref{indTU2k1n1} also satisfies all of the necessary conditions for the enhanced $\CN=5$ supersymmetry discussed in \cite{Evtikhiev:2017heo}, including that the coefficient of $x$ must be $1$.

In fact, if we view \eref{indTU2k1n1} as an $\CN=2$ index, we see that the negative terms at order $x^2$ indicate that the theory has an $SU(2) \cong Spin(3)$ global symmetry, whose character of the adjoint representation is $1+ \omega q^{-1} + \omega^{-1} q$.  This $Spin(3)$ symmetry is indeed the commutant of the $\CN=2$ $R$-symmetry $U(1) \cong Spin(2)$ in the $\CN=5$ $R$-symmetry $Spin(5)$.

\subsection{$U(N)_0$ gauge group and $n$ flavour} \label{sec:nflvTUNzeroCS}
This is also known as the $S$-flip theory \cite{Assel:2018vtq}.
%The superpotential is the same as \eref{supk0}:
%\be
%W= \tr((\mu_C +\mu_H +\mu_Q) \varphi)~.
%\ee
%The $F$-term with respect to $\varphi$ gives \eref{Ftermk0}:  
%\be
%\mu_Q= -\mu_C -\mu_H~.
%\ee
%Since $\tr(\mu_H)=\tr(\mu_C)=0$, it follows that
%\be \label{trQ0k0}
%\tr(\mu_Q) =M= 0~.
%\ee
%From \eref{hatM2}, \eref{Ftermk0} and \eref{trQ0k0}, we have
%\be \label{Mhat2k0}
%(\hat{M}^2)^i_i = \tr \left[  (\mu_H+\mu_C)^2 \right] = 2 \tr(\mu_H \mu_C)~.
%\ee
For $n \geq 3$, from \eref{indexSfoldTUNwnflv}, the indices for $N=2$ and $N=3$ read
\bes{ \label{k0ngeq3N2N3}
n\geq 3: \quad  &1+ x {\blue \left(1+ \chi^{SU(n)}_{[1,0,\ldots,0,1]} (\vec f) \right)}+ x^2 \Big[ \chi^{SU(n)}_{[2,0,\ldots,0,2]} (\vec f) +  \chi^{SU(n)}_{[0,1,0\ldots,0,1,0]} (\vec f)  \\
&\qquad + 3   \chi^{SU(n)}_{[1,0,\ldots,0,1]} (\vec f)  + s  - {\blue \left(1+ \chi^{SU(n)}_{[1,0,\ldots,0,1]} (\vec f) \right)} \Big] \\
&\qquad + \left(\omega \chi^{SU(n)}_{[0,\ldots,2]}  + \omega^{-1}  \chi^{SU(n)}_{[2,0,\ldots,0]} \right) x^{1+\frac{n}{2}} + \ldots~.
}
where we highlight the contribution of the $\CN=3$ flavour currents in blue and $s$ is defined as
\be
s = \begin{cases} 2 & \qquad N=2 \\ 3& \qquad N=3~. \end{cases}
\ee
On the other hand, for $n=2$, the indices are
\be
\scalebox{0.9}{$
\begin{split}
(N=2, n=2): \quad& 1+ x {\blue \left(1+ \chi^{SU(2)}_{[2]} (f) \right)}+ x^2 \Big[ \chi^{SU(2)}_{[4]} (f) + (2 +\omega + \omega^{-1})   \chi^{SU(2)}_{[2]} (f)  \\
&\qquad  + 2  - {\blue \left(1+ \chi^{SU(2)}_{[2]} (f) \right)} \Big] + \ldots \\
(N=3, n=2): \quad& 1+ x {\blue \left(1+ \chi^{SU(2)}_{[2]} (f) \right)}+ x^2 \Big[ \chi^{SU(2)}_{[4]} (f) + (2 +\omega + \omega^{-1})   \chi^{SU(2)}_{[2]} (f)  \\
&\qquad  + 3 + (\omega+\omega^{-1}) - {\blue \left(1+ \chi^{SU(2)}_{[2]} (f) \right)} \Big] + \ldots~.
\end{split}$}
\ee

Note that indices \eref{k0ngeq3N2N3} have the same expressions up to order $x^2$ as the cases of $n \geq 3$ of \eref{indUNknflv}, except that  there are additional terms $\omega \chi^{SU(n)}_{[0,\ldots,0,2]}  + \omega^{-1}  \chi^{SU(n)}_{[2,0,\ldots,0]} $ at order $x^{1+\frac{n}{2}}$.  The latter indicate the presence of the gauge invariant dressed monopole operators with $R$-charge $1+\frac{n}{2}$.  Note that they become marginal for $n=2$.  

For $n\geq 2$, the operators up to $R$-charge $2$ are therefore as described in \eref{MandMhat}--\eref{margTU2singl}\footnote{Curiously, for $(N=3, n=2)$, the index seems to indicate the presence of extra marginal gauge invariant monopole operators with topological fugacities $\omega^{\pm 1}$. These should be identified with the monopole operators $X_{(\pm 1,0,0)}$ with fluxes $(\pm1,0,0)$.  For $n\geq 3$, these operators (if exist) should carry $R$-charge greater than $2$ and is beyond the scope of our analysis.  It would be nice to understand these operators better in the future.}, together with the aforementioned monopole operators in the case of $n=2$.
We do not see the presence of the extra SUSY-current.  We thus conclude that the theory has $\CN=3$ supersymmetry.

\subsubsection*{The special case of $n=1$}
%For $n=1$, in addition to $\CA_H =0$, and $\CA_C=0$, we have 
%\bes{
%M=0
%}
%due to \eref{trQ0k0}.
%  From \eref{Mhat2k0}, we have
%\bes{
%\tr (\mu_H \mu_C) = 0
%}
%for any $N$.  The operator of $R$-charge $1$ is $\tr(\varphi)$, and so the flavour symmetry is $U(1)$.  Regarding the marginal operators, those in items 2 and 3 vanish, and so we have only the operators in item 1, namely those listed in \eref{opssingletsSUn}.

Let us write down explicitly the indices for $N=2$ and $N=3$, which can be computed from \eref{indexSfoldTUNwnflv}:
\bes{
1+ {\blue 1} x +(\omega q^{-2} + \omega^{-1} q^2) x^{\frac{3}{2}} + (s' {\blue -1}) x^2 + \ldots
% + (w^2 q^{-4} + w^{-2} q^4) x^{3}+ \ldots~.\\
%N=3: &\quad 1+ x +(w q^{-2} + w^{-1} q^2) x^{\frac{3}{2}} + 1 x^2  + (w^2 q^{-4} + w^{-2} q^4) x^{\frac{5}{2}}+ \ldots~.
}
where
\be
s' = \begin{cases} 1 & \qquad N=2 \\ 2& \qquad N=3 \end{cases}
\ee
Note that this is similar to the case of $n=1$ in \eref{indUNknflv}, but with additional terms $(\omega q^{-2} + \omega^{-1} q^2)$ at order $x^{\frac{3}{2}}$.
Thus, the $\CN=3$ flavour symmetry in each case is $U(1)$.  The operators up to $R$-charge $2$ are therefore as described in \eref{Rcharge1oneflvTUN} and below, together with the aforementioned dressed monopole operators.  We do not see the presence of the extra SUSY-current.  We thus conclude that the theory has $\CN=3$ supersymmetry, in agreement with the findings in \cite[Section 3.1]{Assel:2018vtq}.

\section{$S$-fold theories with the $T^{[2,1^2]}_{[2,1^2]} (SU(4))$ building block: Preliminary results} \label{sec:SfoldT211}
The purpose of this section is to generalise the previous results on the $S$-fold theories with the $T(U(N))$ building block to those with the $T^{\vec \rho}_{\vec \rho}(SU(N))$ building block.  The $T^{\vec \sigma}_{\vec \rho}(SU(N))$ theories were introduced in \cite{Gaiotto:2008ak}.  They form a large class of 3d $\CN=4$ SCFTs that admits Lagrangian descriptions in terms of linear quivers.  They can also be realised using Type IIB brane configurations, involving D3, D5 and NS5 branes \cite{Hanany:1996ie}. When $\vec \sigma=\vec \rho$ the theory is self-mirror.  We therefore can construct $S$-fold theories by commonly gauging the Higgs and Coulomb branch symmetries of $T^{\vec \rho}_{\vec \rho}(SU(N))$ in the same way as we did for $T(U(N))$.  Due to the technicality of the index computation, we shall restrict ourselves to the $T^{[2,1^2]}_{[2,1^2]} (SU(4))$ theory.

We briefly review important details of the $T^{[2,1^2]}_{[2,1^2]}(SU(4))$ theory in Section \ref{sec:T211rev} and Appendix \ref{app:T211}.  We then construct $S$-fold theories in the subsequent subsections.  As we shall see in Sections \ref{k2zeroflv} and \ref{k1zeroflv}, for some values of CS levels, the theory contains gauge invariant monopole operators in the spectrum.  Although we try to study the chiral ring of such operators using the index and other known theories as a guide, we do not have a full understanding of such a chiral ring.  The results for the $S$-fold theories of this section should therefore be taken as preliminary and we shall not study all possible cases as for the $T(U(N))$ case.  We hope to revisit this problem in the future.

\subsection{The $T^{[2,1^2]}_{[2,1^2]}(SU(4))$ theory} \label{sec:T211rev}
This theory admits the following quiver description \cite{Gaiotto:2008ak}:
\be \label{T211211}
\begin{tikzpicture}[baseline]
\tikzstyle{every node}=[font=\footnotesize]
\node[draw, circle] (node1) at (-1,0) {$1$};
\node[draw, circle] (node2) at (1,0) {$1$};
\node[draw, rectangle] (sqnode1) at (-1,-1.5) {$1$};
\node[draw, rectangle] (sqnode2) at (1,-1.5) {$2$};
\draw[draw=black,solid]  (node1) to (sqnode1);
\draw[draw=black,solid]  (node1) to (node2);
\draw[draw=black,solid]  (node2) to (sqnode2);
% \draw[transform canvas={yshift=-2.5pt}] (node1) to  node[below] {\blue $D$} node[below, xshift=-0.5cm] {$\red F$} node[xshift=-0.5cm] {$ \red \times$} (node2) ;
%  \draw[transform canvas={yshift=2.5pt}] (node1) to node[above] {\blue $U$} (node2);
%\draw[draw=black,solid, -<-=0.5]  (node1) to node[left] {\blue $L$} (sqnode);
%\draw[draw=black,solid, -<-=0.5]  (sqnode) to node[right] {\blue $R$} (node2);
\end{tikzpicture}
\hspace{2cm}
\begin{tikzpicture}[baseline]
\tikzstyle{every node}=[font=\footnotesize]
\node[draw, circle] (node1) at (-1,0) {$1$};
\node[draw, circle] (node2) at (1,0) {$1$};
\node[draw, rectangle] (sqnode1) at (-1,-1.5) {$1$};
\node[draw, rectangle] (sqnode2) at (1,-1.5) {$2$};
\draw[draw=black,solid,  ->-=0.5]  (node1) to  [bend left=20] node[above]{\blue $X$}  (node2);
\draw[draw=black,solid,  -<-=0.5]  (node1) to  [bend right=20] node[below]{\blue $\tilde{X}$}  (node2);
\draw[draw=black,solid, ->-=0.5]  (node1) to  [bend right=20] node[left]{\blue $\tilde{L}$}  (sqnode1);
\draw[draw=black,solid, -<-=0.5]  (node1) to  [bend left=20] node[right]{\blue $L$}  (sqnode1);
\draw[draw=black,solid, ->-=0.5]  (node2) to  [bend right=20] node[left]{\blue $R$}  (sqnode2);
\draw[draw=black,solid, -<-=0.5]  (node2) to  [bend left=20] node[right]{\blue $\tilde{R}$}  (sqnode2);
\draw[black,solid] (node1) edge [out=45,in=135,loop,looseness=4] node[above]{\blue $\varphi_1$}  (node1);
\draw[black,solid] (node2) edge [out=45,in=135,loop,looseness=4] node[above]{\blue $\varphi_2$}  (node2);
\node[draw=none] at (0,-2.5) {$W= L \varphi_1 \tilde{L} + \tilde{X} \varphi_1 X - X \varphi_2 \tilde{X} + \tilde{R} \varphi_2 R$};
\end{tikzpicture}
\ee
where the left diagram is in the 3d $\CN=4$ notation whereas the right diagram is in the 3d $\CN=2$ notation.

\subsection*{The Higgs and Coulomb branch moment maps}

The Higgs branch moment map can be written in terms of the chiral fields in \eref{T211211} as
\be
(\mu_H)^i_j =  \tilde{R}^i R_j~,
\ee
where $i,j=1,2$ are the indices of $U(2)_f$. The $F$-terms with respect to $\varphi_{1,2}$ imply 
\be \label{FtermmuH}
\tr(\mu_H)= \tilde{R}^i R_i= X \tilde{X} = - L \tilde{L}~.
\ee
As a result, $\mu_H$ satisfies the following conditions
\be \label{muHcond}
\mathrm{rank}(\mu_H) \leq 1~, \qquad (\mu^2_H)^i_j = (\mu_H)^i_j  \tr(\mu_H)~, \qquad \tr(\mu_H^2) = (\tr \mu_H)^2~.
\ee

The Coulomb branch moment map can be written as
\be
\mu_C = \begin{pmatrix} \varphi_1 & V_{(1;0)} \\ V_{(-1;0)} & \varphi_2 \end{pmatrix}~.
\ee
where $V_{(m; n)}$ denotes the monopole operator carrying flux $m$ under the left $U(1)$ gauge group in \eref{T211211} and flux $n$ under the right $U(1)$ gauge group in \eref{T211211}.
Since $T^{[2,1^2]}_{[2,1^2]}(SU(4))$ is self-mirror, the Coulomb branch moment map also satisfies the same conditions as \eref{muHcond} with $H$ replaced by $C$:
\be \label{muCcond}
\mathrm{rank}(\mu_C) \leq 1~, \qquad (\mu^2_C)^{i'}_{j'} = (\mu_C)^{i'}_{j'}  \tr(\mu_C)~, \qquad \tr(\mu_C^2) = (\tr \mu_C)^2~.
\ee
where $i' ,j'=1,2$ are the $U(2)_w$ indices.
It then follows that
\be
V_{(1;0)} V_{(-1;0)}= \varphi_1 \varphi_2~.
\ee
Moreover, from the superpotential \eref{T211211}, the $F$-terms with respect to $\tilde{L}$, $L$, $\tilde{R}$ and $R$ give
\be\label{FtermsR1}
L \varphi_1 =0~, \qquad \tilde{L} \varphi_1 =0~, \qquad R_i \varphi_2 = 0~,\qquad \tilde{R}^i \varphi_2 = 0~.
\ee
It then follows that
\be \label{Rphi}
0= \tilde{R}^i R_j \varphi_2  = (\mu_H)^i_j \varphi_2~, \qquad 0=-(L\tilde{L}) \varphi_1 \overset{\eref{FtermmuH}}{=} (\tr \mu_H) \varphi_1~.
\ee
We can rewrite the Coulomb branch symmetry algebra as $SU(2) \times U(1)$, where the $SU(2)$ factor corresponds to the (enhanced) topological symmetry of the left gauge group in \eref{T211211} and the $U(1)$ factor corresponds to that of the right one.  Indeed, the superpartners of the $SU(2)$ current are the triplet $(V_{(1;0)}, \varphi_1, V_{(-1;0)})$, each of which can be constructed from the fields in the vector multiplet of the left gauge group in \eref{T211211} in the UV.  On the other hand, the field $\varphi_2$ is the superpartner of the aforementioned $U(1)$ symmetry current.  Since $V_{(1;0)}$, $V_{(-1;0)}$ and $\varphi_1$ transform in the adjoint representation of an unbroken $SU(2)$ symmetry, it follows that the second equality of \eref{Rphi} has to hold also for $V_{(\pm1;0)}$, namely:
\be \label{trmuHVzero}
 (\tr \mu_H) V_{(1,0)}=0~, \qquad   (\tr \mu_H) V_{(-1,0)}=0~.
\ee
We will see that these quantum relations are also consistent with the index.

Contracting the indices $i$ and $j$ in the first equation of \eref{Rphi}, we have $(\tr\mu_H) \varphi_2=0$. Combining this result with \eref{trmuHVzero}, we obtain
\be \label{HCzero}
 (\tr \mu_H) (\mu_C)^{i'}_{j'} =0~.
\ee
Using mirror symmetry and the fact that the theory is self-mirror, we also have
\be \label{CHzero}
 (\tr \mu_C) (\mu_H)^{i}_{j} =0~.
\ee
Contracting the indices $i$ and $j$ we obtain\footnote{This result can also be obtained by contracting the indices $i$ and $j$ in the first equation of \eref{Rphi} and then summing it with the second equation in \eref{Rphi}, where we have used the fact that $\tr \mu_C= \varphi_1+\varphi_2$.}
\be \label{trHtrCzero}
(\tr \mu_H)( \tr \mu_C)=0~.
\ee

\subsection*{The relevant and marginal operators}
The index of the $T^{[2,1^2]}_{[2,1^2]}(SU(4))$ theory can be written as (see Appendix \ref{app:T211} for more details)
\bes{
%&I_{\eref{T211211}} (\{ (b u, bu^{-1}) , \vec{0} \} , \{ (q h, q h^{-1}) , \vec{0} \}, d) \\
& 1+ x \left[d^{2} \left( 1+ \chi^{SU(2)}_{[2]} (u) \right) +d^{-2} \left( 1+ \chi^{SU(2)}_{[2]} (h) \right)  \right] \\
& \quad +  x^{\frac{3}{2}} \left[d^{3} (b+b^{-1}) \chi^{SU(2)}_{[1]} (u) +d^{-3} (q+q^{-1}) \chi^{SU(2)}_{[1]} (h)  \right] \\
& \quad + x^2 \Big[ d^{4} \left(  1+ \chi^{SU(2)}_{[2]} (u) + \chi^{SU(2)}_{[4]} (u)  \right) + d^{-4} \left(  1+ \chi^{SU(2)}_{[2]} (h) + \chi^{SU(2)}_{[4]} (h)  \right)  \\
& \qquad  \quad  + \chi^{SU(2)}_{[2]} (u) \chi^{SU(2)}_{[2]} (h) -  \left( \chi^{SU(2)}_{[2]} (h)+1 \right) -  \left( \chi^{SU(2)}_{[2]} (u)+1\right) -1 \Big] \\
& \quad + \ldots~.
}
Let us analyse the operators that contribute to the index up to order $x^2$.  It is convenient to split the Higgs and Coulomb branch moment maps into the trace and the traceless part, where the latter is denoted by
\be
(\hat{\mu}_{H,C})^i_j :=({\mu}_{H,C})^i_j- \frac{1}{2} (\tr \mu_{H,C}) \delta^i_j~.
\ee
Since the rank of $\mu_{H,C}$ is at most one, we have
\be
\tr(\hat{\mu}_{H,C}^2) = \frac{1}{2}( \tr \mu_{H,C})^2~.
\ee
The coefficient of order $x$ of the index corresponds to the following operators:
\bes{
d^{2} \left( 1+ \chi^{SU(2)}_{[2]} (u) \right):   &\qquad\qquad  \tr(\mu_C)~, \quad (\hat{\mu}_C)^{i'}_{j'} \\
d^{-2} \left( 1+ \chi^{SU(2)}_{[2]} (h) \right):  &\qquad\qquad  \tr(\mu_H)~, \quad (\hat{\mu}_H)^i_j
}
The coefficient of order $x^{\frac{3}{2}}$ of the index corresponds to the following operators:
\bes{ \label{opsthreehalf}
d^{3} b\chi^{SU(2)}_{[1]} (u): & \qquad \qquad \CU^{i'} := (V_{(1,1)},  \, V_{(0,1)})^{i'} \\
d^{3} b^{-1}\chi^{SU(2)}_{[1]} (u): & \qquad \qquad  \tilde{\CU}_{i'} := (V_{(-1,-1)},  \, V_{(0,-1)})_{i'} \\
d^{-3} q \chi^{SU(2)}_{[1]} (h): & \qquad \qquad \CH^i:=  \tilde{R}^i  \tilde{X} \tilde{L}  \\
d^{-3} q^{-1} \chi^{SU(2)}_{[1]} (h): & \qquad \qquad \tilde{\CH}_i:= L X R_i~.
}
The terms at order $x^{2}$ with positive sign correspond to the following marginal operators:
\bes{
d^{4},\, d^{-4}: &\qquad  \tr (\hat{\mu}^2_{C}) = \frac{1}{2} (\tr \mu_{C})^2 ~,\,\,  \tr (\hat{\mu}^2_{H}) = \frac{1}{2} (\tr \mu_{H})^2  \\
d^{4} \chi^{SU(2)}_{[2]} (u), \, d^{-4} \chi^{SU(2)}_{[2]} (f): &\qquad (\hat{\mu}_{C})^{i'}_{j'} (\tr\mu_C)~, \, \, (\hat{\mu}_{H})^i_j (\tr\mu_H) \\
d^{4} \chi^{SU(2)}_{[4]} (u), \, d^{-4} \chi^{SU(2)}_{[4]} (f): &\qquad (\hat{\mu}_{C})^{i'}_{j'} (\hat{\mu}_{C})^{k'}_{l'}~, \, \,  (\hat{\mu}_{H})^{i}_{j} (\hat{\mu}_{H})^{k}_{l} \\
\chi^{SU(2)}_{[2]} (u) \chi^{SU(2)}_{[2]} (h) : &\qquad   (\hat{\mu}_{C})^{i'}_{j'}  (\hat{\mu}_{H})^{i}_{j}
}
The terms with minus sign confirms that the theory indeed has a $U(1)_b \times SU(2)_u \times U(1)_q  \times SU(2)_h \times U(1)_d$ global symmetry, as expected.
Note that the terms $+d^0 \chi^{SU(2)}_{[2]} (u)$, $+d^0 \chi^{SU(2)}_{[2]} (h)$ and $+d^0 \chi^{SU(2)}_{[0]} (u)\chi^{SU(2)}_{[0]} (h)$ do not appear at order $x^2$.  The absence of such terms confirms the relations \eref{HCzero}, \eref{CHzero}, \eref{trHtrCzero}, and thus also \eref{trmuHVzero}.

\subsection{$U(2)_k$ gauge group with zero flavour} \label{sec:SfoldT211zeroflv}
We consider the following theory
\be \label{U2kzeroflv}
\begin{tikzpicture}[baseline]
\tikzstyle{every node}=[font=\footnotesize]
\node[draw, circle] (node1) at (0,0) {$2_k$};
\draw[red,thick] (node1) edge [out=45,in=135,loop,looseness=5, snake it]  (node1);
\node[draw=none] at (1.8,0.7) {{\red $T^{[2,1^2]}_{[2,1^2]}(SU(4))$}};
\end{tikzpicture}
\ee
The superpotential for \eref{U2kzeroflv} can be written as \cite{Gaiotto:2007qi, Gang:2018huc}
\be \label{WU2k}
W= -\frac{k}{4 \pi} \tr( \varphi^2) + \tr \left( (\mu_C+\mu_H) \varphi \right)
\ee
where $\varphi$ is a complex scalar in the vector multiplet of the $U(2)$ gauge group, and $\mu_C$ and $\mu_H$ are the Coulomb branch and Higgs branch moment maps of the $T^{[2,1^2]}_{[2,1^2]}(SU(4))$ SCFT.  

Let us assume in the following analysis that $k\neq 0$.  We can integrate out $\varphi$. The $F$-terms with respect to $\varphi$ give
\be \label{relphiCH}
\varphi = \frac{2 \pi}{k} (\mu_C+\mu_H)~.
\ee
Substituting this back to \eref{WU2k}, we obtain the effective superpotential after integrating out $\varphi$ to be
\bes{ \label{effW}
W_{\text{eff}} &= \frac{\pi}{k}  \tr (\mu_C+\mu_H)^2 \\
&=   \frac{\pi}{k}  \left[ \tr(\mu_C^2) + \tr(\mu_H^2) +2 \tr(\mu_C \mu_H)  \right] \\
&=  \frac{\pi}{k}  \left[ (\tr\mu_C)^2 + (\tr\mu_H)^2 +2 \tr(\mu_C \mu_H)  \right]~.
}
where in the last line we have used \eref{muHcond} and \eref{muCcond}.  It should be noted that, on the contrary to the effective superpotential \eref{WeffTUNzeroflv} of the $S$-fold theory with the $T(U(N))$ building block, the $U(1)_d$ axial symmetry is broken in this case\footnote{Recall that under the $U(1)_d$ symmetry, $\mu_C$ carries charge $+2$ and $\mu_H$ carries charge $-2$.}.  The index of this theory is given by \eref{indexU2knoflv}.

%\paragraph{The index.} The index of this theory is
%\bes{ \label{indexU2knoflv}
%&\CI_{\eref{U2kzeroflv}}(k; \{w, n \}) \\
%&= \frac{1}{2} \oint \frac{d z_1}{2 \pi z_1} \oint \frac{d z_2}{2 \pi z_2} \sum_{m_1, m_2 \in \BZ} w^{m_1+m_2} z_1^{k m_1+n} z_2^{k m_2+n}  Z^{U(2)}_{\text{vec}}( \{\vec z, \vec m \}) \times \\ 
%&  \qquad I_{\eref{T211211}} \left(\{ (z_1, z_2),(m_1,m_2) \} , \{ (z_1^{-1}, z_2^{-1}),(-m_1,-m_2)\}, d=1  \right)~,
%}
%where the contribution from the $U(2)$ vector multiplet is given by
%\be
% Z^{U(2)}_{\text{vec}}( \{\vec z, \vec m \})  =x^{|m_1-m_2|}+ x^{-|m_1-m_2|}+(-1)^{1+m_1-m_2} (z_1 z_2^{-1} +z_1^{-1} z_2)~,
%\ee
%and the $U(1)_d$ symmetry is broken as can be see from the effective superpotential \eref{effW}, so its fugacity $d$ is set to $1$ in the above expression.

\subsubsection{The case of $|k| \geq 3$} \label{sec:kgeq3noflv}
Evaluating \eref{indexU2knoflv}, we obtain the indices for $|k| \geq 3$:
\be \label{indkgeq3}
\CI_{\eref{U2kzeroflv}}(|k|\geq 3; \{w, n=0 \})  = 1 + 2 x +0x^2+0x^3 +\ldots~.
\ee
where, for each $k$ such that $|k| \geq 3$, the indices differ at order of $x$ greater than $3$.  For example, 
\bes{
k=3: &\qquad 1+ 2x -2(w+ w^{-1})x^{\frac{7}{2}} + 5x^4+\ldots \\
k \leq -3, \, k\geq 4: &\qquad 1+ 2x + 5 x^4 + \ldots~.
}

The coefficient of $x$ indicates that the theory has a $U(1) \times U(1)$ global symmetry. Due to \eref{relphiCH}, we can write $\varphi$ in terms of $\mu_H$ and $\mu_C$.  As a result, there are only two independent operators with $R$-charge $1$, namely
\be \label{Rcharge1}
\tr(\mu_H)~, \qquad \tr(\mu_C)~,
\ee 
corresponding to the term $2x$ in the index. 

Let us now consider the marginal operators. Taking into account of \eref{relphiCH}, \eref{muHcond} and \eref{muCcond}, we can rewrite any marginal operators in terms of a linear combination of the following quantities:  $(\tr \mu_H)^2$, $(\tr \mu_C)^2$, $(\tr \mu_H)(\tr \mu_C)$ and $\tr(\mu_H \mu_C)$.
%\footnote{For example, $\tr(\varphi \mu_H) = \frac{2 \pi}{k} \left[ \tr (\mu^2_H) + \tr(\mu_H \mu_C) \right] \overset{\eref{muHcond}}{=} \frac{2 \pi}{k} \left[ (\tr \mu_H)^2 + \tr(\mu_H \mu_C) \right]$.}  
However, this set of quantities can be reduced further.  Due to \eref{trHtrCzero}, we have $(\tr \mu_H)( \tr \mu_C)=0$.   %Moreover, $\tr(\mu_H \mu_C)$ is not independent from the others for the following reason.  
%Since $\tr(\mu_H +\mu_C)^2$ appears as the effective superpotential \eref{effW}, it should be set to zero in the space of holomorphic functions over the supersymmetric vacua.  Since $\tr(\mu_H +\mu_C)^2 = (\tr \mu_H^2) +( \tr \mu_C^2) +2 \tr(\mu_H \mu_C)= (\tr \mu_H)^2 +( \tr \mu_C)^2 +2 \tr(\mu_H \mu_C)$ by \eref{muHcond} and \eref{muCcond}, it follows that $\tr(\mu_H \mu_C)$ can be written as a linear combination of $(\tr \mu_H)^2$  and $( \tr \mu_C)^2 $ on such a space.   
Hence, there are three independent marginal operators, which can be taken as
\be \label{marginal}
(\tr \mu_H)^2~, \qquad (\tr \mu_C)^2~, \qquad  \tr(\mu_H \mu_C)~.
\ee
%where we remark that $\tr (\mu_{H,C}^2) =  (\tr \mu_{H,C})^2$ due to \eref{muHcond} and \eref{muCcond}.
%As we have discussed in \eref{muHcond} and \eref{muCcond}, $(\tr \mu_{H,C})^2=\tr (\mu_{H,C}^2)$.  Hence, $\tr (\mu_{H,C}^2)$ are not independent operators from what listed above. Next, we show that $(\tr \mu_H)(\tr \mu_C)$ and $\tr(\mu_H \mu_C)$ are not independent from each other.  To do so, we recall that $\mu_H$ and $\mu_C$ has rank at most one, so each of them can be written as a product of two vectors.  It follows that they satisfy the following identity
%\bes{
%(\tr \mu_H)(\tr \mu_C) - \tr(\mu_H \mu_C) &= \frac{1}{2} \left[ \left[ \tr(\mu_H+\mu_C) \right]^2 - \tr(\mu_H+\mu_C)^2  \right] \\
%&\overset{\eref{relphiCH}}{=} \frac{k^2}{8 \pi^2} \left[ (\tr \varphi)^2 - \tr(\varphi^2)\right]~.
%}
%Since $\varphi$ is integrated out, it is fixed to a particular value.  Thus, the expectation values of $(\tr \mu_H)(\tr \mu_C)$ and $ \tr(\mu_H \mu_C)$ must differ by a fixed number.

%Moreover, since $\mu_H$ and $\mu_C$ can be viewed as $2 \times 2$ matrices, they satisfy the following identity:
%\be
%\left[(\tr \mu_H)(\tr \mu_C) - \tr(\mu_H \mu_C) \right ] \mathbf{1}_{2 \times 2} = \mu_H (\tr \mu_C) + \mu_C (\tr \mu_H)  - (\mu_H \mu_C + \mu_C \mu_H )~,
%\ee
%where $\mathbf{1}_{2 \times 2}$ is the $2 \times 2$ identity matrix.  It follows that $\tr(\mu_H \mu_C)$ is not independent from $(\tr \mu_H)(\tr \mu_C)$, since the former can be  can be constructed from a linear combination of the latter and other products of fields with lower $R$-charges. 

Since the coefficient of $x^2$ in the index is equal to the number of marginal operators minus conserved currents and we have $0 x^2$ in \eref{indkgeq3}, it follows that there are three conserved currents that precisely cancel the contribution of the three marginal operators in \eref{marginal}.  Two of the conserved currents are identified with the $U(1)^2$ flavour currents, as can be seen from order $x$ of the index, and the other one is the extra SUSY current.  We thus conclude that $\CN=3$ supersymmetry of theory \eref{U2kzeroflv}, with $k\geq 3$, is enhanced to $\CN=4$ in the IR.

Finally, let us point out that there is a symmetry that exchanges $\mu_H$ and $\mu_C$ for $|k| \geq 3$.  As we shall discuss shortly, this symmetry is absent for $k=2$ and $k=1$.

\subsubsection{The case of $k =2$} \label{k2zeroflv}
Evaluating \eref{indexU2knoflv}, we obtain the index for $k=2$:
\bes{ \label{indexk2}
&\CI_{\eref{U2kzeroflv}}(k= 2; \{w, n=0 \}) \\
&= 1+x \left(2+w +\frac{1}{w }\right) + x^2 \left(w ^2+\frac{1}{w ^2}\right)+\ldots \\
&= 1+x \left[{\blue 1+\chi^{SU(2)}_{[2]}(\omega)} \right]+x^2 \left[ \left( 1+\chi^{SU(2)}_{[4]}(\omega) \right) - {\blue \left( 1+\chi^{SU(2)}_{[2]}(\omega)  \right)}\right] + \ldots
}
where $\omega = w^{\frac{1}{2}}$ and we highlighted the contribution of the flavour currents in blue.

From the coefficient of $x$ we see that, in addition to the operators listed in \eref{Rcharge1}, there are two gauge invariant monopole operators with $R$-charge $1$ that carry topological fugacities $w^{\pm1} =\omega^{\pm 2}$, denoted by $X_\pm$. Hence the operators with $R$-charge $1$ are 
\be \label{Rcharge1k2}
1, \, \omega^2, \, 1, \, \omega^{-2}: \qquad \qquad \tr(\mu_H)~, \qquad X_+~, \qquad \tr(\mu_C) ~, \qquad X_-~,
\ee
The $\CN=3$ flavour symmetry of the SCFT is therefore $SU(2) \times U(1)$. Note that this is larger than that of the case of $|k| \geq 3$, due to the presence of the monopole operators $X_\pm$ with $R$-charge $1$.  Here we have to make a choice whether to take $(X_+, (\tr\mu_H), X_-)$ or $(X_+, (\tr\mu_C), X_-)$ to be a moment map of $SU(2)$.  Whatever choice we make will break the symmetry that exchanges $\mu_H$ and $\mu_C$.  This is a crucial difference between this case and the previously discussed case of $|k| \geq 3$.  For definiteness, let us take the triplet $(X_+, (\tr\mu_C), X_-)$ to be the moment map of $SU(2)$ and $(\tr \mu_H)$ to be that of $U(1)$.\footnote{Of course, we may as well take $(X_+, (\tr\mu_H), X_-)$ to be the moment map of $SU(2)$ and $(\tr \mu_C)$ to be that of $U(1)$.  The arguments below still hold with $H$ interchanged with $C$. \label{footnoteHC}}

Let us consider the marginal operators.  These contribute to order $x^2$ in the index.  We first examine those in the representation $[4]$ of $SU(2)$, whose character is $\chi^{SU(2)}_{[4]}(\omega)= \omega^4+\omega^2+1+\omega^{-2}+\omega^{-4}$.    The terms $\omega^{\pm4}$ should correspond to the operators $X_\pm^2$.  In contrast to \eref{margk2n2}, there is no gauge invariant monopole operator $X_{++}$ or $X_{--}$ with fluxes $(1,1)$ or $(-1,-1)$.  It is also interesting to contrast to the 3d $\CN=4$ $U(2)$ gauge theory with four flavours of fundamental hypermultiplets \eref{Rcharge2U24flv} that there are no operators in the representation $[2]$ of $SU(2)$ in this case.  The candidates for the operators that carry fugacities $\omega^{\pm 2}$ are $X_{\pm} (\tr \mu_H)$ and $X_{\pm} (\tr \mu_C)$.  However, we argue that the former vanishes for the following reason.   Since from \eref{trHtrCzero} we have $(\tr \mu_H)(\tr \mu_C) =0$, we must also have 
\be
(\tr\mu_H) X_{\pm}=0~,
\ee
due to the fact that $(X_+, (\tr\mu_C), X_-)$ transform in the adjoint representation of an unbroken $SU(2)$ flavour symmetry.  We thus conclude that the marginal operators carrying fugacities $\omega^{\pm 2}$ are $X_\pm (\tr \mu_C)$.  At this point, it is also worth comment that, in contrast to \eref{Rcharge2U24flv} and to \eref{margk2n2}, there is no dressed monopole operators, like $X_{(\pm1,0); (0,1)}$, in this case.   Finally, let us discuss the marginal operators that carry zero charge under the topological symmetry, \ie~ those with $\omega^0$.  
The candidates for these are as follows: 
\be \label{possiblemargneutral}
(\tr \mu_H)^2~, \quad (\tr \mu_C)^2~, \quad \tr(\mu_H \mu_C)~, \quad X_+ X_-~.
\ee
From order $x^2$ in the index, there are the following possibilities:
\ben
\item Among \eref{possiblemargneutral}, there are only two independent operators.  There is no $\CN=3$ extra SUSY-current.
\item Among \eref{possiblemargneutral}, there are three independent operators. There is one $\CN=3$ extra SUSY-current.
\item All of the four operators in \eref{possiblemargneutral} are independent from each other. There are two $\CN=3$ extra SUSY-currents.
\een
Let us discuss each of these possibilities in more detail.

Possibility 1 is the most unlikely.  This is because we do not have two relations that reduce four quantities in \eref{possiblemargneutral} to two independent quantities.

Possibility 2 is possible {\it if} we postulate a relation like 
\bes{ \label{assumpmono}
X_+X_-= (\tr\mu_C)^2~.
}
We will shortly comment on the validity of this assumption.
As a result, the marginal operators transforming under the representation $[4]$ of $SU(2)$ are
\be
X_+^2~, \quad  X_+(\tr \mu_C)~, \quad X_+X_-= (\tr\mu_C)^2~, \quad  X_-(\tr \mu_C)~, \quad X_-^2~,
\ee
whereas those transforming as singlets are 
\be
(\tr \mu_H)^2~, \qquad \tr(\mu_H \mu_C)~.
\ee 
In this possibility, the terms at order $x^2$ should be rewritten as
\be
x^2 \left[ \left( 2+\chi^{SU(2)}_{[4]}(\omega) \right) - {\blue \left( 1+\chi^{SU(2)}_{[2]}(\omega)  \right) {\purple -1} }\right] 
\ee
where the term $-1$, highlighted in purple, indicates the presence of an extra SUSY-current.  If this were true, we would conclude that the theory flows to an SCFT with enhanced $\CN=4$ supersymmetry.  We emphasise again that this conclusion relies heavily on assumption \eref{assumpmono}. It may be argued that this cannot be true because if $X_\pm$ correspond to the monopole operators with fluxes $(\pm 1,0)$, then $X_+ X_-$ carries flux $(1,-1)$\footnote{After applying the Weyl symmetry, the flux $(m,n)$ of the monopole operator $X_{(m,n)}$ should be written such that $m \geq n > -\infty$. The flux of of $X_-$ should thus be written as $(0,-1)$. Since $X_+$ has flux $(+1,0)$, it follows that $X_+X_-$ has flux $(1,-1)$.} and not $(0,0)$; hence it should not be equated to $(\tr \mu_C)^2$.  Indeed, the relation of type \eref{assumpmono} does not hold for the 3d $\CN=4$ $U(2)$ gauge theory with $4$ flavours; see \eref{Rcharge2U24flv}.  It would hold if we had an abelian gauge group, like 3d $\CN=4$ $U(1)$ gauge theory with $2$ flavours.

Possibility 3 is the most likely.  In this possibility, the marginal operators transforming under the representation $[4]$ of $SU(2)$ are
\be \label{margquartk2}
X_+^2~, \quad  X_+(\tr \mu_C)~, \quad X_+X_- ~, \quad  X_-(\tr \mu_C)~, \quad X_-^2~,
\ee
whereas those transforming as singlets are 
\be
(\tr \mu_H)^2~, \qquad (\tr \mu_C)^2~, \qquad \tr(\mu_H \mu_C)~.
\ee 
The terms at order $x^2$ should then be rewritten as
\be
x^2 \left[ \left( 3+\chi^{SU(2)}_{[4]}(\omega) \right) - {\blue \left( 1+\chi^{SU(2)}_{[2]}(\omega)  \right) {\purple -2} }\right] 
\ee
where the term $-2$, highlighted in purple, indicates the two $\CN=3$ extra SUSY-currents.  Note that supersymmetry cannot get enhanced to $\CN=5$, since this would violate a necessary condition for $\CN=5$ supersymmetry which states that the coefficient of $x$ has to be $1$ \cite{Evtikhiev:2017heo}.  We are obliged to conclude that the theory flows to a product of two SCFTs, each with $\CN=4$ supersymmetry.  This situation is similar to that studied in \cite{Gang:2018huc}.  It would be interesting to verify this conclusion using other methods and, if it were true, it would be also nice to identify such $\CN=4$ SCFTs.  We leave this for future work.

%\subsubsection{The case of $k =-2$} \label{k2zeroflv}
%In this case, the index is
%\be
%1 + 2 x + 0 x^2 - 2 x^3 \ldots~.
%\ee
%We interpret the the result of the index as follows.  The operators in this case are similar to the case of $k=2$, except that $X_+$ and $X_-$ are no longer gauge invariant.  The operators with $R$-charge $1$ are therefore
%\be \label{Rcharge1km2}
%\tr(\mu_H)~, \qquad \tr(\mu_C)~.
%\ee
%The $\CN=3$ flavour symmetry is then $U(1)^2$.  For the marginal operators, we consider those in the possibility 3 of the case of $k=2$ and remove those with $X_\pm$.  As a result, the operators in \eref{margquartk2} are absent in the case of $k=-2$, and we are left with
%\bes{
%(\tr \mu_H)^2~, \qquad (\tr \mu_C)^2~, \qquad \tr(\mu_H \mu_C)~.
%}
%The term at order $x^2$ in the index should then be written as $(3{\blue -2} {\brown -1})x^2$, where the term $-2$ is the contribution of the $U(1)^2$ flavour current and the term $-1$ is the contribution of the extra SUSY-current. The analysis of operators up to $R$-charge $2$ is indeed very similar to that of the case of $|k| \geq 3$.  We conclude that supersymmetry gets enhanced to $\CN=4$.

\subsubsection{The case of $k =1$} \label{k1zeroflv}
Evaluating \eref{indexU2knoflv}, we obtain the index for $k=1$ as
\bes{ \label{indexk1}
&\CI_{\eref{U2kzeroflv}}(k= 1; \{w, n=0 \}) \\
&=  1+ x \left(2+w^2+\frac{1}{w^2}\right) +x^2  \left(-1+w^4+\frac{1}{w^4}\right) +  x^{\frac{5}{2}} \left(-2 w-\frac{2}{w}\right)+ \ldots \\
&= 1+x \left[{\blue 1+\chi^{SU(2)}_{[2]}(w)} \right]+x^2 \left[\chi^{SU(2)}_{[4]}(w)  - {\blue \left( 1+\chi^{SU(2)}_{[2]}(w)  \right)}\right] \\
& \qquad  -2 x^{\frac{5}{2}} \chi^{SU(2)}_{[1]}(w) +\ldots~.
}

We {\it propose} that the gauge invariant operators with $R$-charge 1 that carry fugacities $w^{\pm 2}$ are the monopole operators with fluxes $\pm (1,1)$, denoted by $X_{++} := X_{(1,1)}$ and $X_{--}:=X_{(-1,-1)}$.  It is interesting to point out that there is no gauge invariant monopole operator with fluxes $\pm(1,0)$ in this theory, since there are no terms $w^{\pm1}$ at order $x$.  The operators with $R$-charge $1$ are
\be \label{Rcharge1k1}
1, \, w^2, \, 1, \, w^{-2}: \qquad \qquad \tr(\mu_H)~, \qquad X_{++}~, \qquad \tr(\mu_C) ~, \qquad X_{--}~,
\ee
corresponding to the coefficient of $x$.  The $\CN=3$ flavour symmetry of the SCFT is therefore $SU(2) \times U(1)$.  Similarly to the case of $k=2$, we have to make a choice whether to take $(X_{++}, (\tr\mu_C), X_{--})$ or $(X_{++}, (\tr\mu_H), X_{--})$ to be a moment map of $SU(2)$.  Picking any of these choices amounts to breaking the symmetry that exchanges $\mu_H$ and $\mu_C$.  For definiteness, we take the triplet $(X_{++}, (\tr\mu_C), X_{--})$ to be the moment map of $SU(2)$ and $(\tr \mu_H)$ to be that of $U(1)$.\footnote{Similarly to footnote \ref{footnoteHC}, we may as well take $(X_{++}, (\tr\mu_H), X_{--})$ to be the moment map of $SU(2)$ and $(\tr \mu_C)$ to be that of $U(1)$.  The arguments below still hold with $H$ interchanged with $C$.}

Let us now examine the marginal operators of this theory. It is convenient to start from those in the representation $[4]$ of $SU(2)$. Those carrying fugacities $w^{\pm 4}$ are $X_{++}^2$ and $X_{--}^2$.  Those carrying fugacities $w^{\pm 2}$ are $X_{++} (\tr \mu_C)$ and $X_{--} (\tr \mu_C)$. It should be noted that $X_{++} (\tr \mu_H)$ and $X_{--} (\tr \mu_H)$ vanish due to the following argument (very similar to that of the case of $k=2$). Since $(\tr \mu_C)(\tr \mu_H)=0$ due to \eref{trHtrCzero} and $(X_{++}, (\tr\mu_C), X_{--})$ transforms as a triplet under an unbroken $SU(2)$ flavour symmetry, we have
\be X_{++} (\tr \mu_H)=X_{--} (\tr \mu_H)=0~.
\ee The marginal operators carrying fugacity $w^0$ are 
\be
(\tr \mu_H)^2~, \qquad (\tr \mu_C)^2~, \qquad  \tr(\mu_H \mu_C)~, \qquad X_{++} X_{--}~.
\ee  
Analogously to \eref{relmonU2w1flv1adj} of the $U(2)$ gauge theory with one adjoint and one fundamental hypermultiplet, we propose that $X_{++} X_{--}$ satisfies a quantum relation:
\bes{
X_{++} X_{--} = (\tr \mu_C)^2~.
}
Note that both left and right hand sides of this equation have magnetic flux $(0,0)$.  In summary, the marginal operators in the representation $[4]$ of $SU(2)$ are
\bes{
X_{++} ^2~, \quad  X_{++} (\tr \mu_C)~, \quad X_{++} X_{--} = (\tr\mu_C)^2~,\quad X_{--} (\tr \mu_C)~, \quad X_{--} ^2~,
}
and those transforming as singlets under $SU(2)$ are
\bes{
(\tr\mu_H)^2~, \quad \tr(\mu_H \mu_C)~.
}
These operators contribute to the terms $\left( 2+\chi^{SU(2)}_{[4]}(w) \right)$ at order $x^2$ in the index.
As a result, the $x^2$ term in \eref{indexk1} should be rewritten as
\bes{
x^2 \left[ \left( 2+\chi^{SU(2)}_{[4]}(w) \right) - {\blue \left( 1+\chi^{SU(2)}_{[2]}(w)  \right) {\purple - 2}}\right]~. 
}
The extra $-2$, highlighted in purple, indicates the presence of two extra SUSY-currents.   The same remark for the case of $k=2$ applies here.  Supersymmetry cannot get enhanced to $\CN=5$, since it would violate a necessary condition for $\CN=5$ supersymmetry which states that the coefficient of $x$ has to be $1$ \cite{Evtikhiev:2017heo}.  We are again obliged to conclude that the theory flows to a product of two SCFTs, each with $\CN=4$ supersymmetry, similarly to the situation encountered in \cite{Gang:2018huc}.  It would be interesting to verify this conclusion using other methods and, if it were true, it would be also nice to identify such $\CN=4$ SCFTs. We leave this for future work.

%\subsubsection{The case of $k =-1$} \label{km1zeroflv}
%In this case, the index is
%\be
%1 + 4 x +x^2 + 2 x^3 + \ldots~.
%\ee
%The number of the operators with $R$-charges up to $2$ is the same as that of the case of $k=1$.  However, the symmetry of the theory and descriptions of some operators are different from the $k=1$ case.  
%
%There are four operators with $R$-charge $1$, contributing at order $x$.  On the contrary to the $k=1$ case, none of them carries a topological fugacity.  The $\CN=3$ flavour symmetry is therefore $U(1)^4$.  Two of such operators are $\tr(\mu_H)$ and $\tr(\mu_C)$.  The other two operators cannot be $X_{++}$ and $X_{--}$, since they carry non-trivial topological charges.  We propose to replace $X_{++}$ and $X_{--}$ by 
%\bes{
%(V_+)^a (V_-)_a~, \qquad (W_+)^a (W_-)_a~.
%}

%%\subsubsection{The case of $k=0$} \label{sec:k0noflv}
%In this case, the index diverges and the theory is `bad' in the sense of \cite{Gaiotto:2008ak}.  This is due to the existence of gauge invariant monopole operators with $R$-charge 0.  We shall discuss in section \ref{sec:higherflv} that when $n$ flavours of fundamental hypermultiplets are added to this theory, such monopole operators have $R$-charge $\frac{n}{2}$.  Thus, for $n \geq 1$ the index of theory ceases to diverge.

\subsection{$U(2)_k$ gauge group with $n$ flavour} \label{sec:higherflv}
Let us now couple to theory \eref{U2kzeroflv} $n$ flavours of hypermultiplets in the fundamental representation of $U(2)$ and obtain
\be \label{U2koneflv}
\begin{tikzpicture}[baseline]
\tikzstyle{every node}=[font=\footnotesize]
\node[draw, circle] (node1) at (0,0) {$2_k$};
\node[draw, rectangle] (node2) at (2.5,0) {$n$};
\draw[red,thick] (node1) edge [out=45,in=135,loop,looseness=5, snake it]  (node1);
\draw[thick,solid] (node1) to (node2);
\node[draw=none] at (1.8,0.7) {{\red $T^{[2,1^2]}_{[2,1^2]}(SU(4))$}};
\end{tikzpicture}
\ee
We propose that the superpotential for this theory is the same as \eref{suponeflva}, namely
\bes{ 
W&= -\frac{k}{4 \pi} \tr( \varphi^2) + \tr \left( (\mu_C+\mu_H) \varphi \right) +\tQ^i_b\varphi^b_a  Q^a_i  \\
&=  -\frac{k}{4 \pi} \tr( \varphi^2) + \tr \left( (\mu_C+\mu_H +\mu_Q) \varphi \right)~,
}
The $F$-terms are the same as \eref{Ftermsmatter} and the consequences of them are as analysed in Appendix \ref{app:Fterms}.  The index of this theory is discussed in Appendix \ref{app:SfoldT211nflv}.

\subsubsection{The case of $n\geq2$} \label{sec:k2ngeq2T211}
We focus on the cases of $(n\geq3$, $|k| \geq 1)$ and $(n=2$, $|k| \geq 3)$.  Evaluating \eref{indexU2knflv} with the background fluxes for the flavour symmetry being set to zero, $\vec{n_f}=0$, we obtain the indices, up to order $x^2$, as follows:
\bes{ \label{ngeq3modkgeq3}
(n\geq3,\, |k| \geq 1): &~  1+ x \left[ {\blue 3+ \chi^{SU(n)}_{[1,0, \ldots, 0,1]} (\vec f)}  \right]+ x^2 \Big[ 2 q  \chi^{SU(n)}_{[1,0,\ldots,0]} (\vec f)  + 2 q^{-1}  \chi^{SU(n)}_{[0,\ldots,0,1]} (\vec f)   \\
&\qquad + \chi^{SU(n)}_{[2, 0, \ldots,0 ,2]} (\vec f) +5  \chi^{SU(n)}_{[1, 0, \ldots,0 ,1]} (\vec f)+ \chi^{SU(n)}_{[0,1,0, \ldots,0,1 ,0]} (\vec f)  +7\\
&\qquad  -{\blue \left(3+ \chi^{SU(3)}_{[1,0, \ldots,0,1]} (\vec f) 
\right) } \Big] +\ldots
}
\bes{ \label{n2modkgeq3}
(n=2, \, |k| \geq 3): &~ 1+ x \left[ {\blue 3+ \chi^{SU(2)}_{[2]} (\vec f)}  \right]+ x^2 \Big[ 2 q  \chi^{SU(2)}_{[1]} (\vec f)  + 2 q^{-1}  \chi^{SU(2)}_{[1]} (\vec f)  \\
& \qquad  +\chi^{SU(2)}_{[4]} (\vec f) +4  \chi^{SU(2)}_{[2]} (\vec f)  +7 \\
&\qquad -{\blue \left(3+ \chi^{SU(2)}_{[2]} (\vec f) 
\right) } \Big] +\ldots
}
where we highlight the contribution of the $U(1)^3 \times SU(n)$ flavour symmetry current in blue.  Let us now analyse the operators with $R$-charges $1$ and $2$.

The operators with $R$-charge $1$ are
\be \label{opRcharge1}
\tr \mu_H~, \qquad \tr \mu_C~, \qquad M^k_k =\tr \mu_Q~, \qquad \hat{M}^i_j
\ee
where we remark that $\hat{M}^i_j$ transforms in the adjoint representation $[1,0,\ldots,0,1]$ of $SU(n)$, and that we can always rewrite $\varphi$ in terms of $\mu_H$, $\mu_C$ and $\mu_Q$ due to \eref{Ftermsmatter}.

Let us now discuss about the marginal operators.  These contribute to positive terms at order $x^2$ of the index.  The terms $ 2 q  \chi^{SU(n)}_{[1,0,\ldots,0]} (\vec f)$ and $2 q^{-1}  \chi^{SU(n)}_{[0,\ldots,0,1]} (\vec f) $ correspond to the gauge invariant combinations constructed by ``dressing'' $Q$ or $\tQ$ to the operators in \eref{opsthreehalf}:
\bes{  \label{dressedQ}
2 q  \chi^{SU(n)}_{[1,0,\ldots,0]} (\vec f) : \qquad Q^a_i  \tilde{\CH}_a ~, &\qquad Q^a_i \tilde{\CU}_a ~, \\
2 q^{-1}  \chi^{SU(n)}_{[0,\ldots,0,1]} (\vec f): \qquad   \tilde{Q}^i_a {\CH}^a~, &\qquad \tQ^i_a \CU^a~.
}
The term $5  \chi^{SU(n)}_{[1, 0, \ldots,0 ,1]} (\vec f)$ corresponds to
\be \label{fiveinadj}
\begin{array}{lll}
\hat{M}^i_j (\tr \mu_H)~, &\qquad \hat{M}^i_j (\tr \mu_C) ~, & \qquad \hat{M}^i_j (\tr \mu_Q) = \hat{M}^i_j  (M^k_k)~,  \\
 (\CA_H)^i_j ~, &\qquad    (\CA_C)^i_j ~, & %\qquad (\underline{\hat{M}^2})^i_j 
\end{array}
\ee
where we have defined $\underline{\hat{M}^2}$ in \eref{defMhat} and $\CA_{H,C}$ in \eref{defCAHC}.  It should be noted that, from \eref{trlessMhat2}, the quantity $(\underline{\hat{M}^2})^i_j$ can be written in terms of a linear combination of $(\CA_H)^i_j$, $(\CA_C)^i_j$ and $\hat{M}^i_j  (M^k_k)= \hat{M}^i_j (\tr \mu_Q)$.   For the special case of $n=2$, we have an extra relation \eref{relAHACMn2}:
\bes{ \label{relAHACMn2}
 (\CA_H)^i_j+  (\CA_C)^i_j = - \hat{M}^i_j (\tr \mu_Q) = - \hat{M}^i_j  (M^k_k) \qquad \text{(for $n=2$)}
}
and so we have only four independent quantities, which correspond to the term $4  \chi^{SU(2)}_{[2]} (\vec f)$ in the index.  The term $ \chi^{SU(n)}_{[0,1,0, \ldots,0,1 ,0]} (\vec f)$ corresponds to
\be \label{marg01010}
\epsilon^{i_1 i_2 \ldots i_n} \epsilon_{j_1 j_2 \ldots j_n} \hat{M}^{j_1} _{i_1} \hat{M}^{j_2}_{i_2} ~.
\ee
The term $ \chi^{SU(n)}_{[2, 0, \ldots,0 ,2]} (\vec f)$ corresponds to the quantity 
\be \label{marg2002}
R^{ik}_{jl} 
\ee
which is a linear combination $ \hat{M}^i_{j} \hat{M}^{k}_{l}$ and other quantities such that any contraction between an upper index and a lower index yields zero; for example, for $n=2$, where $\hat{M}^2$ satisfies \eref{specialn2}, the marginal operators in $[4]$ are
\bes{ \label{marginalin4}
R^{ik}_{jl} := \hat{M}^i_{j} \hat{M}^{k}_{l}  +\frac{1}{6} (\hat{M}^2)^p_p \delta^i_{j} \delta^k_{l} - \frac{1}{3} (\hat{M}^2)^p_p \delta^i_{l} \delta^k_{j} ~, \qquad \text{for $n=2$}~.
}
 Finally the candidates for the marginal operators that do not carry $q$ and $\vec{f}$ fugacities are 
\be \label{listallmarg}
\begin{array}{ll}
\tr({\mu}_H^2) =(\tr \mu_H)^2~, & \qquad  \tr({\mu}_C^2) = (\tr \mu_C)^2~, \\
\tr({\mu}_Q {\mu}_H)=  (\mu_H)^a_b   \tQ^i_a Q^b_i ~, & \qquad (\tr \mu_Q)( \tr \mu_H)~,  \\
 \tr({\mu}_Q {\mu}_C)=  (\mu_C)^a_b   \tQ^i_a Q^b_i ~, & \qquad (\tr \mu_Q)( \tr \mu_C)~,  \\
(\hat{M}^2)^i_i = \hat{M}^i_j \hat{M}^{j}_i  ~, & \qquad   (\tr\mu_Q)^2 = (M^k_k)^2 \\ 
\tr({\mu}_H {\mu}_C) ~,& \qquad (\tr \mu_H)(\tr \mu_C) \overset{\eref{trHtrCzero}}{=}0 ~. \\
\end{array}
\ee
where we recall from \eref{trM2} that $\tr(\mu_Q^2)$ is not independent from the above quantities, since it can be written as
\bes{
\tr(\mu_Q^2) = M^i_j M^j_i =  \hat{M}^i_j \hat{M}^{j}_i +\frac{1}{n} (\tr \mu_Q)^2 = - \tr(\mu_Q \mu_H) - \tr(\mu_Q \mu_C)~.
}
However, the quantities in \eref{listallmarg} are not all independent from each other.  Let us try to reduce them into a smaller set as follows. From \eref{trHtrCzero}, we see that $(\tr \mu_H)(\tr \mu_C)$ vanishes.   From \eref{hatM2}, we see that $(\tr\mu_Q)^2$ is a linear combination of $\tr({\mu}_Q {\mu}_H)$ and $\tr({\mu}_Q {\mu}_C)$ and $(\hat{M}^2)^i_i $ .  %Moreover, since the quantity $\tr \left( \mu_C+\mu_H  +\mu_Q \right)^2$ appears as the superpotential \eref{Weff1}, it is zero on the space of holomorphic functions over the supersymmetric vacua; hence, we can rewrite $\tr (\mu_H \mu_C)$ as a linear combination of some remaining operators in the list.  
In summary, we have eight of such marginal operators:
\be \label{margnof}
\begin{array}{ll}
\tr({\mu}_H^2) =(\tr \mu_H)^2~, & \qquad  \tr({\mu}_C^2) = (\tr \mu_C)^2~, \\
\tr({\mu}_Q {\mu}_H)=  (\mu_H)^a_b   \tQ^i_a Q^b_i ~, & \qquad (\tr \mu_Q)( \tr \mu_H)~,  \\
 \tr({\mu}_Q {\mu}_C)=  (\mu_C)^a_b   \tQ^i_a Q^b_i ~, & \qquad (\tr \mu_Q)( \tr \mu_C)~,  \\
 (\hat{M}^2)^i_i  =  \hat{M}^i_j \hat{M}^{j}_i ~, &  \qquad \tr(\mu_H \mu_C) \\ 
\end{array}
\ee
As a result, the $x^2$ term in \eref{ngeq3modkgeq3} and \eref{n2modkgeq3} should be rewritten as
\bes{
x^2 \left[ \ldots +8 -{\blue \left(3+ \chi^{SU(3)}_{[1,0, \ldots,0,1]} (\vec f) \right) } {\brown -1}  \right]~. 
}
where the term $-1$, highlighted in brown, indicates the presence of an extra SUSY-current.  We conclude that supersymmetry gets enhanced to $\CN=4$.

We also observe that, for $k=2$, the coefficient of $x^{\frac{n}{2}+1}$ in the index contains the terms $w+ w^{-1}$.  Similarly, for $k=1$, the coefficient of $x^{n+1}$ in the index contains the terms $w^2+w^{-2}$.  These indicate that
\bi
\item for $k=2$, there are gauge invariant monopole operators $X_\pm$ with topological charges $\pm1$ with $R$-charge ${\frac{n}{2}+1}$; and
\item for $k=1$, there are gauge invariant monopole operators $X_{++}$ and $X_{--}$ with topological charges $\pm 2$ with $R$-charge ${n+1}$.
\ei
In fact, we have encountered such monopole operators for the case of zero flavour ($n=0$) in sections \ref{k2zeroflv} and \ref{k1zeroflv}.  The above statements generalise the previous results to any $n$.  In particular, for $(n=2, \, k=2)$, the gauge invariant monopole operator $X_\pm$ are marginal operators. This can be seen from the index that can be computed from \eref{indexU2knflv} with $\vec{n_f}=0$:
\bes{
(n=2, \, k= 2): &\quad  1+ x \left[ {\blue 3+ \chi^{SU(2)}_{[2]} (\vec f)}  \right]+ x^2 \Big[ 2 q  \chi^{SU(2)}_{[1]} (\vec f)  + 2 q^{-1}  \chi^{SU(2)}_{[1]} (\vec f)  \\
& \qquad  +\chi^{SU(2)}_{[4]} (\vec f) +4  \chi^{SU(2)}_{[2]} (\vec f)  +w +w ^{-1}+7 \\
&\qquad -{\blue \left(3+ \chi^{SU(2)}_{[3]} (\vec f) 
\right) } \Big] +\ldots~,
}
where there are extra terms $w+w^{-1}$ at order $x^2$ in comparison to \eref{n2modkgeq3}.  
%We shall discuss the case of $(n=1,k=1)$ in the next subsection.

\subsubsection{The case of $n=1$}
In this subsection, we discuss the special case of $n=1$.  The operators are as discussed in the previous subsection, but with the flavour indices $i, j, k=1$, and so they can be dropped.  As a result, we have
\be \label{hatM0n1}
\hat{M} = 0~, \qquad \CA_H=0~, \qquad \CA_C=0~.
\ee

\subsubsection*{The cases of $|k| \geq 3$}
For $|k|\geq 3$, the index can be computed from \eref{indexU2knflv} with $n=1$ and $n_{f_1}=0$:
\bes{ \label{indkgeq3oneflv}
&1+3 x + \left( 3+ 2 q +2 q^{-1} \right) x^2 - x^3+\ldots \\
&=  1+{\blue 3} x + \left( 6+ 2 q +2 q^{-1} - {\blue 3} \right) x^2 - x^3+\ldots~,
}
where we highlight the contribution of the flavour currents in blue, and rewrite the fugacity $f_1$ as $q$.

From \eref{opRcharge1} and \eref{hatM0n1},  we see that the three independent operators with $R$-charge $1$ are
\be  \label{opRcharge1oneflv}
\tr \mu_H~, \qquad \tr \mu_C~, \qquad M=\tr \mu_Q~.
\ee
The flavour symmetry of this theory is therefore $U(1)^3$.

Let us now discuss the marginal operators.  The terms $2q+2q^{-1}$ in \eref{indkgeq3oneflv} correspond to the operators in \eref{dressedQ}, namely
\bes{ \label{margwithf}
2q: \qquad Q^a  \tilde{\CH}_a ~, &\qquad Q^a \tilde{\CU}_a ~, \\
2q^{-1}: \qquad   \tilde{Q}_a {\CH}^a~, &\qquad \tQ_a \CU^a~.
}
Note that all of the operators in \eref{fiveinadj} vanish identically for $n=1$, due to \eref{hatM0n1} and the fact that the flavour indices can be dropped. The marginal operators that do not carry fugacity $q$ are as listed in \eref{margnof}; %however, for $n=1$ we also have 
%\be
%(\tr \mu_Q)^2 = \tr (\mu_Q^2) \overset{\eref{trM2}}{=} -\tr(\mu_Q \mu_H) - \tr(\mu_Q \mu_C) ~, 
%\ee
since $\hat{M}=0$, there are 7 independent quantities:
\be  \label{margnofoneflv}
\begin{array}{ll}
\tr(\mu_H^2) =(\tr \mu_H)^2~, & \qquad  \tr(\mu_C^2) =(\tr \mu_C)^2~, \\
\tr(\mu_Q \mu_H)~, & \qquad (\tr \mu_Q)( \tr \mu_H)~,  \\
 \tr(\mu_Q \mu_C)~, & \qquad (\tr \mu_Q)( \tr \mu_C)~, \\
 \tr(\mu_H \mu_C)
\end{array}
\ee
These operators, together with \eref{margwithf}, contribute $7+2q+2q^{-1}$ to order $x^2$ in the index.  The $x^2$ term of the index should then be rewritten as $(7+2q+2q^{-1}) {\blue - 3} {\brown - 1} $, where the term $-1$ indicates the presence of the extra SUSY-current.  Hence we conclude that supersymmetry gets enhanced to $\CN=4$.

\subsubsection*{The cases of $k= 2$}
The index in this case can be computed from \eref{indexU2knflv} with $k=2$, $n=1$ and $n_{f_1}=0$:
\bes{ \label{indk2oneflv}
&1+{\blue 3} x +(w+w^{-1}) x^{\frac {3}{2}} + \left( 7+ 2 q +2 q^{-1}  - {\blue 3} - {\brown 1}\right) x^2  \\
& \quad +(w+w^{-1}) x^{\frac {5}{2}} + (-1 + w^2 +w^{-2}) x^3\ldots~.
}
where we rewrite the fugacity $f_1$ as $q$.

As can be seen from order $x$, the $\CN=3$ flavour symmetry of the theory is $U(1)^3$. The operators with $R$-charge $1$ are \eref{opRcharge1oneflv}.  In this case, there are also gauge invariant monopole operators $X_\pm$, carrying topological fugacities $w^{\pm 1}$, with $R$-charge $3/2$.  (This is consistent with the observation that the theory with $k=2$ and $n$ flavours, there are gauge invariant monopole operators with $R$-charge $\frac{1}{2}n+1$; see section \ref{sec:higherflv}). The marginal operators are listed in \eref{margwithf} and \eref{margnofoneflv}.  Again, the term $-1$ at order $x^2$ of the index indicates the presence of the extra SUSY-current, and we conclude that supersymmetry gets enhanced to $\CN=4$.

\subsubsection*{The cases of $k= 1$}
The index can be computed from \eref{indexU2knflv} with $k=1$, $n=1$ and $n_{f_1}=0$:
\bes{ \label{indk1oneflv}
&1+{\blue 3} x +x^2 \left(7+2 q+2q^{-1}+w^2+w^{-2} {\blue-  3}{\brown-  1}\right) \\
& \quad -x^3 \left[2 \left(q+q^{-1} \right)\left( w+w^{-1} \right)+4 \left(w+w^{-1}\right)+2\right] +\ldots~.
}
with the fugacity $f_1$ being rewritten as $q$.

The $\CN=3$ flavour symmetry of this theory is $U(1)^3$, and the operators with $R$-charge $1$ are \eref{opRcharge1oneflv}.   The marginal operators are \eref{margwithf} and \eref{margnofoneflv}, together with the gauge invariant monopole operators $X_{++}$ and $X_{--}$, carrying topological fugacities $w^{\pm 2}$.  (This is consistent with the observation that in the theory with $k=1$ and $n$ flavours there are gauge invariant monopole operators with topological charges $\pm2$ and $R$-charge $n+1$; see section \ref{sec:higherflv}).   The term $-1$ at order $x^2$ of the index indicates the presence of the extra SUSY-current, and we conclude that supersymmetry gets enhanced to $\CN=4$.

\acknowledgments
We are indebted to Alessandro Tomasiello for useful discussions. M.S. is partially supported by the ERC-STG grant 637844-HBQFTNCER, by the University of Milano-Bicocca grant 2016-ATESP0586, by theMIUR-PRIN contract 2017CC72MK003, and by the INFN.

\appendix
\section{Expressions of superconformal indices} \label{app:index}
In this section, we summarise the expressions of the superconformal indices of the theories discussed in this paper.  We follow the convention adopted in \cite{Aharony:2013dha, Aharony:2013kma}. 
\subsection{The $T(SU(N))$ and $T(U(N))$ theories} \label{app:TUN}
Let us start by discussing the $T(SU(N))$ theory.  It admits the quiver description
\be \label{TSUN}
\begin{tikzpicture}[baseline]
\tikzstyle{every node}=[font=\footnotesize, minimum size=2em]
\node[draw, circle] (node1) at (0,0) {$1$};
\node[draw, circle] (node2) at (1.5,0) {$2$};
\node[draw=none] (node3) at (3,0) {$\cdots$};
\node[draw, circle] (node4) at (4.5,0) {\tiny $N-1$};
\node[draw, rectangle] (node5) at (6,0) {$N$};
\draw[draw=black,solid]  (node1) to (node2);
\draw[draw=black,solid]  (node2) to (node3);
\draw[draw=black,solid]  (node3) to (node4);
\draw[draw=black,solid]  (node4) to (node5);
\end{tikzpicture}
\ee
The index is given by 
\be
\scalebox{0.9}{$
\begin{split}
&\CI_{T(SU(N))} ( \{(u_1 \ldots, u_N), (n_{u_1}, \ldots, n_{u_N}) \}, \{(h_1 \ldots, h_N), (n_{h_1}, \ldots, n_{h_N}) \},d ) \\
&= \sum_{m^{(1)}_1 \in \BZ} \, \, \sum_{m^{(2)}_1, m^{(2)}_2  \in \BZ} \cdots \sum_{m^{(N-1)}_1, \ldots, m^{(N-1)}_{N-1}  \in \BZ}  \,\, \prod_{j=1}^{N-1}  \frac{1}{j!} \prod_{k=1}^j  \oint \frac{d z^{(j)}_k}{2 \pi i z^{(j)}_k} \,\, u_j^{m^{(j)}_k}  \left( z^{(j)}_k \right)^{n_{u_j}} \times \\
& \quad \prod_{j=1}^{N-1} \, Z_{(j)-(j+1)} (\{ \vec{z}^{(j)}, \vec{m}^{(j)} \}, \{ \vec{z}^{(j+1)}, \vec{m}^{(j+1)} \}, d) \,\, Z_{\varphi_j}(\{ \vec{z}^{(j)}, \vec{m}^{(j)} \}, d) \times \\
& \quad Z_{(N-1)-(N)} (\{ \vec{z}^{(N-1)}, \vec{m}^{(N-1)} \}, \{(h_1 \ldots, h_N), (n_{h_1}, \ldots, n_{h_N}) \}, d) \times \\
& \quad \prod_{j=1}^N Z_{\text{vec}; \, U(j)}  (\{ \vec{z}^{(j)}, \vec{m}^{(j)} \}  )~,
\end{split}$}
\ee
where $ \{(u_1 \ldots, u_N), (n_{u_1}, \ldots, n_{u_N}) \}$ are the fugacities for the (enhanced) $SU(N)$ topological fugacities and the corresponding fluxes, $\{(h_1 \ldots, h_N), (n_{h_1}, \ldots, n_{h_N}) \}$ are the fugacities for the $SU(N)$ flavour symmetries and the corresponding background fluxes.  We also use the shorthand notation:
\bes{
\{ \vec{z}^{(j)}, \vec{m}^{(j)} \} =  \left\{ \left( {z}_1 ^{(j)}, \cdots, {z}_j ^{(j)} \right),  \left( {m}_1 ^{(j)}, \cdots, {m}_j ^{(j)} \right) \right\} ~.
}
They are subject to the conditions
\bes{ \label{condTSUN}
\prod_{i=1}^N u_i =\prod_{i=1}^N h_i=1~, \qquad \sum_{i=1}^N n_{u_i} =  \sum_{i=1}^N n_{h_i} =0~.
}
The fugacity $d$ is that corresponds to the axial symmetry $U(1)_C-U(1)_H$, where $U(1)_C$ and $U(1)_H$ are the Cartan subalgebras of $SU(2)_C$ and $SU(2)_H$ of the $R$-symmetry of the $\CN=4$ $R$-symmetry $SU(2)_C \times SU(2)_H$.  For convenience, we shall henceforth refer to the axial symmetry as $U(1)_d$. Here we do not turn on the background magnetic flux for $U(1)_d$. The contributions of the chiral fields in the theory are as follows:
\be
\scalebox{0.95}{$
\begin{split}
&Z_{(j)-(j+1)} (\{ \vec{z}^{(j)}, \vec{m}^{(j)} \}, \{ \vec{z}^{(j+1)}, \vec{m}^{(j+1)} \}, d) \\
&= \CZ_\chi\left( \{ \vec{z}^{(j)}, \vec{m}^{(j)} \} ,  \{ \vec{z}^{(j+1)}, \vec{m}^{(j+1)} \} , d^{-1}; \frac{1}{2} \right) \times \left( z^{(l)}_i \leftrightarrow 1/z^{(l)}_i,~ m^{(l)}_i \leftrightarrow -m^{(l)}_i \right) \\
&Z_{\varphi_j}(\{ \vec{z}^{(j)}, \vec{m}^{(j)} \}, d) = \CZ_\chi \left(\{ \vec{z}^{(j)}, \vec{m}^{(j)} \}, \{ \vec{z}^{(j)}, \vec{m}^{(j)}\}, d^{2};  1  \right) \\
\end{split}$}
\ee
where we define
\bes{
&\CZ_\chi (\{(a_1, \ldots, a_{\ell_1}) ,(m_1, \ldots, m_{\ell_1})\} , \{(b_1, \ldots, b_{\ell_2}) ,(n_1, \ldots, n_{\ell_2})\}, d; r) \\
&=  \prod_{i=1}^{\ell_1} \prod_{j=1}^{\ell_2} \Bigg[ (a_i^{} b_j^{-1}  x^{r-1} d)^{-\frac{1}{2}(|m_i-n_j|)}  \frac{((-1)^{m_i-n_{j}} a_i^{-1} b_j^{}  x^{2-r+|m_i-n_{j}|} d^{-1}; x^2)}{((-1)^{m_i-n_{j}} a_i^{} b_j^{-1} x^{r+|m_i-n_{j}|} d; x^2)} \Bigg]~.
}
The contribution from the vector multiplet of the $U(r)$ gauge group is given by
\bes{ \label{Zvec}
&Z_{\text{vec}; \, U(r)}  (\{(z_1, \ldots, z_r), (m_1, \ldots,m_r) \}) \\
&= \prod_{1\leq i \neq j \leq r}  x^{-\frac{1}{2} | m_i -m_j|} \left[ 1- (-1)^{m_i-m_j} (z_i z_j^{-1}) x^{|m_i-m_j|} \right] ~.
}
It is interesting to point out that the index of $T(SU(N))$ satisfies the following property
\bes{ \label{TSUNexchange}
&\CI_{T(SU(N))} ( \{\vec h, \vec{n_h} \},  \{\vec u, \vec{n_u} \} ,d )\\
&=  {\red \frac{u_N^{n_{h_1} +\cdots + n_{h_N}} (h_1 \cdots h_N)^{n_{u_N}}}{h_N^{n_{u_1} +\cdots + n_{u_N}} (u_1 \cdots u_N)^{n_{h_N}}}}  \times \CI_{T(SU(N))} ( \{\vec u, \vec{n_u} \}, \{\vec h, \vec{n_h} \} ,d^{-1} )
}
where, upon imposing the conditions \eref{condTSUN}, the prefactor indicated in red is equal to unity.

The index of the $T(U(N))$ theory is defined as follows:
\bes{ \label{indexTUN}
& \CI_{T(U(N))} ( \{(u_1 \ldots, u_N), (n_{u_1}, \ldots, n_{u_N}) \}, \{(h_1 \ldots, h_N), (n_{h_1}, \ldots, n_{h_N}) \},d ) \\
&= {\blue u_N^{n_{h_1} +\cdots + n_{h_N}} (h_1 \cdots h_N)^{n_{u_N}}} \times \CI_{T(SU(N))} ( \{\vec u, \vec{n_u} \}, \{\vec h, \vec{n_h} \},d )
}
where we do {\it not} impose the conditions \eref{condTSUN} in this definition.  Since $T(U(N))$ is a product of $T(SU(N))$ and $T(U(1))$ \cite{Gaiotto:2008ak}, where $T(U(1))$ contains only the mixed Chern--Simons term, we regard the blue factor as the index of the $T(U(1))$ theory\footnote{The importance of this contact term for the $T(U(N))$ theory at the level of the $S^3_b$ partition function was already noticed in \cite[(3.26)]{Bullimore:2014awa} and in \cite[(4.6)]{Aprile:2018oau}.}.   It follows from \eref{TSUNexchange} that the index of $T(U(N))$ satisfies
\bes{
\CI_{T(U(N))} ( \{\vec h, \vec{n_h} \},  \{\vec u, \vec{n_u} \} ,d ) = \CI_{T(U(N))} ( \{\vec u, \vec{n_u} \}, \{\vec h, \vec{n_h} \} ,d^{-1} )~.
}
Upon setting the background fluxes to zero, $\vec{n_u} = \vec{n_h} = \vec 0$, the indices of $T(U(N))$ and $T(SU(N))$ are equal.  In the main text, we are interested in the power series of such indices up to order $x^2$.  The explicit expressions for $N=2$ and $N=3$ are given in \eref{indexTUN23}.

\subsection{$S$-fold theories with the $T(U(N))$ building block} \label{app:indSfoldTUN}
We now examine the index of theory \eref{TUNkzeroflv} formed by gauging the diagonal subgroup of the Higgs and Coulomb branch symmetries of $T(U(N))$:
\be \label{indexSfoldTUN}
\scalebox{0.9}{$
\begin{split}
 \CI_{\eref{TUNkzeroflv}; k, N} (\{ \omega, n \}, d)&=\sum_{m_1, m_2 \ldots, m_N \in \BZ} \frac{1}{N!} \prod_{j=1}^N \oint \frac{d z_j}{2\pi i z_j} \, \omega^{m_j} z_j^{k m_j+n} \times  \\
& \quad Z_{\text{vec}; \, U(N)}  (\{(z_1, z_2, \ldots, z_N), (m_1, m_2,  \ldots,m_N) \}) \times  \\
&\quad \CI_{T(U(N))} ( \{ (z_1,z_2 \ldots, z_N), (m_1, m_2,\ldots, m_N) \},  \\
& \quad \qquad  \qquad  \{ (z^{-1}_1,z^{-1}_2 \ldots, z^{-1}_N), (-m_1, -m_2,\ldots, -m_N) \} ,d )~,
\end{split}$}
\ee
where $\omega$ is the topological symmetry.  Here $n$ is the background magnetic flux for the topological symmetry which we turn off (\ie~ by setting $n=0$) in the main text.    Note the convention that we gauge the Higgs and Coulomb branch symmetries of the $T(U(N))$ theory; they come in opposite way $z_j$ and $z^{-1}_j$ (also $m_j$ and $-m_j$) for $j=1,\ldots, N$.  In the notation of \cite{Assel:2018vtq}, this corresponds to the $U(N)_- = \diag (U(N) \times U(N)^\dagger)$ choice of gauging the Higgs and Coulomb branch symmetries of $T(U(N))$.  Another choice of gauging corresponds to the index
\be
\scalebox{0.9}{$
\begin{split}
\hat{\CI}_{\eref{TUNkzeroflv}; k, N} (\{\omega,n\}, d)&=\sum_{m_1, m_2 \ldots, m_N \in \BZ} \frac{1}{N!} \prod_{j=1}^N \oint \frac{d z_j}{2\pi i z_j} \, \omega^{m_j} z_j^{k m_j+n} \times  \\
&\quad Z_{\text{vec}; \, U(N)}  (\{(z_1, z_2, \ldots, z_N), (m_1, m_2, \ldots,m_N) \}) \times  \\
&\quad \CI_{T(U(N))} ( \{ (z_1,z_2 \ldots, z_N), (m_1, m_2,\ldots, m_N) \},  \\
& \quad \qquad  \qquad  \{ (z_1,z_2 \ldots, z_N), (m_1, m_2,\ldots, m_N) \} ,d )~,
\end{split}$}
\ee
where in the notation of \cite{Assel:2018vtq}, this choice corresponds to the $U(N)_+ = \diag (U(N) \times U(N))$ type of gauging.  It follows from \eref{indexTUN} and from the fact that the index of $T(SU(N))$ is invariant under inversion of the $SU(N)$ fugacities because of the Weyl group of $SU(N)$, that the indices corresponding to these two types of gauging are related by the flipping of the sign of $k$ together with the sign of the background topological flux $n$ up to the change of variables $z_i\to z_i^{-1}$:
\bes{
 \CI_{\eref{TUNkzeroflv}; k, N}(\{\omega,n\}, d) =  \hat{\CI}_{\eref{TUNkzeroflv}; -k, N} (\{\omega,-n\}, d)~.
}
For definiteness, we use the convention of \eref{indexSfoldTUN}, namely the $U(N)_-$ type of gauging, throughout the paper.

Let us now examine the $S$-fold theory with $n$ flavours, namely theory \eref{TUNnflv}, whose index is
\be \label{indexSfoldTUNwnflv}
\scalebox{0.9}{$
\begin{split}
 \CI_{\eref{TUNnflv}; k, N, n} (\omega; \vec{n_f})&=\sum_{m_1, m_2 \ldots, m_N \in \BZ} \frac{1}{N!} \prod_{j=1}^N \oint \frac{d z_j}{2\pi i z_j} \, \omega^{m_j} z_j^{k m_j} \times  \\
 &\quad Z_{\text{vec}; \, U(N)}  (\{(z_1,z_2, \ldots, z_N), (m_1, m_2, \ldots,m_N) \}) \times  \\
&\quad \CI_{T(U(N))} ( \{ (z_1,z_2 \ldots, z_N), (m_1, m_2,\ldots, m_N) \},  \\
& \quad \qquad  \qquad  \{ (z^{-1}_1,z^{-1}_2 \ldots, z^{-1}_N), (-m_1, -m_2,\ldots, -m_N) \} ,d=1 ) \times \\
& \quad Z_{\text{fund}} (\{(z_1,z_2 \ldots, z_N), (m_1, m_2,\ldots, m_N) \}, \\
&  \quad \qquad  \qquad  \{ (f_1, f_2, \ldots, f_n), (n_{f_1}, n_{f_2}, \ldots, n_{f_n}) \})~.
\end{split}$}
\ee
Note that the axial $U(1)_d$ symmetry is broken by the fundamental hypermultiplets, as can be seen from effective superpotential \eref{Weff1a}, and so we set $d=1$ in the index of $T(U(N))$. In the above expression, we also turn off the background flux for the topological symmetry. The contribution of the fundamental hypermultiplets is given by
\bes{ \label{Zfund}
Z_{\text{fund}} (\{ \vec z, \vec m \}, \{ \vec f, \vec{n_f} \}) &= \CZ_\chi\left( \{ \vec z, \vec m \}, \{ \vec f, \vec{n_f} \} , d=1; \frac{1}{2} \right) \times  \\
& \quad \left( z_i \leftrightarrow 1/z_i,~ m_i \leftrightarrow -m_i,  f_j \leftrightarrow f^{-1}_j, n_{f_j} \leftrightarrow -n_{f_j}  \right) ~.
}
In the main text, we set the background flavour magnetic fluxes to zero, $\vec{n_f}=\vec{0}$.

\subsection{The $T^{[2,1^2]}_{[2,1^2]}(SU(4))$ theory} \label{app:T211}
The index of this theory can be computed from the quiver description \eref{T211211} as
\bes{
&\hat{\CI}_{\eref{T211211}} ( \{ v_1, n_{v_1}\}, \{ v_2, n_{v_2} \} , \{a, n_a \} , \{(b_1,b_2) , (n_{b_1}, n_{b_2}) \},d ) \\
&= \sum_{m_1, m_2 \in \BZ} \oint \frac{dz_ 1}{2\pi z_1} \oint \frac{dz_2}{2\pi i z_2} v_1^{m_1} z_1^{n_{v_1}}  v_2^{m_2} z_2^{n_{v_2}} \times \\
& \qquad Z_{L,\tilde{L}} (\{ z_1, m_1 \}, \{ a, n_a \}, d)  Z_{R,\tilde{R}} (\{ z_2, m_2 \}, \{ \vec b, \vec{n_{b}} \}, d)  \times \\
& \qquad Z_{X,\tilde{X}} (\{ z_1, m_1 \}, \{ z_2, m_2 \}, d) Z_{\varphi_1}(\{ z_1, m_1 \}, d) Z_{\varphi_2}(\{ z_2, m_2\},d) ~,
}
where $\{ v_1, n_{v_1}\}, \{ v_2, n_{v_2} \}$ are the topological fugacities and the corresponding fluxes for each $U(1)$ gauge group, $\{z_1, m_1\}$, $\{z_2, m_2\}$ are gauge fugacities and fluxes for each $U(1)$ gauge group.  The fugacities and the corresponding background fluxes for the $U(1)$ and $U(2)$ flavour symmetries are denoted by $\{ a, n_a \}$ and $\{(b_1,b_2) , (n_{b_1}, n_{b_2}) \} = \{ \vec b, \vec{n_{b}} \}$ respectively.  The fugacity $d$ corresponds to the axial symmetry, as described above.    The contributions of the chiral fields in the theory are as follows:
\be
\scalebox{0.9}{$
\begin{split}
Z_{L,\tilde{L}} (\{ z_1, m_1 \}, \{ a, n_a \}, d)  &= \CZ_\chi\left(\{ z_1, m_1 \}, \{ a, n_a \}, d^{-1}; \frac{1}{2} \right) \\
& \qquad \times (z_1 \leftrightarrow z_1^{-1},~ a \leftrightarrow a^{-1} ) \\
Z_{R,\tilde{R}} (\{ z_2, m_2 \}, \{ \vec b, \vec{n_{b}} \}, d) &= \CZ_\chi\left(\{ z_2, m_2 \}, \{ (b_1,b_2), (n_{b_1}, n_{b_2}) \}, d^{-1}; \frac{1}{2} \right) \\
& \qquad \times (z_2 \leftrightarrow z_2^{-1},~ b_{1,2} \leftrightarrow b_{1,2}^{-1}) \\
Z_{X,\tilde{X}} (\{ z_1, m_1 \}, \{ z_2, m_2 \}, d  ) &= \CZ_\chi \left(\{ z_1, m_1 \}, \{ z_2, m_2 \}, d^{-1}; \frac{1}{2}  \right) \times (z_{1,2} \leftrightarrow z_{1,2}^{-1} )~, \\
Z_{\varphi_1} (\{ z_1, m_1 \}, d  ) &= \CZ_\chi \left(\{ z_1, m_1 \}, \{ z_1, m_1 \}, d^{2};  1  \right) \\
Z_{\varphi_2} (\{ z_2, m_2 \}, d  ) &= \CZ_\chi \left(\{ z_2, m_2 \}, \{ z_2, m_2 \}, d^{2}; 1 \right) 
\end{split} $}
\ee
Setting the background magnetic fluxes to zero, $n_{v_1} = n_{v_2} = n_a = n_{b_1} =  n_{b_2} =0$, and setting $d=1$, we obtain the following series expansion of $\hat{\CI}_{\eref{T211211}}$ in $x$:
\be
\scalebox{0.9}{$
\begin{split}
&\hat{\CI}_{\eref{T211211}} ( \{ v_1, 0\}, \{ v_2, 0 \} , \{a, 0 \} , \{(b_1,b_2) , (0,0) \}, d=1 ) \\
& = 1+x \left(\frac{b_1}{b_2}+\frac{b_2}{b_1}+v_1+\frac{1}{v_1}+4\right) \\
&\qquad + x^{\frac{3}{2}} \left(\frac{a}{b_1}+\frac{a}{b_2}+\frac{b_1}{a}+\frac{b_2}{a}+v_1 v_2+v_2+\frac{1}{v_2}+\frac{1}{v_1 v_2}\right) \\
& \qquad +x^2 \Big(\frac{b_1 v_1}{b_2}+\frac{b_1}{b_2 v_1}+\frac{b_2}{b_1 v_1}+\frac{b_2 v_1}{b_1} \\
& \qquad  \qquad \quad +\frac{b_1^2}{b_2^2}+\frac{2 b_1}{b_2}+\frac{b_2^2}{b_1^2}+\frac{2 b_2}{b_1}+v_1^2+2 v_1+\frac{2}{v_1}+\frac{1}{v_1^2}+2\Big) + \ldots~.
\end{split} $}
\ee

Since the $T^{[2,1^2]}_{[2,1^2]}(SU(4))$ theory is self-mirror, the Higgs and Coulomb branch symmetries are equal, each of which is $\left(\frac{U(2) \times U(1)}{U(1)}\right)$.  We rewrite $\hat{\CI}_{\eref{T211211}}$ in such a way that the fugacities and the correponding background fluxes of such symmetries appear on equal footing.  For this purpose, we make the following reparametrisation:
\be
\begin{array}{lllll}
v_1 = w_1 w_2^{-1}~, &\qquad v_2 = w_2~, & \qquad b_1 = a f_1~, & \qquad b_2 = a f_2 \\
n_{v_1} = n_{w_1} - n_{w_2} ~, & \qquad n_{v_2} = n_{w_2} ~, & \qquad n_{b_1} = n_a + n_{f_1}~, & \qquad n_{b_2} = n_a + n_{f_2}~.
\end{array} 
\ee
Let us also define
\bes{
&\CI_{\eref{T211211}} (\{ \vec{w}, \vec{n_w} \} , \{ \vec{f}, \vec{n_f} \},  \{a, n_a \},d  ) \\
&:= \hat{\CI}_{\eref{T211211}} ( \{ w_1 w_2^{-1}, n_{w_1} - n_{w_2} \}, \{ w_2, n_{w_2} \} , \{a, n_a \} , \\
&  \qquad \quad \{(a f_1,a f_2) , (n_a + n_{f_1},n_a + n_{f_2}) \},d )~.
}
The function $\CI_{\eref{T211211}}$ has the following properties:
\bes{
&\CI_{\eref{T211211}} (\{ \vec{w}, \vec{n_w} \} , \{ \vec{f}, \vec{n_f} \},  \{a, n_a \}, d  ) \\
&= w_1^{n_a} a^{n_{w_1}} \CI_{\eref{T211211}} (\{ \vec{w}, \vec{n_w} \} , \{ \vec{f}, \vec{n_f} \},  \{1, 0 \} , d ) ~.
}
and
\bes{ \label{propCIexch}
&\CI_{\eref{T211211}} (\{ \vec{w}, \vec{n_w} \} , \{ \vec{f}, \vec{n_f} \},  \{a, n_a \} ,d )  \\
&= \left( \frac{w_1^{n_a} a^{n_{w_1}}}{f_1^{n_a} a^{n_{f_1}}}  \right) \CI_{\eref{T211211}} ( \{ \vec{f}, \vec{n_f} \}, \{ \vec{w}, \vec{n_w} \},  \{a, n_a \}, d^{-1}  ) ~.
}
If we define
\bes{ \label{defIhat}
& \hat{I}_{\eref{T211211}} (\{ \vec{w}, \vec{n_w} \} , \{ \vec{f}, \vec{n_f} \}, \{a, n_a\} ,d )\\
& :=  {\blue f_1^{n_a} a^{n_{f_1}} } \times \CI_{\eref{T211211}} (\{ \vec{w}, \vec{n_w} \} , \{ \vec{f}, \vec{n_f} \},  \{a, n_a \},d  )~,
}
then the identity \eref{propCIexch} implies that , the index $\hat{I}_{\eref{T211211}}$ satisfies the following condition
\bes{
\hat{I}_{\eref{T211211}} (\{ \vec{w}, \vec{n_w} \} , \{ \vec{f}, \vec{n_f} \}, \{a, n_a\} ,d ) = \hat{I}_{\eref{T211211}} (\{ \vec{f}, \vec{n_f} \}, \{ \vec{w}, \vec{n_w} \} , \{a, n_a\} ,d^{-1}) ~.
}  
Note that the prefactor indicated in blue in \eref{defIhat} indicates a mixed Chern--Simons term, similarly to the $T(U(N))$ theory\footnote{The importance of contact terms for the $T^\sigma_\rho[SU(N)]$ theory at the level of the $S^3_b$ partition function was noticed in \cite{Hwang:2020wpd}, see for example equation (2.56) of that reference for the case of $\sigma=[2,1^2]$ and $\rho=[1^4]$.}.

For simplicity, in the main text, we focus on the case $\{ a, n_a \}=\{ 1,0 \}$ and define
\be
I_{\eref{T211211}} (\{ \vec{w}, \vec{n_w} \} , \{ \vec{f}, \vec{n_f} \},d  ) := \hat{I}_{\eref{T211211}} (\{ \vec{w}, \vec{n_w} \} , \{ \vec{f}, \vec{n_f} \}, \{1,0\} ,d ) ~,
\ee
and so it satisfies the following property:
\be
I_{\eref{T211211}} (\{ \vec{w}, \vec{n_w} \} , \{ \vec{f}, \vec{n_f} \}, d  ) = I_{\eref{T211211}} (\{ \vec{f}, \vec{n_f} \},\{ \vec{w}, \vec{n_w} \}, d^{-1}   )~.
\ee
The series expansion of $I_{\eref{T211211}}$ in $x$ when $\vec{n_w} = \vec{n_f} = (0,0)$ is as follows:
\be \label{indflux0}
\scalebox{0.9}{$
\begin{split}
& I_{\eref{T211211}} (\{ \vec{w}, \vec{0} \} , \{ \vec{f}, \vec{0} \}, d) \\
&=1+ x \left[ d^{-2} \left( \frac{f_1}{f_2}+\frac{f_2}{f_1} +2\right)+d^2 \left( \frac{w_2}{w_1}+\frac{w_1}{w_2} +2\right) \right]+ \\
& \quad + x^{3/2} \left[ d^{-3} \left( f_1+f_2+\frac{1}{f_2}+\frac{1}{f_1} \right)+ d^3 \left( w_1+w_2+\frac{1}{w_1}+\frac{1}{w_2} \right) \right] \\
& \quad + x^2 \Big[\frac{f_1 w_2}{f_2 w_1}+\frac{f_1 w_1}{f_2 w_2}+\frac{f_2 w_2}{f_1 w_1}+\frac{f_2 w_1}{f_1 w_2}+d^{-4} \Big( \frac{f_1^2}{f_2^2} +\frac{f_2^2}{f_1^2}+\frac{2 f_1}{f_2}+\frac{2 f_2}{f_1}+3 \Big) \\
& \qquad \qquad +d^4 \Big(\frac{w_1^2}{w_2^2} + \frac{w_2^2}{w_1^2}+\frac{2 w_1}{w_2}+\frac{2 w_2}{w_1}+3\Big) -4 \Big] +\ldots~.
\end{split}$
}
\ee
Setting  $\{ a, n_a \}=\{ 1,0 \}$ amounts to modding out the $U(1)$ factor in the numerator of the symmetry $\frac{U(2) \times U(1)}{U(1)}$ by the $U(1)$ in the denominator; the result is then identified with the $U(2)$ symmetry for the Higgs or the Coulomb branch.  
%Observe that \eref{indflux0} is invariant under the interchange of $\vec f$ and $\vec w$, together with $d \rightarrow d^{-1}$, as required.

It is convenient to rewrite the index \eref{indflux0} by setting
\be
w_1 = b u~, \quad w_2 = b u^{-1}~,\quad  f_1 = q h~, \quad  f_2 = q h^{-1}
\ee
so that
\bes{
&I_{\eref{T211211}} (\{ (b u, bu^{-1}) , \vec{0} \} , \{ (q h, q h^{-1}) , \vec{0} \}, d) \\
&= 1+ x \left[d^{2} \left( 1+ \chi^{SU(2)}_{[2]} (u) \right) +d^{-2} \left( 1+ \chi^{SU(2)}_{[2]} (h) \right)  \right] \\
& \quad +  x^{\frac{3}{2}} \left[d^{3} (b+b^{-1}) \chi^{SU(2)}_{[1]} (u) +d^{-3} (q+q^{-1}) \chi^{SU(2)}_{[1]} (h)  \right] \\
& \quad + x^2 \Big[ d^{4} \left(  1+ \chi^{SU(2)}_{[2]} (u) + \chi^{SU(2)}_{[4]} (u)  \right) + d^{-4} \left(  1+ \chi^{SU(2)}_{[2]} (h) + \chi^{SU(2)}_{[4]} (h)  \right)  \\
& \qquad  \quad  + \chi^{SU(2)}_{[2]} (u) \chi^{SU(2)}_{[2]} (h) {\blue -  \left( \chi^{SU(2)}_{[2]} (h)+1 \right) -  \left( \chi^{SU(2)}_{[2]} (u)+1\right)} {\brown -1} \Big] \\
& \quad + \ldots
}
where the blue terms denote the contribution of the $U(2) \times U(2)$ global symmetry of the theory and the brown term $-1$ denotes the contribution of the $U(1)_d$ axial symmetry.

\subsection{$S$-fold theories with the $T^{[2,1^2]}_{[2,1^2]}(SU(4))$ building block} \label{app:SfoldT211nflv}
We now examine the index of theory \eref{U2kzeroflv}  formed by gauging the diagonal subgroup of the Higgs and Coulomb branch symmetries of $T^{[2,1^2]}_{[2,1^2]}(SU(4))$:
\be \label{indexU2knoflv}
\scalebox{0.9}{$
\begin{split}
 \CI_{ \eref{U2kzeroflv}} (k; \{w, n \})&=\sum_{m_1, m_2 \in \BZ} \frac{1}{2!}  \left[ \prod_{j=1}^2 \oint \frac{d z_j}{2\pi i z_j} \, w^{m_j} z_j^{k m_j+n} \right] Z_{\text{vec}; \, U(2)} (\{(z_1,z_2), (m_1,m_2) \} )\times  \\
&\quad I_{\eref{T211211}}( \{ (z_1,z_2), (m_1, m_2) \}, \{ (z^{-1}_1,z^{-1}_2), (-m_1, -m_2)\} ,d=1 )~,
\end{split}$}
\ee
where $\omega$ is the topological symmetry and the contribution $Z_{\text{vec}; \, U(2)}$ of the $U(2)$ vector multiplet is given by \eref{Zvec}.  Here $n$ is the background magnetic flux for the topological symmetry which we turn off (\ie ~ by setting $n=0$) in the main text.  Due to the effective superpotential \eref{effW}, the axial symmetry $U(1)_d$ is broken and so we set $d=1$ in the above expression.

Similarly to the case of $T(U(N))$, we can couple $n$ flavours of the fundamental hypermultiplets to the $U(2)$ gauge group of theory \eref{U2kzeroflv}.  This results in theory \eref{U2koneflv}, whose index is  
\bes{ \label{indexU2knflv}
&\CI_{ \eref{U2kzeroflv}} (n, k; w, \{h, \vec{n_h} \}) \\
&=\sum_{m_1, m_2 \in \BZ} \frac{1}{2!}  \left[ \prod_{j=1}^2 \oint \frac{d z_j}{2\pi i z_j} \, w^{m_j} z_j^{k m_j} \right] Z_{\text{vec}; \, U(2)} (\{(z_1,z_2), (m_1,m_2) \} )\times  \\
&\quad I_{\eref{T211211}}( \{ (z_1,z_2), (m_1, m_2) \}, \{ (z^{-1}_1,z^{-1}_2), (-m_1, -m_2)\} ,d=1 ) \times \\
& \quad Z_{\text{fund}} (\{(z_1,z_2), (m_1, m_2) \}\{ (h_1, h_2, \ldots, h_n), (n_{h_1}, n_{h_2}, \ldots, n_{h_n}) \})~.
}
where the contribution $Z_{\text{fund}}$ of the fundamental hypermultiplet is given by \eref{Zfund}.  We also turn off the background magnetic flux for the topological symmetry in the above expression.  In the main text, we also set the background fluxes for the flavour symmetries to zero, $\vec{n_h}=0$, and use the fugacity map:
\bes{
h_1= q f_1, \quad h_2= q f_2 f_1^{-1}~, \quad h_3 = q f_3 f_2^{-1}~, \quad \ldots~, \quad h_n = q f_{n-1}^{-1}~,
}
where $f_1, \ldots, f_n$ are the fugacities of the $SU(n)$ flavour symmetry and $q$ is the fugacity for the $U(1)$ flavour symmetry.

\section{$S$-fold theories with $T(U(1))$: $U(1)_{k-2}$ gauge theory} \label{app:TU1}
In this section, we briefly review $S$-fold theories with the $T(U(1))$ building block.  Although it turns out that these theories are simply ordinary 3d $\CN=3$ Chern--Simons matter theories\footnote{In fact, the pure $S$-fold theories (\ie~ those without hypermultiplet matter) of this type were considered in \cite{Ganor:2014pha, Ganor:2019nnv}. These are simply pure abelian Chern--Simons theories with several $U(1)$ gauge groups, with mixed Chern--Simons couplings between them.}, they are useful for comparing and contrasting with those constructed using the $T(U(N))$ theory with $N >1$.  

The $T(U(1))$ theory is an almost trivial theory with a recipe for coupling external abelian vector multiplets containing gauge fields $A_1$ and $A_2$ \cite{Gaiotto:2008ak}. Such a coupling is the supersymmetric completion of the following Chern--Simons term:
\be \label{CSTU1}
-\frac{1}{2\pi} \int A_1 \wedge dA_2~.
\ee
In an $S$-fold theory, the $U(N)\times U(N)$ symmetry of the $T(U(N))$ theory is commonly gauged, say with a Chern-Simons level $k$.  For $N=1$, the term \eref{CSTU1} gives rise to a Chern--Simons level $-2$ to the $U(1)$ gauge group.  After combining with the Chern-Simons level $k$, we see that the $S$-fold theory in question is nothing but the $U(1)_{k-2}$ gauge theory.  

From the perspective of the index, the mixed Chern--Simons term in $T(U(N))$ contributes $u_N^{n_{h_1} +\cdots + n_{h_N}} (h_1 \cdots h_N)^{n_{u_N}}$, where $(u_1, \ldots, u_N; n_{u_1}, \ldots, n_{u_N})$ are the $U(N)$ topological fugacities and the associated background fluxes and $(h_1, \ldots, h_N; n_{h_1}, \ldots, n_{h_N})$ are the $U(N)$ flavour fugacities and the associated background fluxes.  When both $U(N)$ are commonly gauged, we set $h_i = z_i$, $u_i = z^{-1}_i$, $n_{h_i} =m_i$, $n_{u_i} = -m_i$, for $i=1,\ldots, N$, where $z_i$ are the gauge fugacities and $m_i$ are the corresponding gauge fluxes. This results in $(z_1\cdots z_{N})^{-m_N} z_N^{-m_1-\ldots-m_N}$.  In the case of $N=1$, this is simply $z_1^{-2m_1}$, which is the contribution of the $U(1)$ gauge group with Chern--Simons level $-2$.  Together with the term $z_1^{km_1}$ due to Chern--Simons level $k$ of the $U(1)$ gauge group, we have $z_1^{(k-2) m_1}$, which is the contribution of the $U(1)$ gauge group with Chern--Simons level $k-2$, as expected.

The superpotential for the 3d $\CN=3$ $U(1)_{k-2}$ pure gauge theory is 
\bes{
W = - \frac{k-2}{4 \pi}\varphi^2~.
}
For $k\neq 2$, $\varphi$ can be integrated out, and we are left with a topological field theory. For $k=2$, we have the theory of a free $\CN=4$ abelian vector multiplet.

We can also couple $n$ flavours of hypermultiplets to this theory and obtain the 3d $\CN=3$ $U(1)_{k-2}$ gauge theory with $n$ flavours, whose superpotential is
\bes{
W=- \frac{k-2}{4 \pi} \varphi^2+ \tQ^i \varphi Q_i~,
\label{superpotu1k-2}
}
with $i=1,\ldots, n$.  Note that, for $k=2$, this is in fact the 3d $\CN=4$ $U(1)$ gauge theory with $n$ flavours.   

\subsubsection*{The case of $n\geq 3$}

Let us focus on the case of $n\geq 3$ for the moment. The index of this theory, for $n\geq 3$, is
\bes{ \label{indexUkm}
k=2: \quad & 1+ x {\blue \left(1+ \chi^{SU(n)}_{[1,0,\ldots,0,1]} (\vec f) \right)}+ x^2 \Big[ \chi^{SU(n)}_{[2,0,\ldots,0,2]} (\vec f)   - {\blue \left(1+ \chi^{SU(n)}_{[1,0,\ldots,0,1]} (\vec f) \right)} \Big] \\
&\qquad + \ldots+ (\omega + \omega^{-1})x^{\frac{n}{2}} + \ldots  \\
k\neq 2: \quad & 1+ x {\blue \left(1+ \chi^{SU(n)}_{[1,0,\ldots,0,1]} (\vec f) \right)}+ x^2 \Big[ \chi^{SU(n)}_{[2,0,\ldots,0,2]} (\vec f)   - {\blue \left(1+ \chi^{SU(n)}_{[1,0,\ldots,0,1]} (\vec f) \right)} \Big] \\
&\qquad + \ldots \\
}
We remark that the crucial difference between the cases of $k=2$ and $k \neq 2$ are the terms $(\omega + \omega^{-1})x^{\frac{n}{2}}$ due to the presence of the gauge invariant monopole operators $X_\pm$ with $R$-charge $\frac{n}{2}$.   For $n=3,\,4$, these monopole operators contribute with the terms at order $x^\frac{3}{2}$ and $x^2$ respectively.  For $n \geq 5$, the index up to order $x^2$ of these cases are equal.   Despite this equality, we emphasise that the operators in the cases of $k=2$ and $k \neq 2$ are different.  We will shortly describe these in detail.

For $k=2$, the term $\tr (\varphi^2)$ in \eqref{superpotu1k-2} is absent and the $F$-terms are 
\bes{
\tQ^i \varphi =0~, \qquad \varphi Q_i =0~, \qquad \tQ^i Q_i =0~.
}
Due to the last equality, the mesons $M^i_j = \tQ^i Q_j$ satisfy
\bes{ \label{mesonscond}
M^i_i=0~, \qquad (M^2)^i_j = M^i_k M^k_j = 0~.
}
Moreover, we have
\bes{ \label{phiMzero}
\varphi M^i_j =0~.
}
The operators with $R$-charge $1$ are
\bes{
\varphi~, \qquad  M^i_j
}
contributing $1+ \chi^{SU(n)}_{[1,0,\ldots,0,1]} (\vec f)$ at order $x$.  The operators at order $x^2$ that contribute $\chi^{SU(n)}_{[2,0,\ldots,0,2]} (\vec f)$ are
\bes{
M^i_j M^k_l
}
satisfying \eref{mesonscond}.  There is, however, another marginal operator, namely
\bes{
\varphi^2~.
}
The order $x^2$ of the index in the first line of \eref{indexUkm} should be rewritten as
\bes{
\ldots+ x^2 \Big[ 1+\chi^{SU(n)}_{[2,0,\ldots,0,2]} (\vec f)  {- \blue \left(1+ \chi^{SU(n)}_{[1,0,\ldots,0,1]} (\vec f) \right)} {\brown -1} \Big] +\ldots
}
where the contribution from the $\CN=3$ extra SUSY-current is highlighted in brown\footnote{From the perspective of the $\CN=2$ index, this $-1$ can be viewed as the contribution of the axial symmetry, denoted by $U(1)_d$ in the main text, under which $\varphi$ carries charge $+2$ and each of $Q_i, \tQ^j$ carries charge $-1$. Note that this symmetry is broken when $k\neq 2$.}.  Due to the presence of this current, the corresponding IR SCFT has $\CN=4$ supersymmetry, as expected. 

Let us now assume that $k\neq 2$.  The $F$-terms are
\bes{
 \varphi Q_i = 0~, \qquad \tQ^i \varphi=0~, \qquad \varphi = \frac{2 \pi}{k-2} \tQ^i Q_i~.
}
The meson matrix $M^i_j= \tQ^i Q_j$ thus satisfies the conditions
\bes{ \label{condphiM}
\varphi M^i_j =0~, \qquad \varphi = \frac{2\pi}{k-2} M^i_i~.
}
Note that $\varphi$ can be integrated out using the last equality, after which the effective superpotential is
\bes{ 
W_{\text{eff}} = \frac{\pi}{k-2} (\tQ^i Q_i)^2 =  \frac{\pi}{k-2} (M^i_i)^2 ~.
}
Multiplying $M^j_k$ to both sides of the second equation of \eref{condphiM} and using the first equation of \eref{condphiM}, we obtain
\bes{ \label{condMM1}
 (M^i_i) M^j_k=0~.
}
Contracting the indices $j$ and $k$, we see that $M^i_i$ is nilpotent:
\bes{ \label{condMM2}
(M^i_i)^2=0~.
}
The operators with $R$-charge $1$ are
\bes{
M^i_i~, \qquad \hat{M}^i_j := M^i_j - \frac{1}{n} (M^k_k) \delta^i_j~.
}
Using the identity
\bes{
(\hat{M}^2)^i_j &= (M^2)^i_j - \frac{2}{n} (M^k_k) M^i_j +\frac{1}{n^2} (M^k_k)^2 \delta^i_j~,
}
and the conditions \eref{condMM1} and \eref{condMM2}, we obtain
\bes{ \label{condMhatMhat}
(\hat{M}^2)^i_j &= (M^2)^i_j  = \tQ^i Q_k \tQ^k Q_j = (M^k_k) M^i_j \overset{\eref{condMM1}}{=}0~.
}
Thus, the marginal operators are
\bes{
\hat{M}^i_j  \, \hat{M}^k_l
}
satisfying \eref{condMhatMhat}.  These contribute the term $\chi^{SU(n)}_{[2,0,\ldots,0,2]} (\vec f)$ at order $x^2$ in the index.  In this case, we do not see the presence of an extra SUSY-current.  The corresponding IR SCFT thus has $\CN=3$ supersymmetry.

\subsubsection*{The case of $n=2$}
The case of $k=2$ is simply the 3d $\CN=4$ $U(1)$ gauge theory with $2$ flavours or the $T(SU(2))$ theory, whose index is
\bes{
&1+ x \left( {\blue \chi^{SU(2)}_{[2]}(\omega) + \chi^{SU(2)}_{[2]}(f)}   \right) +  x^2 \Big[\Big(\chi^{SU(2)}_{[4]}(\omega) + \chi^{SU(2)}_{[4]}(f)   \\
& \quad  - {\blue \left( \chi^{SU(2)}_{[2]}(\omega) + \chi^{SU(2)}_{[2]}(f)   \right) } {\brown -1} \Big]  +\ldots~,
}
where we set the topological fugacity $w$ to $w=\omega^2$.  The operators with $R$-charge $1$ are $M^i_j$, satisfying \eref{mesonscond}, together with
\bes{
C = \begin{pmatrix} \varphi & X_+ \\ X_- & -\varphi \end{pmatrix}~,
}
satisfying $(C^2)^{i'}_{j'} = C^{i'}_{k'} C^{k'}_{j'} =0$.  Due to \eref{phiMzero}, we also have
\bes{
C^{i'}_{j'} M^i_j = 0~.
}
The marginal operators are
\bes{
C^{i'}_{j'} C^{k'}_{l'} ~, \qquad M^{i}_{j} M^{k}_{l}~. 
}
The contribution of the $\CN=3$ extra SUSY-current is highlighted above in brown.

The index for the case of $k\neq 2$ is simply \eref{indexUkm} with $n=2$:
\bes{ \label{indexUkm2}
1+ x {\blue \left(1+ \chi^{SU(2)}_{[2]} (\vec f) \right)}+ x^2 \Big[ \chi^{SU(2)}_{[4]} (\vec f)   - {\blue \left(1+ \chi^{SU(2)}_{[2]} (\vec f) \right)} \Big]~.
}
The operators with $R$-charges up to $2$ are as described previously.

\subsubsection*{The case of $n=1$}
For $k=2$, we have the 3d $\CN=4$ $U(1)$ gauge theory with $1$ flavour, which flows to the theory of a free hypermultiplet.  

For $k\neq 2$, the operator with $R$-charge $1$ is $M$, satisfying $M^2=0$ due to \eref{condMM2}.  There is no marginal operator in this case. The indices are
\bes{
k \neq 1, \,  2, \, 3: \quad &  1+{\blue 1}x {\blue-1}x^2 +2x^3 +\ldots \\
k=1: \quad & 1+{\blue 1}x+ ({\blue -1} {\brown - \omega q^{-1} - \omega^{-1} q}) x^2 + (2+ \omega q^{-1} + \omega^{-1} q) x^3 +\ldots \\
k=3: \quad & 1+{\blue 1}x+ ({\blue -1} {\brown - \omega q - \omega^{-1} q^{-1}}) x^2 + (2+ \omega q + \omega^{-1} q^{-1}) x^3 +\ldots
}
For $k \neq 1, \,  2, \, 3$, we don't see the presence of an extra SUSY-current, and so we conclude that the theory has $\CN=3$ supersymmetry.  On the other hand, for $k=1, 3$, where the theory is simply the $U(1)_{\pm 1}$ gauge theory with $1$ flavours, we found two $\CN=3$ extra SUSY-currents, and so we conclude that the theory has enhanced $\CN=5$ supersymmetry, as proposed in \cite{Garozzo:2019ejm}.  From the perspective of the $\CN=2$ index, the negative terms at order $x^2$ correspond to the conserved current, which indicates that the theory has an $SU(2) \cong Spin(3)$ global symmetry. This is a commutant of the $Spin(2)$ $R$-symmetry of $\CN=2$ supersymmetry in the $Spin(5)$ $R$-symmetry of $\CN=5$ supersymmetry.

\section{Monopole operators in some 3d $\CN=4$ gauge theories} \label{app:3dN4}
In this section, we analyse the Coulomb branch operators of two 3d $\CN=4$ gauge theories, namely the $U(N)$ gauge theory (with $N=2, \, 3$) with one adjoint and one fundamental hypermultiplets and the $U(2)$ gauge theory with four flavours, using the indices and Coulomb branch Hilbert series.   The aim is to write down explicitly the Coulomb branch operators with $R$-charges up to $2$ and their relations.  These turn out to be extremely useful in drawing an analogy with operators in the $S$-fold theories discussed in the main text.

\subsection{$U(2)$ and $U(3)$ gauge theories with one adjoint and one fundamental hypermultiplets}
Let us first consider the $U(2)$ gauge group.
The index of this theory is
\bes{ \label{U2oneadjonefund}
& 1+ x^{\frac{1}{2}} (d [1]_w + d^{-1} [1]_c) + x  (2 d^2 [2]_w +2 [1]_w [1]_c + 2 d^{-2} [2]_c)  \\
& +x^{\frac{3}{2}} \Big[ d^3 (2[3]_w + [1]_w)  + 3d [2]_w [1]_c + 3d^{-1} [2]_c [1]_w + d^{-3} (2[3]_c + [1]_c) \Big] \\
& + x^2 \Big[ d^4( 3 [4]_w + [2]_w+1) +4 d^2 [3]_w [1]_c + (d \rightarrow d^{-1}, w \leftrightarrow c)+5 [2]_w [2]_c  \\
& \qquad \qquad- [2]_c-[2]_w -2 \Big] + \ldots \\
}
The terms at order $x^{\frac{1}{2}}$ indicate that the theory contains two free hypermultiplets, and so the above expression can be rewritten as
\bes{ \label{U2oneadjonefund1}
&\CI_{\text{free}}(x; c d^{-1}) \, \CI_{\text{free}} (x; c^{-1} d^{-1})\, \CI_{\text{free}}(x; w d)\,  \CI_{\text{free}} (x; w^{-1} d)  \\
&  \times \Big[ 1+ x \left( d^2 [2]_w + [1]_w [1]_c + d^{-2} [2]_c  \right)  + x^2 \Big( d^4 [4]_w + d^2 [3]_w [1]_c + \\ 
& \quad  +d^{-4} [4]_c + d^{-2}  [3]_c [1]_w +[2]_w[2]_c \\
& \quad  -d^2  [1]_w[1]_c  -d^{-2} [1]_w[1]_c   - [2]_w - [2]_c -1  \Big)+\ldots \Big]
}
where $\CI_{\text{free}}(x; \omega) $ is defined in \eref{freeindex}.  In fact, this index can be rewritten in terms of characters of $SU(4)$ representations as
\bes{ \label{U2oneadjonefund2}
&\CI_{\text{free}}(x; c d^{-1}) \, \CI_{\text{free}} (x; c^{-1} d^{-1})\, \CI_{\text{free}}(x; w d)\,  \CI_{\text{free}} (x; w^{-1} d)  \\
&  \times \Big[ 1+  [2,0,0] x +  \Big( [4,0,0] -[1,0,1]  \Big) x^2 + \ldots \Big]~,
}
where we have used the following decompositions of representations of $SU(4)$ into $SU(2)_w \times SU(2)_c \times U(1)_d$:
\bes{
[2,0,0] \quad &\longrightarrow \quad [2;0]_{+2} + [1;1]_{0} +[ 0;2]_{-2} \\
[4,0,0] \quad &\longrightarrow \quad [4;0]_{+4}+ [3;1]_{+2} +[2;2]_{0}+ [1;3]_{-2} +[0;4]_{-4} \\
[1,0,1] \quad &\longrightarrow \quad [1;1]_{+2} + [2;0]_{0}+ [0;0]_{0}  +[0;2]_{0} + [1;1]_{-2}~.
}

Let us discuss \eref{U2oneadjonefund1} from the perspective of the $\CN=3$ index, in which case we have to set $d=1$.  The index can then be rewritten in terms of characters of $USp(4) \cong Spin(5)$ representations as follows:  
\bes{
&\CI_{\text{free}}(x; c) \, \CI_{\text{free}} (x; c^{-1})\, \CI_{\text{free}}(x; w)\,  \CI_{\text{free}} (x; w^{-1})  \\
& \times \Big[ 1+ {\blue [0,2] }x + x^2 \left( [0,4] {\blue -  [0,2]} {\brown - [1,0]}  \right)+\ldots \Big]~.
}
The $\CN=3$ flavour current is in the adjoint representation $[0,2]$ of $Spin(5)$.  We indicate its contribution to the index in blue. The brown negative term at order $x^2$ in \eref{U2oneadjonefund1} implies that there are five extra SUSY conserved currents in the vector representation $[1,0]$ of $Spin(5)$.  We thus conclude that the interacting SCFT part of this theory has $\CN = 3+5=8$ enhanced supersymmetry, in agreement with \cite[Section 5.1]{Kapustin:2010xq}.  Indeed, the symmetry $Spin(5)$ is the commutant of the $\CN=3$ $R$-symmetry $Spin(3)$ in the $\CN=8$ $R$-symmetry $Spin(8)$.  Another way to see this is to view \eref{U2oneadjonefund2} as an $\CN=2$ index, in which the $SU(4) \cong Spin(6)$ global symmetry is manifest.  This is actually the commutant of the $\CN=2$ $R$-symmetry $Spin(2)$ in $Spin(8)$, which is the $R$-symmetry of an $\CN=8$ SCFT.

We remark that, in \eref{U2oneadjonefund}, we include the contribution from the free hypermultiplets.  In particular they contribute negative terms $-(d [1]_w + d^{-1} [1]_c)$ at order $x^{3/2}$ and $-([2]_w + d^2[1]_w[1]_c+ d^{-2}[1]_w[1]_c + [2]_c+2)$ at order $x^2$; see \eref{freeindex}.  These can combine with the contribution of the interacting SCFT part and cancel that of the operators constructed from products with the aforementioned free fields.

We denote the monopole operator with flux $(m,n)$ by $X_{(m,n)}$, which carries topological charge $m+n$ and $R$-charge $\frac{1}{2}(|m|+|n|)$.   Note that one can always use the Weyl symmetry of $U(2)$ to arrange the flux into the form $m \geq n > -\infty$.   As in the main text, we use the following shorthand notations below:
\bes{
X_{\pm} :=X_{(\pm 1,0)}~, \qquad  X_{++} :=X_{(1, 1)}~, \qquad  X_{--} :=X_{(-1, -1)}~.
}

In the following analysis we focus on the Coulomb branch operators. Up to order $x^2$, these correspond to the terms with the highest power of $d$ in \eref{U2oneadjonefund}.  Another convenient way is to compute a quantity that counts such operators, known as Coulomb branch Hilbert series, which can be regarded as a limit of the index (see (3.41) of \cite{Razamat:2014pta}).  For the theory in question, the Hilbert series is computed in section 4.1 of \cite{Cremonesi:2013lqa}:
\bes{ \label{HSU2oneadjonefund}
& \sum_{m \geq n > -\infty} x^{\frac{1}{2}(|m|+|n|)} P_{U(2)} (x; m,n) w^{m+n} \\
 &=\PE \left[ x^{\frac{1}{2}} [1]_w + x [2]_w - x^2  \right] \\
&= 1+ x^{\frac{1}{2}} [1]_w  +2 x [2]_w +x^{\frac{3}{2}} (2[3]_w + [1]_w) + x^2 ( 3 [4]_w + [2]_w+1)  +\ldots~,
}
with 
\be
P_{U(2)} (x; m,n) = \begin{cases} (1-x)^{-2}~, & \qquad m\neq n \\ (1-x)^{-1}(1-x^2)^{-1}~, & \qquad m =n \end{cases}
\ee
The second line of \eref{HSU2oneadjonefund} indicates that the Coulomb branch is isomorphic to $\BC^2 \times (\BC^2/\BZ_2)$.

The Coulomb branch operators that carry $R$-charge $1/2$ are the monopole operators with fluxes $(\pm1, 0)$
\be \label{freehyper}
[1]_w: \qquad X_{+}~, \quad X_{-}
\ee 
They parametrise the $\BC^2$ factor of the Coulomb branch and decouple as a free hypermultiplet.  These correspond to the term $x^{\frac{1}{2}} [1]_w$ inside the $\PE$ in \eref{HSU2oneadjonefund}.

The Coulomb branch operators with $R$-charge $1$ are
\be \label{Rcharge1U2w1adj1flv}
\begin{array}{lll}
~[2]_w: \qquad X_{++}~,& \qquad (\tr \varphi)~, &\qquad X_{--} \\
~[2]_w: \qquad X_{+}^2~ ,& \qquad X_{+} X_{-}~, & \qquad X_{-}^2 ~.
\end{array}
\ee
It should be noted that $X_{+} X_{-} = X_{(1,0)} X_{(-1,0)} = X_{(1,0)} X_{(0,-1)}$ is not subject to any relation and is an independent operator; it can be identified with the monopole operator with flux $(1,-1)$.  The quantities in the first line are generators of the Coulomb branch, corresponding to the term $x [2]_w$ inside the $\PE$ in \eref{HSU2oneadjonefund}. 

The Coulomb branch operators with $R$-charge $3/2$ are
\be\label{R3halfU21adj1flv}
\begin{array}{llll}
~[3]_w: \quad X_{+}^3~, & \quad X_{+}^2 X_{-}~, & \quad X_{+} X_{-}^2~, & \quad X_{-}^3 \\
~[3]_w: \quad X_{++} X_{+}~, & \quad X_{++} X_{-}~,& \quad X_{--} X_{+}~,  & \quad X_{--} X_{-} \\
~[1]_w: \quad X_{+} (\tr \varphi) ~, & \quad X_{-} (\tr \varphi)~.
\end{array}
\ee

The Coulomb branch operators with $R$-charge $2$ are
\be \label{margU21adj1flv}
\scalebox{0.9}{$
\begin{array}{lllll}
~[4]_w: \quad X_{+}^4~, & \quad X_{+}^3 X_{-}~, & \quad X_{+}^2 X_{-}^2~, &   \\
&\quad X_{+} X_{-}^3~, &\quad  X_{-}^4 & & \\
~[4]_w: \quad X_{++}^2~, & \quad X_{++} (\tr \varphi)~, & \quad X_{++} X_{--}=(\tr \varphi)^2 ~,& \\
& \quad X_{--} (\tr \varphi)~, & \quad X_{--}^2  & & \\
~[4]_w: \quad  X_{++} X_{+}^2~, & \quad X_{++} (X_{+} X_{-})~, & \quad X_{++} X_{-}^2 = X_{+}^2 X_{--}\\
& \quad X_{--} (X_{+} X_{-})~, & \quad X_{--} X_{-}^2 & & \\
~[2]_w: \quad X_{+}^2 (\tr \varphi)~, & \quad  X_{+} X_{-} (\tr \varphi)~, & \quad X_{-}^2 (\tr \varphi) \\
~[0]_w: \quad \tr (\varphi^2) 
\end{array}$}
\ee
where the relation
\be \label{relmonU2w1flv1adj}
X_{++} X_{--}=(\tr \varphi)^2
\ee
is the defining equation of the factor $\BC^2/\BZ_2$ of the Coulomb branch.  Notice that the left hand side $X_{++} X_{--} = X_{(1,1)} X_{(-1,-1)}$ occupies the point $(0,0)$ on the magnetic lattice and so as the right hand side.  This relation corresponds to the term $-x^2$ inside the $\PE$ in \eref{HSU2oneadjonefund}.  Moreover, the relation 
\be
X_{++} X_{-}^2 = X_{+}^2 X_{--}
\ee
follows from the fact that the monopole operators on the left and right hand sides of the equation occupy the same point $(1,-1)$ in the magnetic lattice.

In the case of the $U(3)$ gauge group, the Coulomb branch Hilbert series reads
\bes{ \label{U3w1adj1fund}
&\PE \left[ x^{\frac{1}{2}} [1]_w + x [2]_w + x^{\frac{3}{2}}  [3]_w - x^{\frac{5}{2}} [1]_w - x^3 [2]_w +\ldots  \right] \\
&= 1+ x^{\frac{1}{2}} [1]_w  +2 x [2]_w +x^{\frac{3}{2}} (3[3]_w + [1]_w) + x^2 ( 4 [4]_w +2 [2]_w+2)  +\ldots~.
}
The notations need to be slightly modified as follows:
\bes{
X_{\pm} :=X_{(\pm 1,0,0)}~, \qquad  X_{\pm \pm} :=X_{\pm (1, 1,0)}~, \qquad  X_{\pm \pm \pm} :=X_{\pm (1, 1, 1)}~.
}
As we can see from the above Hilbert series, the generators of the Coulomb branch are the same as for $N=2$, except that there are additional ones with $R$-charge $3/2$ in the representation $[3]_w$:
\bes{
[3]_w:  \quad  X_{+++}~,  \quad X_{+; (0,1)}~,  \quad X_{-;(0,1)} ~, \quad X_{---}~.
}
The dressed monopole operators $X_{\pm; (0,1)}$ are as discussed in (5.4) of \cite{Cremonesi:2013lqa}:
\be
X_{\pm; (r,s)} := X_{(\pm 1,0,0); (r,s)} = (\pm 1, 0,0) \phi_1^r (\phi_2^s+ \phi_3^s) + \text{permutations}~,
\ee
where along the Coulomb branch $\varphi$ can be diagonalised as $\diag(\phi_1, \phi_2, \phi_3)$.

\subsection{$U(2)$ gauge theory with four flavours of fundamental hypermultiplets}
The index of this theory reads
\bes{
&1+ x \left(d^2[2]_w + d^{-2} [1,0,1]_{\vec f} \right) + x^2 \Big[ d^4 ([4]_w + [2]_w + 1) + [2]_w [1,0,1]_{\vec f} \\
& \qquad + d^{-4} ([2,0,2]_{\vec f} + [0,2,0]_{\vec f} )  - [2]_w  - [1,0,1]_{\vec f} -1 \Big]+ \ldots~.
}
The monopole operator $X_{(m,n)}$ with flux $(m,n)$ carries the topological charge $m+n$ and $R$-charge $2(|m|+|n|)-|m-n|$.  The Coulomb branch operators are captured by the highest powers of $d$ at each order of $x$ in the index.  The information about the Coulomb branch chiral ring is contained in the Hilbert series, which was discussed in (5.6) of \cite{Cremonesi:2013lqa}:
\bes{ \label{HSU2w4flv}
& \sum_{m \geq n > -\infty} x^{2(|m|+|n|)-|m-n|} P_{U(2)} (x; m,n) w^{2(m+n)} \\
 &=\PE \left[ x[2]_w + x^2[2]_w - x^3 -x^4  \right] \\
&= 1+ x [2]_w + x^2  ([4]_w + [2]_w + 1) +\ldots~.
}

The Coulomb branch operators with $R$-charge $1$ are
\be
[2]_w: \qquad X_{(1,0)}~, \quad (\tr \varphi)~, \quad X_{(-1,0)}~.
\ee
These correspond to the term $x[2]_w$ in the $\PE$ in \eref{HSU2w4flv}.

The Coulomb branch operators with $R$-charge $2$ are
\bes{ \label{Rcharge2U24flv}
[4]_w: &\quad X^2_{(1,0)}~, \quad  X_{(1,0)} (\tr \varphi)~, \quad X_{(1,0)} X_{(-1,0)}~, \quad  X_{(-1,0)} (\tr \varphi)~, \quad X^2_{(-1,0)} \\
[2]_w: &\quad X_{(1,0); (0,1)} ~, \quad \tr(\varphi^2) ~, \qquad  X_{(-1,0); (0,1)} \\
[0]_w: &\quad (\tr \varphi)^2
}
The second line contains the dressed monopole operators, as discussed in (5.4) of \cite{Cremonesi:2013lqa}:
\be \label{dressedmonopoles}
X_{(\pm 1,0); (r,s)} = (\pm 1, 0) \phi_1^r \phi_2^s+ (0, \pm 1) \phi_2^r \phi_1^s~,
\ee
where along the Coulomb branch $\varphi$ can be diagonalised as $\diag(\phi_1, \phi_2)$.  The quantities in the second line correspond to the term $x^2 [2]_w$ inside the $\PE$ in \eref{HSU2w4flv}.  The quantities in the first and third lines of \eref{Rcharge2U24flv} correspond to the symmetric product $\Sym^2 [2] = [4]+[0]$.

In order to understand the relations at order $x^3$ and $x^4$, as indicated by the Hilbert series \eref{HSU2w4flv}, it is convenient to define the following traceless matrices, containing the generators of the Coulomb branch:
\be
\mathcal{X}_1 := \begin{pmatrix} \tr \varphi & X_{(1,0)}\\ X_{(-1,0)} & - \tr \varphi \end{pmatrix}~, \qquad
\mathcal{X}_2 := \begin{pmatrix} \tr(\varphi^2) & X_{(1,0); (0,1)}\\ X_{(-1,0); (0,1)} & - \tr (\varphi^2) \end{pmatrix}~,
\ee
each of which transforms in the adjoint representation of $SU(2)$.  Similarly to (4.19) and (4.20) of \cite{Hanany:2011db}, the relations at order $x^3$ and $x^4$ can be written respectively as
\bes{
x^3: \quad &\tr( \mathcal{X}_1 \mathcal{X}_2) = 0 \\
 &\quad \Leftrightarrow \quad  X_{(1,0)}X_{(-1,0); (0,1)} + X_{(-1,0)}X_{(1,0); (0,1)} + 2(\tr \varphi)\tr(\varphi^2) =0~, \\
x^4: \quad &\tr(\mathcal{X}_2^2)+ \alpha (\tr \mathcal{X}^2_1)^2 =0 \\
&\quad \Leftrightarrow \quad  X_{(1,0); (0,1)} X_{(-1,0); (0,1)} + [ \tr(\varphi^2)]^2 \\
& \qquad \qquad\qquad  + 2\alpha [X_{(1,0)} X_{(-1,0)}+ (\tr \varphi)^2]^2 =0~,
}
where $\alpha$ is a non-zero constant, which can be absorbed by a redefinition of $\mathcal{X}_1$ or $\mathcal{X}_2$.

\section{Consequences of the $F$-term equations \eref{Ftermsmatter}} \label{app:Fterms}
In this appendix, we discuss consequences of the $F$-term equations \eref{Ftermsmatter} on gauge invariant quantities.
It is convenient to define
\be \label{defmuQ}
M^i_j := \tilde{Q}^i_a  Q^a_j~, \qquad (\mu_Q)^a_b = \tilde{Q}^i_b  Q^a_i
\ee
so that we have
\be
M^i_i = \tr \mu_Q~.
\label{Mii}
\ee
It then follows that
\be \label{QQQ}
\scalebox{1}{$
\begin{split}
Q^b_j M^j_i &\overset{\eref{defmuQ}}{=} Q^a_i (\mu_Q)^b_a \overset{\eref{Ftermsmatter}}{=}   Q^a_i \left( \frac{k}{2\pi} \varphi -\mu_C-\mu_H \right)^b_{~a} \overset{\eref{Ftermsmatter}}{=} - Q^a_i (\mu_C+\mu_H)^b_{~a}~, \\
\tilde{Q}^j_b M^i_j &\overset{\eref{defmuQ}}{=} \tilde{Q}^i_a (\mu_Q)^a_b \overset{\eref{Ftermsmatter}}{=}   \tilde{Q}^i_a\left( \frac{k}{2\pi} \varphi  - \mu_C -\mu_H \right)^a_{~b} \overset{\eref{Ftermsmatter}}{=} - \tilde{Q}^i_a(\mu_C+\mu_H)^a_{~b}~,
\end{split}$}
\ee
or, equivalently,
\be \label{Qsummu}
Q^a_j \left[ (\mu_H+\mu_C)^b_a \delta^j_i + M^j_i \delta^b_{a}  \right]=0~, \qquad \tilde{Q}^j_a  \left[ (\mu_H+\mu_C)^a_b \delta^i_j +M^i_j \delta^a_{b} \right] =0~.
\ee
Multiplying $\tilde{Q}^k_b$ to both sides of the first equation in \eref{Qsummu}, we obtain 
\be  \label{MtrM}
M^k_j M^j_i = (M^2)^k_i= -(\mu_H+\mu_C)^b_a \tQ^k_b Q^a_i~.
\ee
Contracting the indices $k$ and $i$, we obtain
\bes{ \label{trM2}
(M^2)^l_l = \tr (\mu_Q^2)= -(\mu_H+\mu_C)^a_b \tQ^l_a Q^b_l=  -\tr \left[  ( \mu_H+\mu_C) \mu_Q \right]~. 
}
%We have the following matrix equations:
%\be \label{relmuQ2}
%(\mu_Q^2)^a_c \overset{\eref{defmuQ}}{=} M^j_i ( \tQ^i_c Q^a_j)  \overset{\eref{muQrel2}}{=} -( \mu_H+\mu_C) \mu_Q ~.
%\ee
%Substituting this into \eref{Weff1}, we obtain the effective superpotential
%\be \label{Weffoneflv}
%W_{\text{eff}} = \frac{\pi}{k} \tr(\mu_H+\mu_C)(\mu_H+\mu_C+\mu_Q) ~.
%\ee}

For $n \geq2$, it is convenient to define 
\be
\hat{M}^i_j = M^i_j - \frac{1}{n} (M^k_k) \delta^i_j = M^i_j - \frac{1}{n} (\tr \mu_Q) \delta^i_j ~.
\ee
It satisfies the following identifies:
\bes{\label{idenM2}
(\hat{M}^2)^i_j &= (M^2)^i_j - \frac{2}{n} (M^k_k) M^i_j +\frac{1}{n^2} (M^k_k)^2 \delta^i_j~, \\
(\hat{M}^2)^i_i  &= (M^2)^j_j -\frac{1}{n} (M^k_k)^2~.
}
In the special case of $n=2$, due to the Hamilton--Cayley theorem\footnote{For a $2 \times 2$ matrix $A$, it satisfies $A^2 -(\tr A)A+\frac{1}{2}\left[ (\tr A)^2 - \tr(A^2) \right] \BU_{2 \times 2} =0$.}, we also have
\be \label{specialn2}
(\hat{M}^2)^i_j = \frac{1}{2}  (\hat{M}^2)^k_k \, \delta^i_j~, \qquad \text{for $n=2$}~.
\ee
Using \eqref{Mii}, \eqref{MtrM}, \eref{trM2} and \eref{idenM2}, we obtain
\bes{ \label{hatM2}
(\hat{M}^2)^i_j &=  -(\mu_H+\mu_C)^b_a \tQ^i_b Q^a_j -\frac{2}{n}  \hat{M}^i_j (\tr \mu_Q) -\frac{1}{n^2} (\tr\mu_Q)^2 \delta^i_j~, \\
(\hat{M}^2)^i_i &=  -\tr \left[  ( \mu_H+\mu_C) \mu_Q \right] -\frac{1}{n} (\tr\mu_Q)^2~. 
}
It is also convenient to define
\be \label{defMhat}
 (\underline{\hat{M}^2})^i_j := (\hat{M}^2)^i_j  - \frac{1}{n} (\hat{M}^2)^k_k \delta^i_j ~.
\ee
Then, from \eref{hatM2}, we have
\bes{ \label{trlessMhat2}
 (\underline{\hat{M}^2})^i_j  &= -(\mu_H+\mu_C)^a_b   \tQ^i_a Q^b_j  + \frac{1}{n} \tr(\mu_H\mu_Q+\mu_C \mu_Q) \delta^i_j -\frac{2}{n} \hat{M}^i_j (\tr \mu_Q) \\
&= - (\CA_H)^i_j -(\CA_C)^i_j -\frac{2}{n} \hat{M}^i_j (\tr \mu_Q)~, \\
}
where we define
\bes{ \label{defCAHC}
 (\CA_H)^i_j &:= (\mu_H)^a_b   \tQ^i_a Q^b_j -\frac{1}{n}\tr (\mu_H \mu_Q) \delta^i_j~, \\  
 (\CA_C)^i_j &:= (\mu_C)^a_b  \tQ^i_a Q^b_j -\frac{1}{n}\tr (\mu_C \mu_Q) \delta^i_j~.
}
Using \eref{specialn2}, we also have
\be\label{trlessMhat2vanishn2}
 (\underline{\hat{M}^2})^i_j   = 0~, \quad \text{for $n=2$}~,
\ee
and so it follows from \eref{trlessMhat2} that
\bes{ \label{relAHACMn2}
 (\CA_H)^i_j+  (\CA_C)^i_j = - \hat{M}^i_j (\tr \mu_Q) = - \hat{M}^i_j  (M^k_k)~, \quad \text{for $n=2$}~.
}

\bibliographystyle{ytphys}
\bibliography{ref}
\end{document}